\documentclass[acmsmall]{acmart}
\usepackage{xcolor}
\usepackage{placeins}
\AtBeginDocument{%
  }

\acmJournal{TOSEM}


\usepackage{adjustbox}
\usepackage{algorithm}
\usepackage{algorithmicx}
\usepackage{algpseudocode}
\usepackage{amsmath}
\usepackage{array}
\usepackage{balance}
\usepackage{booktabs}
\usepackage[skip=1pt,labelfont=bf]{caption}
\usepackage{calc}
\usepackage{calligra}
\usepackage{color}
\usepackage{colortbl}
\usepackage{courier}
\usepackage{csvsimple}
\usepackage{enumitem}
\usepackage{fancybox}
\usepackage{fontenc}
\usepackage{fontawesome5}
\usepackage{graphicx}
\usepackage{listings}
\usepackage{longtable}
\usepackage{lscape}
\usepackage{makecell}
\usepackage{marvosym}
\usepackage{moreverb}
\usepackage{multicol}
\usepackage{multirow}
\usepackage{pifont}
\usepackage{pgfplots}
\usepackage[figuresright]{rotating}
\usepackage{setspace}
\usepackage{siunitx}
\usepackage{soul}
\usepackage{subcaption}
\usepackage{tablefootnote}
\usepackage{tabularx}
\usepackage[most]{tcolorbox}
\usepackage{threeparttable}
\usepackage{tikz}
\usepackage[normalem]{ulem}
\usepackage{url}
\usepackage{wasysym}
\usepackage{xspace}
\usepackage{xcolor}

\definecolor{aprcolor}{RGB}{55,126,184}

\definecolor{vulncolor}{RGB}{228,112,28}

\colorlet{bothcolor}{aprcolor!50!vulncolor}

\newcommand{\aprsys}[1]{\textcolor{aprcolor}{#1}}
\newcommand{\vulnsys}[1]{\textcolor{vulncolor}{#1}}
\newcommand{\bothsys}[1]{\textcolor{bothcolor}{#1}}

\usepgfplotslibrary{statistics}
\usetikzlibrary{matrix, shapes.geometric, arrows}
\usetikzlibrary{shapes, arrows, positioning}

\algnewcommand\algorithmicforeach{\textbf{for each}}
\algdef{S}[FOR]{ForEach}[1]{\algorithmicforeach\ #1\ \algorithmicdo}

\newcolumntype{L}[1]{>{\raggedright\let\newline\\\arraybackslash\hspace{0pt}}m{#1}}
\newcolumntype{C}[1]{>{\centering\let\newline\\\arraybackslash\hspace{0pt}}m{#1}}
\newcolumntype{R}[1]{>{\raggedleft\let\newline\\\arraybackslash\hspace{0pt}}m{#1}}

\definecolor{codegreen}{rgb}{0,0.6,0}
\definecolor{codered}{rgb}{1,0,0}
\definecolor{codegray}{rgb}{0.5,0.5,0.5}
\definecolor{codepurple}{rgb}{0.58,0,0.82}
\definecolor{backcolour}{rgb}{0.95,0.95,0.92}
\definecolor{lightgray}{gray}{0.9}

 { \newcommand{\mynote}[2]{
      \fbox{\bfseries\sffamily\scriptsize#1}
        {\small$\blacktriangleright$\textsf{\emph{#2}}$\blacktriangleleft$}}}
        { \newcommand{\mynote}[2]{}}
        
\definecolor{DarkOrange}{rgb}{0.8,0.3,0.0}
\definecolor{DarkCyan}{rgb}{0.0, 0.55, 0.55}
\definecolor{DarkCyel}{rgb}{1.0, 0.49, 0.0}
\definecolor{yellow-green}{rgb}{0.6, 0.8, 0.2}

\newcolumntype{?}{!{\vrule width 1pt}}

\newcommand{\etal}{\emph{et~al.}\xspace}

\lstdefinelanguage{mymarkdown}{
    morekeywords={*,\#, \#\#, \#\#\#},
    sensitive=false,
    morecomment=[l]{//},
    morestring=[b]",
    commentstyle=\color{codegreen},
    keywordstyle=\color{magenta},
    numberstyle=\tiny\color{codegray},
    stringstyle=\color{codepurple},
    basicstyle=\small,
    breakatwhitespace=false,         
    breaklines=true,
    breakindent=0pt,
    keepspaces=true,                 
    numbers=left,                    
    numbersep=5pt,                  
    showspaces=false,                
    showstringspaces=false,
    showtabs=false,                  
    tabsize=2,
}

\lstdefinestyle{mystyle}{
    commentstyle=\color{codegreen},
    keywordstyle=\color{magenta},
    numberstyle=\small\color{black},
    stringstyle=\color{codepurple},
    basicstyle=\scriptsize\ttfamily,
    breakatwhitespace=false,
    breaklines=true,
    captionpos=b,
    keepspaces=true,
    showspaces=false,
    showstringspaces=false,
    showtabs=false,
    tabsize=2
}

\lstdefinelanguage{diff}{
  morecomment=[f][\color{blue}]{@@},     
  morecomment=[f][\color{red}]-,         
  morecomment=[f][\color{codegreen}]+,       
  morecomment=[f][\color{red}]{---}, 
  morecomment=[f][\color{codegreen}]{+++},
  numberstyle=\tiny\color{codegray},
  numbers=left,                    
  numbersep=5pt,         
}

\setlist{noitemsep} 

\definecolor{darkpastelred}{rgb}{0.76, 0.23, 0.13}
\definecolor{ao(english)}{rgb}{0.0, 0.5, 0.0}

\definecolor{darkpastelred}{rgb}{0.76, 0.23, 0.13}
\definecolor{ao(english)}{rgb}{0.0, 0.5, 0.0}

\newboolean{useblue}
\setboolean{useblue}{false} 

\newcommand{\maybeblue}[1]{%
    \ifthenelse{\boolean{useblue}}%
    {\textcolor{blue}{#1}}%
    {#1}%
}

\colorlet{rqaccent}{blue!60}
\colorlet{rqframe}{blue!35}
\colorlet{rqback}{blue!3}

\newtcolorbox{rqbox}[1]{enhanced,breakable,
  lines before break=4, pad at break*=4pt,
  colback=rqback, colframe=rqframe,
  boxrule=0.5pt, arc=1.2mm,                 
  borderline west={2pt}{0pt}{rqaccent},     
  left=6pt,right=6pt,top=6pt,bottom=6pt,
  fonttitle=\bfseries, title={Answer to #1},
  before skip=6pt, after skip=6pt}

\newtcolorbox{overviewbox}[1][Overview]{%
  enhanced,
  breakable,
  colback=gray!2,        
  colframe=gray!40,      
  boxrule=0.4pt,
  arc=1mm,               
  left=6pt,right=6pt,
  top=5pt,bottom=5pt,
  borderline west={2pt}{0pt}{gray!60}, 
  before skip=10pt,
  after skip=10pt,
  coltitle=black,
  fonttitle=\bfseries,
  colbacktitle=gray!8,
  title=#1,
  attach boxed title to top left={yshift=-1pt,xshift=4pt},
  boxed title style={%
    boxrule=0pt,
    colframe=gray!8,
    interior empty,
    right=4pt,
    left=0pt,
    top=1pt,
    bottom=1pt
  }
}
\begin{document}
\newboolean{showcomments}
\setboolean{showcomments}{true}

\definecolor{DarkOrange}{rgb}{0.8,0.3,0.0}
\definecolor{DarkCyan}{rgb}{0.0, 0.55, 0.55}
\definecolor{DarkCyel}{rgb}{1.0, 0.49, 0.0}
\definecolor{yellow-green}{rgb}{0.6, 0.8, 0.2}

\newcommand{\todoc}[2]{{\textcolor{#1} {\textbf{#2}}}}
\newcommand{\todoblue}[1]{\todoc{blue}{\textbf{#1}}}
\newcommand{\todogreen}[1]{\todoc{yellow-green}{\textbf{#1}}}
\newcommand{\todored}[1]{\todoc{red}{\textbf{#1}}}

\newcommand{\bachle}[1]{{\color{red}\textbf{Bach:}} {\todoblue{#1}}}
\newcommand{\lingming}[1]{{\color{red}\textbf{Lingming:}} {\todoblue{#1}}}
\newcommand{\yang}[1]{{\color{red}\textbf{Boyang:}} {\todogreen{#1}}}

\title{A Survey of LLM-based Software Repair: Taxonomies, Design Paradigms, and Applications}
\author{Boyang Yang}
\affiliation{%
  \institution{School of Artificial Intelligence (School of Software), Yanshan University}
  \city{Qinhuangdao}
  \country{China}}
\email{yby@ieee.org}

\author{Zijian Cai}
\affiliation{
  \institution{School of Artificial Intelligence (School of Software), Yanshan University}
  \city{Qinhuangdao}
  \country{China}
}
\email{zijiancai6@gmail.com}

\author{Fengling Liu}
\affiliation{%
  \institution{School of Computer Science, China University of Mining and Technology}
  \city{Xuzhou}
  \country{China}
}
\email{lainring@gmail.com}

\author{Xuan Bach D. Le}
\affiliation{%
  \institution{School of Computing and Information Systems, University of Melbourne}
  \city{Melbourne}
  \country{Australia}
}
\email{bach.le@unimelb.edu.au}

\author{Lingming Zhang}
\affiliation{%
  \institution{Department of Computer Science, University of Illinois Urbana-Champaign}
  \city{Urbana}
  \country{USA}
}
\email{lingming@illinois.edu}

\author{Tegawendé F. Bissyandé}
\affiliation{%
  \institution{SnT, University of Luxembourg}
  \city{Luxembourg}
  \country{Luxembourg}
}
\email{tegawende.bissyande@uni.lu}

\author{Yang Liu}
\affiliation{%
  \institution{School of Computer Science and Engineering, Nanyang Technological University}
  \city{Singapore}
  \country{Singapore}
}
\email{yangliu@ntu.edu.sg}

\author{Haoye Tian}
\affiliation{%
  \institution{Department of Computer Science, Aalto University}
  \city{Espoo}
  \country{Finland}
}
\email{tianhaoyemail@gmail.com}
\thanks{\textsuperscript{*}Corresponding author.}

\renewcommand{\shortauthors}{Yang et al.}

\begin{abstract}
Large language models (LLMs) are reshaping automated program repair. We present a reproducible hierarchical taxonomy that organizes 66 LLM-based repair systems according to where repair capability and control logic principally reside: task-adapted parameters, prompt and context design, designer-specified workflows, or LLM-directed runtime control. Adaptation, generation pattern, runtime control, and auxiliary evidence are preserved as separate coded dimensions. This representation exposes control distinctions hidden by utilization labels and supports cross-paradigm analysis of system design and evaluation evidence. To the best of our knowledge, it is the first publicly available LLM-based software repair survey to operationalize Fine-Tuning, Prompting, Procedural, and Agentic as one corpus-wide primary classification.
Our hierarchy complements prior surveys through one corpus-wide primary decision rule, while the result-level protocol audit bounds which reported results are defensibly comparable.
To analyze how repair systems use benchmarks, we record benchmark variants, metrics, base models, and fault-localization assumptions for each system's primary reported result, identifying protocol-aligned comparison windows and benchmark-family fragments. We clarify paradigm trade-offs in task alignment, deployment cost, controllability, and multi-hunk or cross-file repair. We outline open challenges and research directions. Our artifacts and scripted survey pipeline are publicly available at \url{https://github.com/GLEAM-Lab/ProgramRepair}.
\end{abstract}

\maketitle

\section{Introduction}
\begin{figure}[h]
    \centering
    \includegraphics[width=\linewidth]{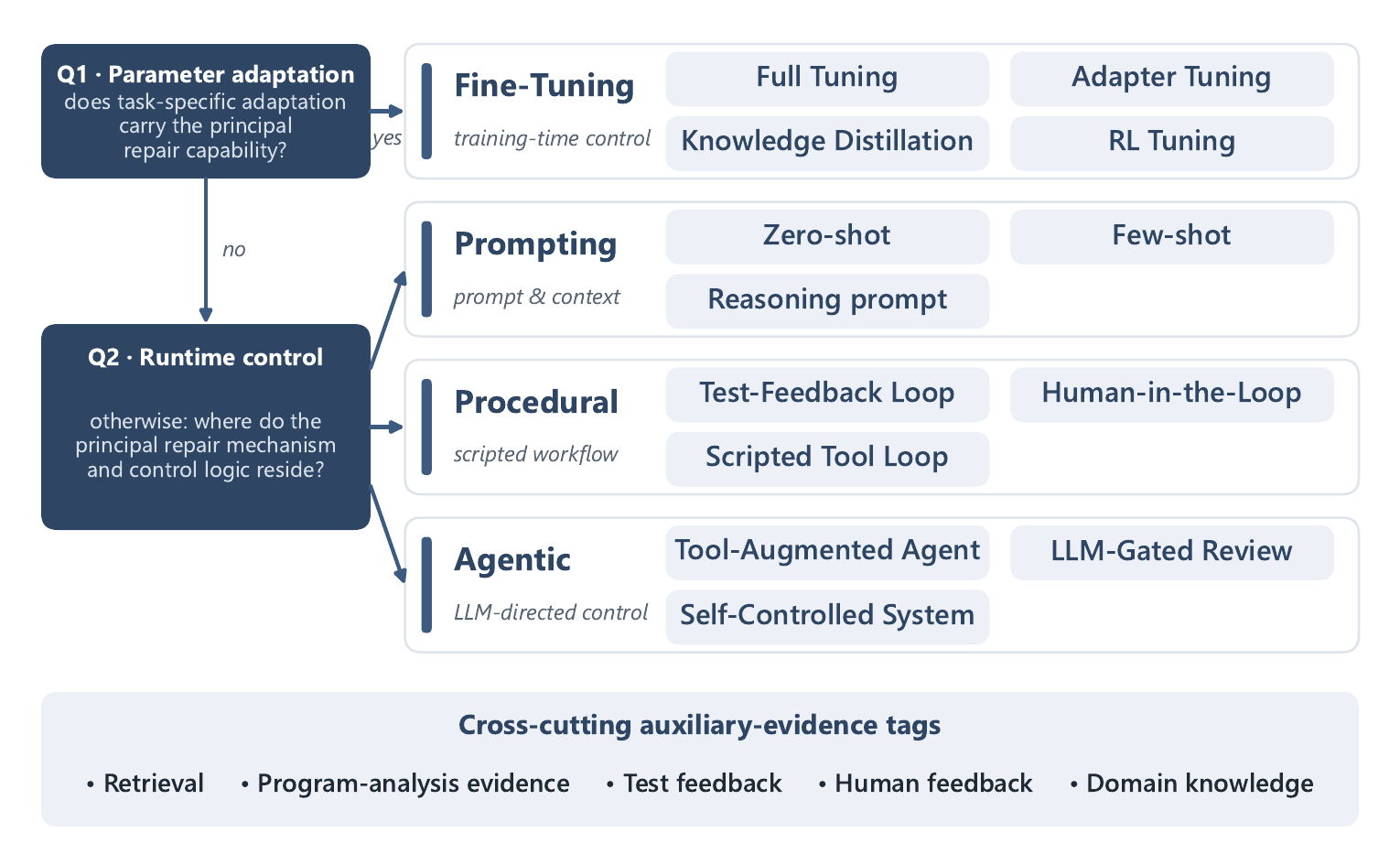}
    \caption{Hierarchical taxonomy of LLM-based software repair. Fine-Tuning applies when task-specific adaptation carries the principal repair capability; otherwise, Prompting, Procedural, and Agentic distinguish prompt and context design, designer-specified workflows, and LLM-directed runtime control. Retrieval, program-analysis evidence, test feedback, human feedback, and domain knowledge are non-exclusive auxiliary-evidence tags.}
    \label{fig:taxonomies}
\end{figure}

LLM-based software repair has become a promising research direction in software engineering~\cite{zhang2024systematic,huang2024evolving,zhang2023critical,fan2023large.shaheen2026ai}. Recent code-focused LLMs, including CodeT5~\cite{wang2021codet5,wang2023codet5+}, Codex~\cite{chen2021evaluatinglargelanguagemodels}, CodeLlama~\cite{roziere2023code}, StarCoder~\cite{li2023starcoder,lozhkov2024starcoder}, Qwen-Coder~\cite{hui2024qwen25codertechnicalreport}, and DeepSeek-Coder~\cite{guo2024deepseek,zhu2024deepseek}, have demonstrated capabilities to understand and generate code, enabling them to fix bugs and vulnerabilities in programs with minimal or without human input. Prior studies report strong repair performance on established benchmarks~\cite{sobania2023analysis,prenner2022can,xia2024automated,yang2024cref,jiang2023impact}. The research frontier has quickly moved from zero-shot prompting on single-function benchmarks~\cite{pearce2023examining,xia2022less} to parameter-adapted~\cite{silva2025repairllama,yangmorepair}, domain-knowledge-guided vulnerability repair~\cite{zhang2025acfix,shao2026pailgen}, or tool-augmented workflows that operate at repository-level and on real vulnerabilities~\cite{xia2025demystifying,yang2025enhancing}, making it timely to introduce a clear taxonomy that relates these emerging paradigms and their augmentation strategies.

This rapid shift has also motivated several surveys. Zhang~\etal provide a critical study of ChatGPT’s repair behavior on a contamination-resistant benchmark and later a systematic review that organizes LLM-based software repair by utilization mode and repair scenario rather than by jointly modeling runtime control authority and parameter adaptation~\cite{zhang2023critical,zhang2024systematic}.
Zhou~\etal provide a security-oriented overview focused on vulnerability detection and repair~\cite{zhou2024large}.
He~\etal focus on LLM-as-judge protocols for patch assessment rather than full repair pipelines~\cite{he2025code}.
Haque~\etal survey parameter-efficient tuning for large code models and only briefly touch on repair tasks~\cite{haque2025systematic}.
Anand~\etal discuss AI-driven software repair at a high level without a taxonomy centered on control paradigms~\cite{anand2024comprehensive}.
Existing surveys organize LLM-based repair by utilization mode, repair scenario, security scope, tuning method, evaluation role, or broad AI-assisted repair, but they do not provide the same combination of a corpus-wide primary-locus rule and a protocol-level comparison audit. The published TOSEM version of Zhang~\etal~\cite{zhang2024systematic} includes Agent and a repository-level Procedural/Agentic split while retaining utilization modes and repair scenarios as its corpus-level organization; our rule instead assigns one corpus-wide primary locus while preserving adaptation, control, and auxiliary evidence as separate coded dimensions.

Taken together, these studies clarify many aspects of modern software repair but leave three gaps. First, none offers a unified taxonomy that treats fine-tuning, prompting, procedural pipelines, and agentic frameworks as first-class paradigms tied to where repair capability and control over the repair loop reside. Second, auxiliary context sources such as retrieval, analysis, tests, human feedback, and domain knowledge are not modeled as separate evidence layers that can combine with any paradigm. Third, prior reviews largely summarize techniques, benchmark datasets, or headline repair scores, but they do not analyze reported repair-system results at the protocol level: the same benchmark name can refer to different variants, base models, pass@k budgets, fault-localization assumptions, or oracle strength, making cross-paper score comparison unreliable.

\begin{figure}[h]
    \centering
    \includegraphics[width=\linewidth]{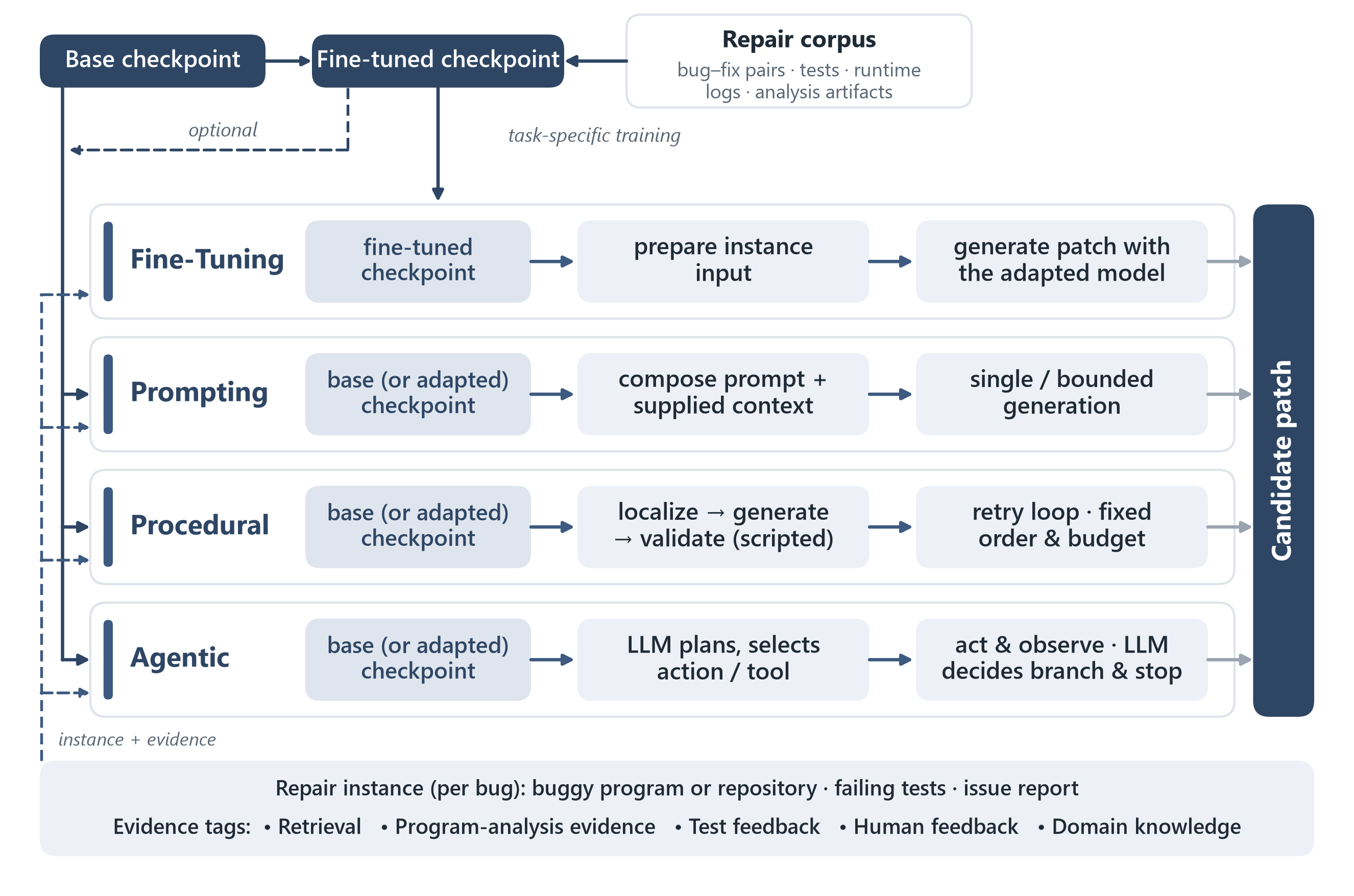}
    \caption{Overall repair flows. Each lane is an alternative route from available evidence and model artifacts to a candidate patch. Retrieval, program-analysis evidence, test feedback, human feedback, and domain knowledge are cross-cutting inputs.}
    \label{fig:taxonomies_relationship}
\end{figure}

To address these gaps, we present a systematic survey of 66 LLM-based software repair systems published or released between January 2022 and April 2026. Rather than grouping systems only by how they invoke an LLM, we apply a hierarchical primary-locus rule: we first identify systems whose principal repair capability depends on task-specific parameter adaptation and then distinguish prompt- and context-centered repair, designer-specified workflows, and LLM-directed runtime control. This yields Fine-Tuning, Prompting, Procedural, and Agentic as four primary paradigms (Figure~\ref{fig:taxonomies}) while preserving auxiliary evidence as separate, non-exclusive fields. Retrieval supplies external knowledge such as relevant code, documentation, or historical fixes~\cite{lewis2021retrievalaugmentedgenerationknowledgeintensivenlp,yang2025enhancing}; program-analysis evidence supplies compiler diagnostics, fault-localization scores, constraints, failing-test logs, or execution traces~\cite{dolcetti2025helpingllmsimprovecode,xiao2025predicatefixrepairingstaticanalysis,li2024hybrid,jain2023staticfixer}. Either source can support any primary paradigm without determining that paradigm.
Figure~\ref{fig:taxonomies_relationship} shows the corresponding repair flows: each lane is an alternative route from available evidence and model artifacts to a candidate patch, and the shared inputs at the bottom can support any lane.

\textbf{Our contributions.}
\begin{enumerate}[label=(\roman*),leftmargin=*,itemsep=2pt]
  \item We provide a reproducible hierarchical taxonomy that explains where repair capability and control logic principally reside while preserving adaptation, generation pattern, runtime control, and auxiliary evidence as separate coded dimensions for hybrid systems. This representation exposes control distinctions hidden by utilization labels and supports cross-paradigm analysis; to the best of our knowledge, it is the first publicly available LLM-based software repair survey to operationalize Fine-Tuning, Prompting, Procedural, and Agentic as one corpus-wide primary classification.

  \item We map 66 systems into this taxonomy and analyze evaluation at the level of reported repair-system results. For each system, we record benchmark variant, pass@k or task-specific metric, base model, and fault-localization assumptions, then separate protocol-aligned comparison windows from same-family but non-comparable benchmark fragments. This result-level view explains when shared benchmark names do or do not support score comparison.
  
  \item We analyze how the paradigms behave in different scenarios, such as function-level benchmarks, repository-level repair, vulnerability patching, educational tutoring, and industrial continuous integration, and derive design guidelines that relate paradigm choices to defect scope, data availability, latency, and compute budgets.
  
  \item We maintain a living, versioned bibliography and release an end-to-end scripted pipeline that reproduces our literature retrieval, screening, and coding process, together with all extracted metadata, so that other researchers can audit our survey and reuse the artifact at \url{https://github.com/GLEAM-Lab/ProgramRepair}.
\end{enumerate}

\section{Related Work}

Earlier surveys either predate LLM-based repair or mention LLMs only briefly, so they do not analyze LLM-centered control structures. Huang~\etal~\cite{huang2023survey} updated the traditional landscape with a four-way taxonomy over search, constraint, template, and learning techniques, but relegated LLMs to a short remark about future potential. Zhang~\etal~\cite{zhang2023survey} presented a thorough synthesis of learning-based software repair, framing the workflow around localization, generation, ranking, and validation; because their literature search concluded before mid-2022, large-scale LLMs such as Codex receive only brief mention, and prompt-driven repair workflows are not yet covered in depth. Winter~\etal~\cite{winter2022let} shifted focus to developer perceptions and usability, without a technical taxonomy. Dikici~\etal~\cite{dikici2025advancements} grouped tools into template, machine-learning, and deep-learning categories, implicitly folding GPT-style systems into the deep-learning bin without analyzing their distinctive traits or their impact on the repair loop.

More recent work concentrates on LLMs and related evaluation issues. Anand~\etal~\cite{anand2024comprehensive} explicitly surveyed LLM-based software repair and code generation but covered only a small set of software repair papers and did not propose a dedicated taxonomy. Haque~\etal~\cite{haque2025systematic} conducted a systematic literature review on parameter-efficient fine-tuning for LLMs, covering software repair but focusing on fine-tuning methods rather than repair workflows. He~\etal~\cite{he2025code} examined the emerging LLM-as-judge paradigm for code evaluation, offering insights into assessing LLM-generated patches while omitting a repair-oriented taxonomy. Zhou~\etal~\cite{zhou2024large} published a review that catalogs work on vulnerability detection and repair and discusses adaptation techniques and dataset usage, but its analysis is limited to security-related bugs, and it treats retrieval-augmented and agentic methods mainly as future directions. Zhang~\etal conducted a literature review focused on LLM-based software repair~\cite{zhang2024systematic}, collecting a broader corpus of LLM-based repair studies and classifying them by LLM utilization modes and repair scenarios, rather than coding runtime control authority, parameter adaptation, auxiliary evidence, and protocol comparability as separate dimensions. Abdollahi~\etal~\cite{abdollahi2026benchmarking} surveyed benchmarks for LLM-based code intelligence across many tasks, emphasizing dataset construction, task coverage, benchmark quality, contamination risk, and evaluation metrics. We cite it as a benchmark context; our analysis is narrower and system-level, asking when published repair-system results can be compared under compatible protocols. Zhang~\etal provide a critical review that uses ChatGPT as a case study on a contamination-resistant benchmark, examining prompt design and data-contamination risks~\cite{zhang2023critical}. Table~\ref{tab:survey_comparison} summarizes these surveys and their coverage of LLM-based techniques.

\begin{table*}[t]
\small
\caption{Recent surveys on automated program repair and their coverage of LLM-based techniques.}
\resizebox{\textwidth}{!}{%
\begin{tabular}{@{}p{3.0cm} p{2.2cm} p{0.9cm} p{2.4cm} p{3.2cm} p{4.4cm}@{}}
\toprule
\textbf{Survey} & \textbf{Venue} & \textbf{Year} & \textbf{LLM Coverage} & \textbf{Classification Scheme} & \textbf{Key Limitations} \\
\midrule
Winter~\etal~\cite{winter2022let} & \emph{IEEE TSE} & 2022 & None & Human-factor perspective & Lacks technical classification; predates LLM era. \\
Huang~\etal~\cite{huang2023survey} & arXiv & 2023 & Mentioned as future work & Search, constraint, template, learning & No category devoted to LLM repair. \\
Zhang~\etal~\cite{zhang2023survey} & \emph{ACM TOSEM} & 2023 & Moderate & Pipeline stages & Excludes non-ML approaches and prompt-based repair. \\
Zhang~\etal~\cite{zhang2023critical} & arXiv & 2023 & Strong for ChatGPT case & ChatGPT-centered critical review and prompt variants & Single-model case study; not a general software repair taxonomy or broad LLM-based software repair survey. \\
Anand~\etal~\cite{anand2024comprehensive} & arXiv & 2024 & Strong & LLM-based software repair and code generation & software repair coverage limited; no fine-grained taxonomy. \\
Zhang~\etal~\cite{zhang2024systematic} & \emph{ACM TOSEM} & 2026 & Strong & LLM utilization modes and repair scenarios & Emphasizes how LLMs are used rather than a primary-locus rule over control and adaptation; limited treatment of retrieval/analysis as cross-cutting evidence layers and system-level protocol-comparability analysis. \\
Abdollahi~\etal~\cite{abdollahi2026benchmarking} & \emph{ACM TOSEM} & 2026 & Benchmark-focused & Code-intelligence benchmark landscape & Covers benchmark datasets across many coding tasks; does not analyze published repair-system results as protocol-compatible or non-compatible comparison windows. \\
Dikici~\etal~\cite{dikici2025advancements} & \emph{KIS} & 2025 & Partial & Template, ML, deep learning & LLM tools subsumed under deep learning. \\
Haque~\etal~\cite{haque2025systematic} & arXiv & 2025 & Moderate & Fine-tuning strategy taxonomy & Focused on fine-tuning methods, not repair workflows. \\
He~\etal~\cite{he2025code} & arXiv & 2025 & Evaluation only & Evaluation role taxonomy & Covers judging, not generation or workflow design. \\
Zhou~\etal~\cite{zhou2024large} & \emph{ACM TOSEM} & 2025 & Strong & Security-oriented vulnerability repair & Focuses only on vulnerabilities; no agentic approaches. \\
\midrule
This work & -- & 2026 & Strong & Control authority and parameter adaptation, with auxiliary evidence tags and protocol-aligned benchmark-result analysis & Focuses on a criteria-bounded 66-system corpus; does not jointly re-run all systems under one benchmark protocol. \\
\bottomrule
\end{tabular}}
\label{tab:survey_comparison}
\end{table*}

Across these studies, three limitations remain. First, most taxonomies either predate LLM-based repair or retain a pipeline- or model-centric view and therefore do not separate parameter adaptation, runtime control authority, and auxiliary evidence as independent dimensions. Second, LLM-based works are omitted in some surveys or folded into broad deep-learning bins, which obscures how foundation models reshape software repair and how external tools, retrieval modules, and analysis components are orchestrated around them. Third, evaluation challenges specific to LLM-based repair, including weak test suites, heterogeneous pass@$k$ settings, differing fault-localization assumptions, and benchmark-family fragmentation, are not aligned at the level of individual repair systems in the comparison surveys listed in Table~\ref{tab:survey_comparison}. We complement these surveys by offering a multidimensional taxonomy centered on parameter adaptation and runtime control authority with explicit auxiliary-evidence tags, by mapping 66 recent systems into this design space together with their defect scope and deployment scenario, and by analyzing repair-system results through protocol-aligned comparison windows rather than treating benchmark names alone as comparable evidence.

The public version history records the chronology of the two classifications. The arXiv v2 of Zhang~\etal~\cite{zhang2024systematic} (12 May 2024) organized systems by fine-tuning, few-shot, and zero-shot utilization strategies, without an Agent category or a Procedural/Agentic split. The first arXiv version of this survey (30 June 2025) already operationalized Fine-Tuning, Prompting, Procedural, and Agentic across the corpus using the four primary loci above. The arXiv v3 of Zhang~\etal (21 October 2025) later added Agent and separately introduced a repository-level Procedural/Agentic distinction. The two surveys thus converged independently on the control distinction and remain complementary: Zhang~\etal retain utilization modes and scenario-specific organization, whereas our rule assigns one corpus-wide primary locus while preserving orthogonal coded dimensions.

\section{Survey Methodology}
\label{sec:method}

We conduct a systematic survey of LLM-based automated program repair. Our goal is to build a high-quality, representative corpus of LLM-based repair systems rather than an exhaustive census of every paper that mentions LLMs and bugs. Following the Goal Question Metric (GQM) approach~\cite{caldiera1994goal}, we first articulate the overall issue, object, viewpoint, and purpose, then derive research questions and analysis procedures. We further use established systematic review and snowballing guidance to structure search, screening, and citation chasing~\cite{keele2007guidelines,wohlin2014guidelines}.

\textbf{Issue.} Fragmented definitions of repair workflows and heterogeneous evaluation protocols make it hard to compare systems, understand their strengths, and accumulate comparable results. \textbf{Object.} We analyze design paradigms and augmentation layers for LLM-based software repair, evaluation protocols, benchmark families and defect scopes, and reported assumptions around correctness and scalability. \textbf{Viewpoint.} We adopt the perspective of researchers and practitioners in software engineering, program analysis, and ML systems. \textbf{Purpose.} We aim to map the design space, characterize how different paradigms and benchmarks are adopted in practice, and systematize evaluation assumptions, thereby highlighting open problems around semantic correctness, repository-scale, and multi-hunk repair. Throughout this paper, we use \emph{LLM-based software repair} as an umbrella term for repair systems whose pipelines centrally invoke a pre-trained neural code model or large language model at inference time.

\subsection{Research Questions}
\begin{itemize}[noitemsep,topsep=0pt,leftmargin=*]

\item \textbf{RQ1}: \textbf{\textit{Which design dimensions characterize LLM-based software repair systems, and how reliably can these dimensions be coded across the surveyed corpus?}} Utilization labels hide where repair capability and runtime control actually reside; answering this question gives readers a reliability-audited vocabulary for placing and comparing hybrid systems without conflating adaptation, orchestration, and auxiliary evidence. If the taxonomy is a primary contribution, it should be explicitly constructed and checked rather than taken for granted. We derive the taxonomy from the 66 representative works through bottom-up coding, separate runtime control from parameter adaptation and evidence augmentation, and report independent coding reliability assessment results for the resulting taxonomy.

\item \textbf{RQ2}: \textbf{\textit{How do model adaptation, runtime orchestration, evidence augmentation, and validation mechanisms combine across repair scopes and deployment scenarios, and what trade-offs do these combinations reveal?}} The answer turns the coded corpus into design guidance: which combinations of adaptation, control, and evidence are used at each repair scope, and the trade-offs each choice incurs in cost, controllability, and context dependence. Existing surveys do not code parameter adaptation, runtime control, and auxiliary evidence as separate dimensions, which makes design trade-offs harder to compare across systems. We map the 66 representative systems to this taxonomy and analyze how different combinations relate to defect scope, deployment setting, controllability, external-tool reliance, and cost.

\item \textbf{RQ3}: \textbf{\textit{Which reported results are actually comparable across common repair benchmarks once differences in benchmark variant, underlying model, pass@$k$, fault-localization assumption, oracle strength, and contamination control are taken into account?}} Because the same benchmark name can hide different protocols, this question determines which published results can be compared at all, and which same-benchmark scores should not be read as a ranking. Headline pass@$k$ and task-specific repair metrics are reported under heterogeneous protocols, which makes cross-paper comparison unreliable. For each system, we construct a benchmark-specific evaluation profile and compare results only within compatible settings, while also identifying recurring risks from weak test suites, missing validation, and possible data leakage.

\item \textbf{RQ4}: \textbf{\textit{What recurrent failure modes and methodological gaps emerge across the literature, and which research directions are directly supported by this evidence?}} Answering it distills the corpus's recurring, evidence-backed limitations into priorities: the evaluation safeguards worth adopting now, and the research directions the evidence actually supports. Rather than proposing generic future work, we synthesize the limitations repeatedly reported in the surveyed systems and connect them to the design and evaluation patterns identified above. This yields an evidence-based research agenda for semantic correctness, repository-scale repair, and multi-hunk repair.

\end{itemize}

\begin{figure}[h]
    \centering
    \includegraphics[width=\linewidth]{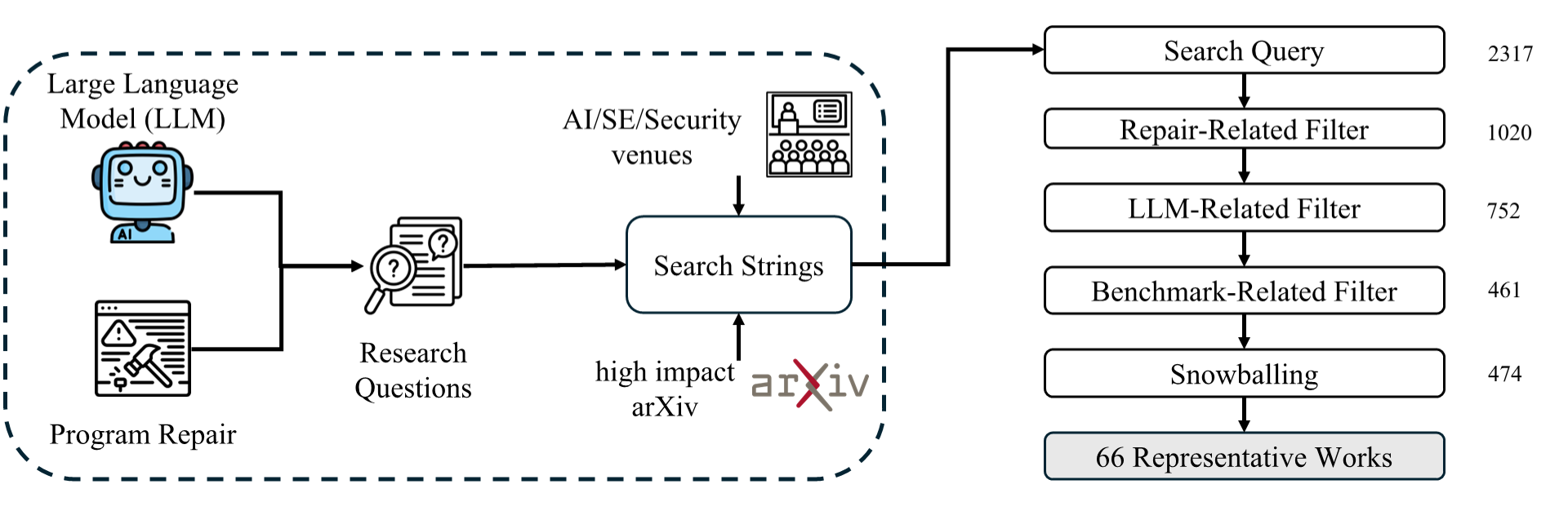}
    \caption{Workflow of literature survey.}
    \label{fig:prisma}
\end{figure}

\subsection{Search Strategy}
We executed structured digital-library queries on \emph{ACM DL}, \emph{IEEE Xplore}, \emph{DBLP}, and \emph{arXiv}. To maximize recall, we first ran a broad query set and then refined it iteratively; the last full query across all sources was run on 30 April 2026. The exact query templates and per-source strings are archived in our replication package. We then applied a year filter ($\ge$2022) to align with the period in which LLM-based software repair became technically and practically feasible. Popular code-targeted LLMs such as Codex were publicly available only in late 2021, so work prior to 2022 falls outside the time frame covered by our search. We also checked the major SE, AI/NLP, and security venue proceedings and journal records via the official proceedings pages and the libraries above. To make the archival search scope explicit and reproducible, we used Category A of the seventh edition of the China Computer Federation (CCF) \emph{Recommended International Academic Conferences and Journals} as one operational reference point.\footnote{CCF, \emph{Recommended International Academic Conferences and Journals}, 7th edition, released 31 March 2026 and updated 9 April 2026: \url{https://www.ccf.org.cn/Academic_Evaluation/By_category/2026-03-31/870181.shtml}.} We selected CCF Category A rather than CORE A because, for this survey, its combined conference-and-journal coverage mapped more directly onto our software-engineering-centered archival target, whereas CORE A is conference-oriented.\footnote{CORE, \emph{ICORE Conference Rankings Portal}: \url{https://portal.core.edu.au/conf-ranks/}.} This choice provides a transparent, reproducible search lens. We then retained the software engineering, AI/NLP, and security venue families directly relevant to LLM-based software repair, including ICSE, TOSEM, NeurIPS, ACL, and USENIX Security. This procedure ensures that influential work from these peer-reviewed venue families enters the raw pool even when it is not in our final representative set.

Our query templates combine (i) core software repair terms, (ii) LLM names or model terms, (iii) benchmark names, and (iv) concrete tool or system names. We formed concrete queries by conjoining at least one software repair term with one or more model, benchmark, or tool keywords. For each data source, we adapted the syntax to its search interface, for example, restricting matches to title, abstract, and keywords where supported. Table~\ref{tab:search-keywords} summarizes the keyword families used in the search. The tool and system names are used as supplementary recall-check and artifact-synchronization terms; they do not by themselves determine inclusion, and every retained record still passes the same eligibility criteria and manual screening protocol.

After collecting the raw results, we applied three rule-based automatic filters, as shown in Figure~\ref{fig:prisma}. A \emph{repair-related} filter retained records whose titles or abstracts mentioned core software repair terms, along with quantitative evaluation metrics such as pass@$k$, accuracy, or fix rate; at this step we also recorded venue information for later use in the eligibility criteria. An \emph{LLM-related} filter then removed records that did not mention any LLM or model keyword from Table~\ref{tab:search-keywords}. A \emph{benchmark-related} filter finally retained records that appeared to report empirical evaluation on a reproducible dataset, either by mentioning at least one established benchmark family or by using phrases that denote a clearly described defect dataset or issue suite. We treat this step as a high-recall prefilter that is subsequently checked during manual screening. To make this filter auditable, our replication package releases the stage counts, the 474-candidate reference-label sheet used for screening checks, the final adjudicated selection sheet, the full 408-record exclusion breakdown, a targeted 29-record exclusion-pattern audit of high-confidence rejected records, and a fixed-seed random audit of 100 records removed by the benchmark-related filter. The random rejected-record audit found 0 potential false negatives and therefore required no corpus change; the audit summary is released in \path{artifact/benchmark_filter_random_audit.csv}. The exact keyword lists and regular expression rules for all filters are documented in \path{artifact/search_keywords_and_filters.md}. To prevent drift between the manuscript and released artifact, \path{artifact/manuscript_artifact_mapping.md} links manuscript-facing screening, taxonomy, benchmark-protocol, evaluation-reliability, and challenge-evidence tables to their released CSV files within the directory \path{artifact}. We complemented this automated filtering with backward citation snowballing and manual full-text screening, as detailed next.

\begin{table}[t]
  \centering
  \caption{Representative keyword families used in our search.}
  \label{tab:search-keywords}
  \tiny
  \begin{tabularx}{.8\linewidth}{lX}
    \toprule
    \textbf{Family} & \textbf{Keywords} \\
    \midrule
    Core software repair terms 
    & \texttt{program repair}, \texttt{automated program repair}, \texttt{automatic program repair},
      \texttt{software repair}, \texttt{code repair}, \texttt{bug fixing}, \texttt{bug repair},
      \texttt{patch generation}, \texttt{patch synthesis}, \texttt{vulnerability repair},
      \texttt{vulnerability patching}, \texttt{security patch}, \texttt{debugging},
      \texttt{bug localization}, \texttt{fault localization} \\[0.3em]
    LLM / model terms
    & \texttt{large language model}, \texttt{code LLM}, \texttt{neural language model},
      \texttt{pre-trained model}, \texttt{foundation model}, \texttt{generative AI},
      \texttt{GPT-2}, \texttt{GPT-3}, \texttt{GPT-3.5}, \texttt{GPT-4}, \texttt{GPT-4o},
      \texttt{Codex}, \texttt{ChatGPT}, \texttt{LLaMA}, \texttt{LLaMA 2}, \texttt{CodeLLaMA},
      \texttt{StarCoder}, \texttt{StarCoder2}, \texttt{CodeGen}, \texttt{CodeGen2},
      \texttt{Mistral}, \texttt{Gemini}, \texttt{PaLM}, \texttt{Qwen}, \texttt{DeepSeek}, \texttt{Phi},
      \texttt{T5}, \texttt{BERT}, \texttt{RoBERTa}, \texttt{CodeBERT}, \texttt{GraphCodeBERT},
      \texttt{PLBART}, \texttt{CodeT5}, \texttt{CodeT5+}, \texttt{UniXcoder}, \texttt{InCoder} \\[0.3em]
    Benchmarks
    & \texttt{Defects4J (v1.x, v2.x)}, \texttt{SWE-bench}, \texttt{SWE-bench Lite},
      \texttt{QuixBugs}, \texttt{ManyBugs}, \texttt{IntroClass}, \texttt{Bugs.jar}, \texttt{Bears},
      \texttt{BugsInPy}, \texttt{CodeFlaws}, \texttt{BugSwarm}, \texttt{CVEfixes},
      \texttt{VulnFix}, \texttt{VJBench}, \texttt{HumanEval}, \texttt{MBPP}, \texttt{DS-1000} \\[0.3em]
    Tools / systems
    & \texttt{AlphaRepair}, \texttt{ChatRepair}, \texttt{ThinkRepair}, \texttt{REx},
      \texttt{ContrastRepair}, \texttt{DRCodePilot}, \texttt{CREF}, \texttt{HULA},
      \texttt{PATCH}, \texttt{KGCompass}, \texttt{Agentless}, \texttt{SWE-agent}, \texttt{SWE-agent M},
      \texttt{SWE-Search}, \texttt{AutoCodeRover}, \texttt{RepairAgent}, \texttt{MAGIS}, \texttt{OpenHands},
      \texttt{LANTERN}, \texttt{VulDebugger}, \texttt{InferFix}, \texttt{TraceFixer}, \texttt{TracePrompt},
      \texttt{KNOD}, \texttt{DistiLRR}, \texttt{NARRepair}, \texttt{TSAPR}, \texttt{SecRepair}, \texttt{AdaPatcher},
      \texttt{D4C}, \texttt{APPATCH}, \texttt{SAN2PATCH}, \texttt{PredicateFix}, \texttt{LLM4CVE},
      \texttt{PailGen}, \texttt{ACFix}, \texttt{Dr.Fix}, \texttt{IntDiagSolver},
      \texttt{RepairLLaMA}, \texttt{MORepair}, \texttt{SWE-RL}, \texttt{Vul-R2}, \texttt{SpecRover},
      \texttt{Abstain and Validate}, \texttt{Learn-by-Interact} \\
    \bottomrule
  \end{tabularx}
\end{table}

\subsection{Eligibility Criteria}

\noindent\textbf{Inclusion criteria.}

A record is retained if it satisfies all of the following:

\begin{enumerate}[label=(\roman*),leftmargin=*]
  \item \textbf{Repair task.} 
  The primary technical contribution targets automated repair or modification of software artifacts (for example, source code, scripts, configuration files, or smart contracts) in response to bugs, test failures, or security vulnerabilities, and includes at least one concrete repair task rather than only detection, localization, or code understanding.

  \item \textbf{Central use of LLMs.}
  The study employs a large language model at \emph{inference time} as a central component of the repair pipeline, such as generating patches, editing code, ranking candidate patches, validating or criticizing patches, or orchestrating multi-step repair workflows, rather than using LLMs only for data synthesis, annotation, or commentary.

  \item \textbf{Empirical evaluation on reproducible benchmarks.}
  The paper reports empirical results on at least one benchmark or dataset whose evaluation setting is documented well, including but not limited to popular benchmarks such as \emph{Defects4J}, \emph{SWE-bench}, and other open-sourced repository-level bug or vulnerability datasets, or carefully described industrial projects. The paper should provide quantitative repair metrics such as the number of correctly fixed bugs, pass@$k$, or patch correctness rate.

  \item \textbf{Publication quality and time frame.}
  Archival works were published between 2022 and April 2026 in one of the retained software engineering, AI/NLP, or security conference or journal families listed below. These families were identified using CCF Category A as the operational lens described above and then screened for direct relevance to LLM-based software repair. CCF listing alone neither guaranteed inclusion nor excluded records found through database search and snowballing; arXiv-only systems were assessed separately under the concrete-system and reproducible-evidence criteria.
  We retained arXiv-only systems as representative records only when they report concrete LLM-based repair pipelines, reproducible benchmark evidence, and no in-scope archival version was found by the search date. Adjacent arXiv/OpenReview records outside this venue/version and representative-record boundary are logged as boundary records rather than inserted one by one into the count-based corpus. This restriction keeps the corpus focused on studies with archival publication records or directly inspectable benchmark evidence within an auditable representative corpus. For retained arXiv-only systems and boundary records, we record the version status and retention rationale in the released artifact.

  \item \textbf{Accessibility, extractability, and basic reporting quality.}
  The full text is available in PDF format and provides sufficient methodological detail to extract key attributes of the repair setting, including LLM family, repair paradigm, benchmarks, and evaluation metrics. When multiple versions of the same work exist, we retain the latest version. During extraction, we also log simple quality indicators, such as whether code or data are publicly released and whether the evaluation setup is described in enough detail to support replication.

\end{enumerate}

\noindent\textbf{Exclusion criteria.}

A record is excluded if it satisfies one of the following:

\begin{enumerate}[label=(\roman*),leftmargin=*]
  \item it is a duplicate of another repair work, for instance, an earlier workshop, technical report, or arXiv version of an included conference or journal paper, in which case we retain only the most complete and latest version;

  \item it does not concern automated software repair, for example, it only addresses bug or vulnerability detection or localization, program understanding, or generic code generation, without any patch generation or patch evaluation component, such as \cite{goyal2026optimizing};

  \item it does not employ a large language model at inference time as a central component to generate, rank, or judge patches, for instance, the LLM is used only for data synthesis, labeling, or commentary, but not in the repair pipeline itself;

  \item it does not report empirical results on at least one benchmark or dataset, or it relies exclusively on proprietary data while omitting sufficient detailed information and experimental settings.
\end{enumerate}

\noindent\textbf{Primary venue scope.}
We use the following conference and journal families as the peer-reviewed archival scope for this survey. The initial families were identified through the CCF Category A lens described above and were then restricted by direct relevance to LLM-based software repair in software engineering, AI/NLP, and security. Together, these families define the survey's operational peer-reviewed archival boundary.
\begin{itemize}[leftmargin=*]
  \item Software engineering venues and journals: ICSE, FSE, ASE, ISSTA, TSE, TOSEM.
  \item AI and NLP venues: NeurIPS, ICML, ICLR, AAAI, ACL.
  \item Security venues: USENIX Security.
\end{itemize}

\subsection{Taxonomy Construction and Reliability Audit}

We constructed the taxonomy through bottom-up coding of the included systems rather than directly importing categories from prior APR surveys. Following established guidance on taxonomy development for software-engineering artifacts~\cite{humbatova2020taxonomy}, two authors first performed an initial calibration pass over representative and boundary papers covering all four primary paradigms and major benchmark families, and recorded recurring design questions: whether the base model is adapted, who controls the outer repair loop, whether runtime generation is single-shot or iterative, which external evidence sources are injected into the repair process, what repair scope is targeted, and which benchmark or deployment scenario is evaluated. We then refined these questions into the taxonomy codebook shown in Table~\ref{tab:taxonomy-codebook}. Here, an evidence tag is a non-exclusive code for a source used by the system, while a scenario tag records its primary evaluation setting. Reasoning prompts, including chain-of-thought-style prompts, are recorded as a prompting attribute; external evidence remains an auxiliary tag.

\begin{table*}[t]
\small
\caption{Taxonomy codebook and reliability checks. The primary taxonomy paradigms used in later tables are derived from these dimensions rather than treated as mutually exclusive descriptions of all system behavior.}
\label{tab:taxonomy-codebook}
\resizebox{\textwidth}{!}{%
\begin{tabular}{p{3.1cm} p{4.0cm} p{5.0cm} p{5.2cm}}
\toprule
\textbf{Coded dimension} & \textbf{Values used in the codebook} & \textbf{Role in the analysis} & \textbf{Reliability evidence} \\
\midrule
Parameter adaptation & none, full tuning, adapter tuning (parameter-efficient fine-tuning, PEFT), knowledge distillation, RL tuning (RLFT) & Distinguishes systems whose main repair result depends on adapting model parameters from systems that keep the base model frozen at inference time. & Included in the final 66-system primary-paradigm reliability audit; independent coders agreed on 65/66 labels, $\kappa=0.98$. \\
Runtime control authority & single-shot human-scripted call, scripted multi-step loop, LLM-selected tool/action, LLM judge, self-controlled agent & Separates prompting, procedural pipelines, and agentic workflows according to who chooses the next repair step. & Control-subtype reliability audit; independent coders agreed on 64/66 labels, $\kappa=0.97$. \\
Generation pattern & single generation, iterative generation, search over candidates & Records whether a system repairs once or conditions later calls on tests, tools, or previous outputs. & Checked against paper algorithms and pipeline figures during extraction. \\
Auxiliary evidence layer & retrieval, static/dynamic analysis, test feedback, human feedback, domain knowledge & Captures cross-cutting context sources without turning retrieval or program-analysis evidence into separate top-level paradigms. & Independent coders had complete raw agreement on retrieval and analysis tags (66/66); test-feedback, human-feedback, and domain-knowledge signals were checked during extraction and adjudication rather than reported as separate kappa-coded tags. \\
Repair scope and scenario & program repair, vulnerability repair, or both; localized benchmark, repository-level issue, vulnerability repair, industrial/practitioner workflow, educational tutoring & Supports RQ2 analysis of where each design is evaluated and prevents benchmark scope from being confused with the design paradigm. & Independent coders reached complete raw agreement on deployment-scenario tags in the 66-system coding reliability assessment. \\
\bottomrule
\end{tabular}}
\end{table*}

The resulting taxonomy therefore records several dimensions for each system. For corpus-level tables, we assign one primary taxonomy paradigm from the parameter-adaptation and runtime-control dimensions, while preserving auxiliary evidence, repair scope, and scenario tags for hybrid behavior. This avoids treating retrieval-enriched or analysis-enriched repair as mutually exclusive top-level buckets: a fine-tuned model, a single-turn prompt, a scripted loop, or an agent can all use retrieved examples, static or dynamic analysis results, test feedback, human feedback, or domain knowledge. Borderline systems were discussed explicitly, and the released coding sheets include evidence locations and rationales for the final assignments. Table~\ref{tab:taxonomy-boundary-cases} shows representative adjudication decisions.

\begin{table*}[t]
\small
\caption{Representative taxonomy boundary-case adjudications. These examples illustrate how hybrid systems are assigned one primary taxonomy paradigm while retaining auxiliary evidence and scenario tags.}
\label{tab:taxonomy-boundary-cases}
\resizebox{\textwidth}{!}{%
\begin{tabular}{p{2.2cm} p{4.4cm} p{4.3cm} p{6.5cm}}
\toprule
\textbf{System} & \textbf{Ambiguity} & \textbf{Coding decision} & \textbf{Rationale} \\
\midrule
SWE-RL & Fine-tuned reinforcement-learning model used with an Agentless-style pipeline & Fine-Tuning / RL Tuning; pipeline evidence retained in protocol notes & The main reported repair capability depends on adapting the model parameters, while the Agentless-style wrapper is recorded as supporting workflow evidence rather than the primary taxonomy paradigm. \\
PATCH & Retrieval and search evidence can be read as either a scripted loop or agentic exploration & Procedural / Scripted Tool Loop with retrieval tag & The evidence-gathering and patch-generation stages are organized by a fixed workflow; retrieval remains an auxiliary evidence source rather than a separate top-level paradigm. \\
TSAPR & Tree search, candidate scoring, and an LLM critic overlap scripted search and judge-based control & Agentic / LLM-Gated Review & The LLM critic gates which candidate patches or search branches survive, so runtime control is partly delegated to an LLM judge. \\
SpecRover & Specification-guided patch review could be treated as post-hoc checking rather than repair control & Agentic / LLM-Gated Review; repository-level scenario & The specification and review model affects candidate acceptance on SWE-bench Lite tasks, so it is coded as judge-controlled repair rather than only as an external metric. \\
NTR & Template guidance and evidence enrichment accompany a trained repair model & Fine-Tuning / Full Tuning with auxiliary evidence & The primary reported result depends on a trained model; templates and evidence guide generation but do not define an outer scripted repair loop. \\
ACFix & Access-control graphs and static checks accompany a GPT-4 repair prompt & Prompting / analysis-enriched & The repair step remains a prompt-based generation step; contract-specific analysis artifacts are coded as auxiliary evidence. \\
\bottomrule
\end{tabular}}
\end{table*}

We assessed taxonomy-coding reliability through an independent coding reliability assessment over the final 66-system corpus. The two independent coders recoded each system with the written decision rules and recorded evidence locations before final reconciliation. They agreed on 65 of 66 primary-paradigm labels (Cohen's $\kappa=0.98$), 64 of 66 control-subtype labels ($\kappa=0.97$), and reached complete raw agreement on retrieval, analysis, and deployment-scenario tags (66/66). The remaining primary-paradigm/control boundary cases were adjudicated against benchmark scope, workflow evidence, and full-text descriptions before the final scenario and taxonomy tables were used. These coding sheets and the final adjudicated labels are released with the artifact.

\subsection{Screening Flow}

Figure~\ref{fig:prisma} shows the screening flow, and Table~\ref{tab:filtering-stats} reports the number of records at each stage, from 2,317 initial hits to 66 representative works. In the first stage, records were filtered fully automatically using the repair-related, LLM-related, and benchmark-related rules described above, without any manual intervention. The benchmark-related filter produced 461 records that mentioned at least one known benchmark or a clearly described defect dataset. This filter is intentionally permissive and serves only to prioritize papers that already signal reproducible evaluation.

We then applied backward citation snowballing from these 461 records, scanning their references for additional LLM-based software repair systems and adding papers that also passed the automatic filters. This step increased the pool to 474 records before manual screening, as shown in Table~\ref{tab:filtering-stats}.

In the second stage, the 474 candidate records were manually assessed against the inclusion and exclusion criteria. For the corpus reported in this paper, two independent coders screened the complete 474-record candidate pool using the same inclusion/exclusion criteria and evidence fields. Before adjudication, their raw include/exclude labels agreed on 467 of 474 records (Cohen's $\kappa=0.94$). We then resolved the seven disagreements through full-text adjudication, covering empirical studies that propose or evaluate key LLM-based repair methods, duplicate versions of the same work, and records selected only in one coding pass. The adjudicated decisions form the final 66-system corpus used in the manuscript. The revised 474-record pool includes four archival-status records added during publication-status synchronization: U-001 (ACFix), U-002 (PailGen), U-003 (IntDiagSolver), and U-004 (Dr.Fix). These additions were treated with the same extraction form as the rest of the corpus and fit within existing taxonomy dimensions and benchmark-protocol groups, rather than introducing a new primary taxonomy paradigm or a protocol-aligned comparison window. The coding guidelines, per-stage record lists, screening templates, raw agreement summary, and final adjudicated selection sheet are included in our replication package. Because our eligibility criteria restrict attention to the listed major archival venue and journal families, selected with reference to CCF Category A and topic relevance, and to retained arXiv systems with recorded benchmark evidence and version-status rationale, we treat these 66 studies as a criteria-bounded corpus of current LLM-based software repair research within this scope rather than a complete survey of all existing systems.

For each included paper, we then extracted a common 15-field form covering the primary taxonomy paradigm, control subtype, auxiliary evidence tags, defect scope, deployment scenario, benchmark family, benchmark name, base model, training dataset, adaptation strategy, metric, reported result, fault-localization assumption, oracle type, and artifact availability. One author populated this form for each retained paper, and a second author checked the taxonomy, scenario, benchmark, model, metric, and assumption fields against the paper and artifact records. Disagreements or missing values were resolved through discussion, with a third author consulted for boundary cases. During extraction, we also recorded basic quality indicators described above. Per-stage record lists, the scripts for fetching and filtering papers from multiple sources, and analysis notebooks that regenerate our tables and figures are provided in our replication package. 

\begin{table}[t]
  \small
  \caption{Screening statistics for the filtering of representative works.}
  \label{tab:filtering-stats}
  \setlength{\tabcolsep}{6pt}
  \begin{tabular}{lrr}
    \toprule
    \textbf{Stage} & \textbf{Records remaining} & \textbf{Notes} \\ \midrule
    Search query hits                      & 2317 & Raw keyword queries \\
    Repair-related filter                  & 1020  & Non-repair topics removed \\
    LLM-related filter                     & 752  & Records without central LLM use removed \\
    Benchmark-related filter               & 461  & Records without benchmarks removed \\
    Snowballing                            & 474  & Backward citation chasing and additions \\
    Representative works                   & 66   & 408 full-text exclusions; 13.9\% retained \\
    \bottomrule
  \end{tabular}
\end{table}

\begin{table}[t]
  \small
  \caption{Full-text exclusion breakdown for the 408 excluded records.}
  \label{tab:exclusion-breakdown}
  \centering
  \begin{tabular}{lr}
    \toprule
    \textbf{Exclusion reason} & \textbf{Records} \\ \midrule
    Not software repair & 141 \\
    LLM not central at inference time & 82 \\
    Benchmark, evaluation, or empirical study only & 65 \\
    No patch-generation or repair pipeline & 58 \\
    Insufficient full-text detail & 43 \\
    Survey or review & 8 \\
    Detection/localization only & 7 \\
    Duplicate or superseded version & 3 \\
    Other boundary exclusion after adjudication & 1 \\
    \bottomrule
  \end{tabular}
\end{table}

\section{Selected Works Overview and Taxonomy Distribution}

\label{sec:corpus-overview}

\subsection{Taxonomy at a Glance}

To make the taxonomy concrete, Table~\ref{tab:taxonomies} presents the four primary taxonomy paradigms in a compact corpus view. The table groups systems by the primary paradigm derived from parameter adaptation and runtime control, and it keeps auxiliary evidence sources, defect scope, and deployment scenario as separate coded fields rather than collapsing them into the group label. We separate systems whose main repair results rely on task-specific adaptation of a base model (\emph{Fine-Tuning}) from those whose principal repair mechanism and control logic reside elsewhere (\emph{Prompting}, \emph{Procedural}, and \emph{Agentic}, distinguished by their control structure). Auxiliary context sources, including retrieval and analysis, are treated as separate tags that can apply to any of the four paradigms. The resulting primary-paradigm label is used for corpus-level aggregation, not as a complete description of hybrid system behavior.

\begin{table}[t]
\small
\caption{Landscape of LLM-based software repair by control subtype. Row groups summarize the primary taxonomy paradigm, while auxiliary evidence sources remain explicit tags rather than top-level buckets. System names in blue denote program repair, orange denote vulnerability repair, and a mixed color denotes systems evaluated on both scopes.}
\label{tab:taxonomies}
\resizebox{\textwidth}{!}{%
\begin{tabular}{@{}p{2.8cm} p{1.8cm} p{1.5cm} p{2.6cm} p{1.0cm} p{7.0cm}@{}}
\toprule
\textbf{Control subtype} & \textbf{Training} & \textbf{Generation} & \textbf{Control} & \textbf{Aux.\ tags} & \textbf{Representative Systems} \\
\midrule
\multicolumn{6}{c}{\textbf{Fine-Tuning}} \\ \midrule
Full Tuning & fine-tuning & single & scripted by human & -- &
  \aprsys{Huang~\etal~(2023)}~\cite{huang2023empirical},
  \aprsys{Jiang~\etal~(2023)}~\cite{jiang2023impact},
  \vulnsys{VulMaster~(2024)}~\cite{zhou2024out},
  \aprsys{RepairCAT~(2024)}~\cite{jiang2024repaircat} \\
Adapter Tuning & fine-tuning & single & scripted by human & -- &
  \aprsys{Li~\etal~(2024)}~\cite{li2024exploring},
  \aprsys{RepairLLaMA~(2025)}~\cite{silva2025repairllama},
  \aprsys{MORepair~(2025)}~\cite{yangmorepair},
  \aprsys{Luo~\etal~(2025)}~\cite{luo2025fine},
  \aprsys{Ruiz~\etal~(2025)}~\cite{ruiz2025art} \\
Knowledge Distillation & fine-tuning & single & scripted by human & -- &
  \aprsys{KNOD~(2023)}~\cite{jiang2023knod},
  \aprsys{NARRepair~(2024)}~\cite{yang2024narrepair},
  \aprsys{DistiLRR~(2025)}~\cite{wong2024distilrr}\\
RL Tuning & fine-tuning & single & scripted by human & -- &
  \aprsys{RePair~(2024)}~\cite{zhao2024repair},
  \vulnsys{SecRepair~(2024)}~\cite{islam2024llm},
  \aprsys{SWE\mbox{-}RL~(2025)}~\cite{wei2025swe},
  \aprsys{AdaPatcher~(2025)}~\cite{dai2025less},
  \vulnsys{Vul\mbox{-}R2~(2025)}~\cite{wen2025vul} \\
Full Tuning & fine-tuning & single & scripted by human & evidence &
  \aprsys{TraceFixer~(2023)}~\cite{bouzenia2023tracefixer},
  \aprsys{InferFix~(2023)}~\cite{jin2023inferfix},
  \aprsys{PyTy~(2024)}~\cite{chow2024pyty},
  \aprsys{NTR~(2025)}~\cite{huang2024template} \\
\midrule
\multicolumn{6}{c}{\textbf{Prompting}} \\ \midrule
Zero-shot & none & single & scripted by human & -- &
  \aprsys{AlphaRepair~(2022)}~\cite{xia2022less},
  \aprsys{Prenner~\etal~(2022)}~\cite{prenner2022can},
  \aprsys{Fan~\etal~(2023)}~\cite{fan2023automated},
  \aprsys{Tian~\etal~(2023)}~\cite{tian2023chatgpt} \\
Few-shot & none & single & scripted by human & -- &
  \aprsys{Xia~\etal~(2023)}~\cite{xia2023automated},
  \aprsys{Gao~\etal~(2023)}~\cite{gao2023makes},
  \aprsys{Ahmed~\etal~(2023)}~\cite{ahmed2023majority},
  \aprsys{CEDAR~(2023)}~\cite{nashid2023retrieval} \\
Zero-/few-shot & none & single & scripted by human & retrieval &
  \aprsys{Ehsani~\etal~(2025)}~\cite{ehsani2025hierarchical},
  \aprsys{RLCE~(2024)}~\cite{chen2024large},
  \aprsys{DSrepair~(2025)}~\cite{ouyang2025knowledge},
  \vulnsys{PailGen~(2026)}~\cite{shao2026pailgen} \\
Zero-/few-shot & none & single & scripted by human & analysis &
  \aprsys{D4C~(2025)}~\cite{xu2025aligning},
  \vulnsys{APPATCH~(2025)}~\cite{nong2025appatch},
  \aprsys{TracePrompt~(2025)}~\cite{haque2025traceprompt},
  \vulnsys{ACFix~(2025)}~\cite{zhang2025acfix},
  \aprsys{Dr.Fix~(2026)}~\cite{zhuo2026apimisuse} \\
\midrule
\multicolumn{6}{c}{\textbf{Procedural}} \\ \midrule
Test-Feedback Loop & none & multiple & scripted by human & test feedback &
  \aprsys{ChatRepair~(2024)}~\cite{xia2024automated},
  \aprsys{ThinkRepair~(2024)}~\cite{yin2024thinkrepair},
  \aprsys{REx~(2024)}~\cite{tang2024code},
  \aprsys{ContrastRepair~(2025)}~\cite{kong2025contrastrepair} \\
Human-in-the-Loop & none & multiple & scripted by human & human feedback &
  \aprsys{CREF~(2024)}~\cite{yang2024cref},
  \aprsys{HULA~(2025)}~\cite{takerngsaksiri2025human},
  \aprsys{DRCodePilot~(2024)}~\cite{zhao2024enhancing} \\
Scripted Tool Loop & none & multiple & scripted by human & retrieval &
  \aprsys{Agentless~(2025)}~\cite{xia2025demystifying},
  \aprsys{PATCH~(2025)}~\cite{zhang2025patch},
  \aprsys{KGCompass~(2025)}~\cite{yang2025enhancing} \\
Scripted Tool Loop & none & multiple & scripted by human & analysis &
  \aprsys{Repilot~(2023)}~\cite{wei2023copiloting},
  \vulnsys{SAN2PATCH~(2025)}~\cite{kim2025logs},
  \vulnsys{LLM4CVE~(2025)}~\cite{fakih2025llm4cve},
  \vulnsys{PredicateFix~(2026)}~\cite{xiao2025predicatefixrepairingstaticanalysis},
  \aprsys{IntDiagSolver~(2026)}~\cite{du2026environmentcrash} \\
\midrule
\multicolumn{6}{c}{\textbf{Agentic}} \\ \midrule
Tool-Augmented Agent & none & multiple & decided by LLM & varies &
  \aprsys{SWE\mbox{-}Agent~(2024)}~\cite{yang2024swe},
  \aprsys{AutoCodeRover~(2024)}~\cite{zhang2024autocoderover},
  \aprsys{RepairAgent~(2025)}~\cite{bouzenia2025repairagent},
  \vulnsys{VulDebugger~(2025)}~\cite{liu2025agent},
  \aprsys{SWE\mbox{-}Agent M~(2025)}~\cite{yang2025swem},
  \aprsys{OpenHands~(2025)}~\cite{wang2024openhands},
  \aprsys{LANTERN~(2026)}~\cite{luo2025unlocking}\\
LLM-Gated Review & none & multiple & decided by LLM & judge module &
  \bothsys{TSAPR~(2025)}~\cite{hu2025tsaprtreesearchframework},
  \aprsys{SpecRover~(2025)}~\cite{ruan2025specrover},
  \aprsys{Abstain and Validate~(2026)}~\cite{cambronero2025abstain} \\
Self-Controlled System & none & multiple & decided by LLM & varies &
  \aprsys{MAGIS~(2024)}~\cite{tao2024magis},
  \aprsys{SWE\mbox{-}Search~(2025)}~\cite{antoniades2025swe},
  \aprsys{Learn\mbox{-}by\mbox{-}Interact~(2025)}~\cite{su2025learn} \\
\bottomrule
\end{tabular}}
\end{table}

\subsection{Classification rules}

Figure~\ref{fig:taxonomies} summarizes one hierarchical primary-locus rule. A primary paradigm records where repair capability and control logic principally reside, while parameter adaptation, prompting, orchestration, and auxiliary evidence may coexist in one system and are preserved in separate coded fields.

First, we ask whether task-specific parameter adaptation carries the principal repair capability and whether the headline repair result depends on the adapted parameters. We assign Fine-Tuning only in that case, including full tuning, adapter or other parameter-efficient tuning, knowledge distillation, and reinforcement-learning fine-tuning. Merely invoking an adapted checkpoint inside a larger workflow is insufficient. For hybrid systems, we determine the principal source from the authors' stated contribution, workflow, headline comparisons, and ablation evidence when available.

For the remaining systems, Prompting applies when the prompt and supplied context carry the principal repair mechanism and neither a designer-specified multi-stage trajectory nor an LLM-directed controller carries the main result. Zero-shot, few-shot, and reasoning prompts, including chain-of-thought-style prompts, remain Prompting under this rule. Multiple independent samples generated solely to estimate pass@$k$ are evaluation replications rather than workflow steps.

Procedural applies when a designer-specified workflow carries the principal runtime control. The designer fixes the stage order, available tools, retry policy, and stopping budget, even if an LLM produces diagnoses, patches, or bounded reviews inside those stages. A bounded analysis, scoring, review, or validation module therefore does not by itself change the primary paradigm.

Agentic applies when LLM-directed action or tool selection, branching, regeneration, or termination is the principal runtime controller. LLM-Gated Review is Agentic only when learned judgment gates acceptance, branching, regeneration, or stopping; an offline score used only for evaluation is not agentic control.

For example, Agentless~\cite{xia2025demystifying}, ChatRepair~\cite{xia2024automated}, and Repilot~\cite{wei2023copiloting} can all be described as zero- or few-shot in utilization terms, yet all three are Procedural under this rule because designer-specified orchestration carries their principal repair logic.

The primary label is accompanied by separate fields for parameter adaptation, generation pattern, secondary control mechanisms, auxiliary evidence, and repair scope or scenario. Thus, a Procedural or Agentic system can still record prompting and fine-tuning components without losing them, while the primary label states which mechanism principally carries the reported repair capability and control logic.

Within Fine-Tuning, the control subtypes are Full Tuning, Adapter Tuning, Knowledge Distillation, and RL Tuning. Within Prompting, we distinguish Zero-shot, Few-shot, and Zero-/few-shot settings according to exemplar bug-fix pairs. Reasoning prompts, including chain-of-thought-style prompts, are recorded as a prompting attribute rather than a separate control subtype. Within Procedural, the control subtypes are Test-Feedback Loop, Human-in-the-Loop, and Scripted Tool Loop. Within Agentic, the control subtypes are Tool-Augmented Agent, LLM-Gated Review, and Self-Controlled System. These subtype labels refine the primary outcome; they do not replace the separate evidence fields.

Retrieval, program-analysis evidence, test feedback, human feedback, and domain knowledge are non-exclusive auxiliary-evidence tags that may support any primary paradigm. The scenario field separately records the primary evaluation setting. Aggregated tables report the primary paradigms, while the released codebook preserves these orthogonal properties for auditing hybrid systems.

\begin{figure}[h]
    \centering
    \includegraphics[width=.4\linewidth]{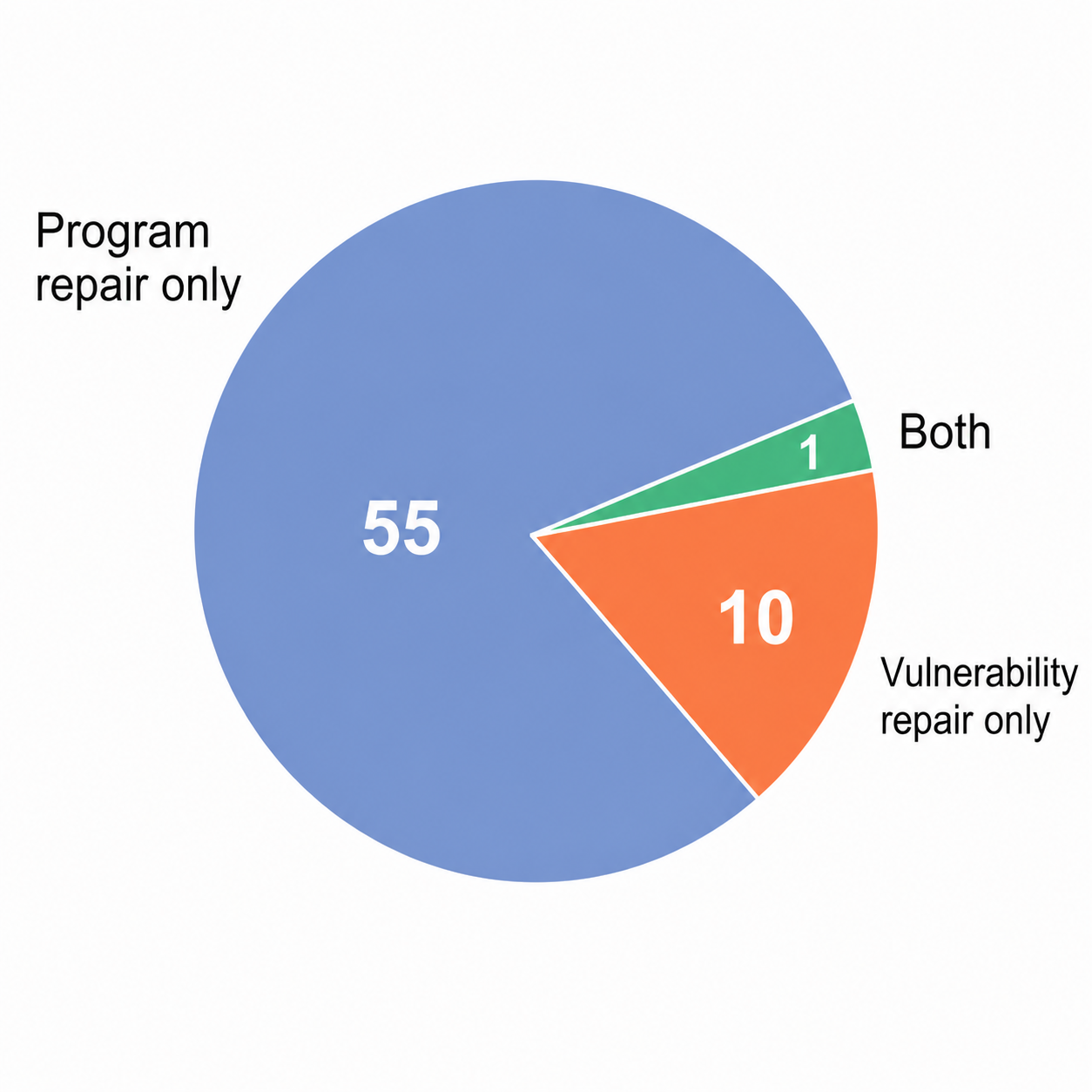}
    \caption{Defect scope distribution of 66 selected works.}
    \label{fig:repair-scope}
\end{figure}

Figure~\ref{fig:repair-scope} summarizes the defect scope of the 66 repair systems: 55 systems target general program repair only, 10 systems focus exclusively on vulnerability repair, and a single system addresses both.

\subsection{Venues of selected works}

\begin{figure}[h]
    \centering
    \includegraphics[width=.75\linewidth]{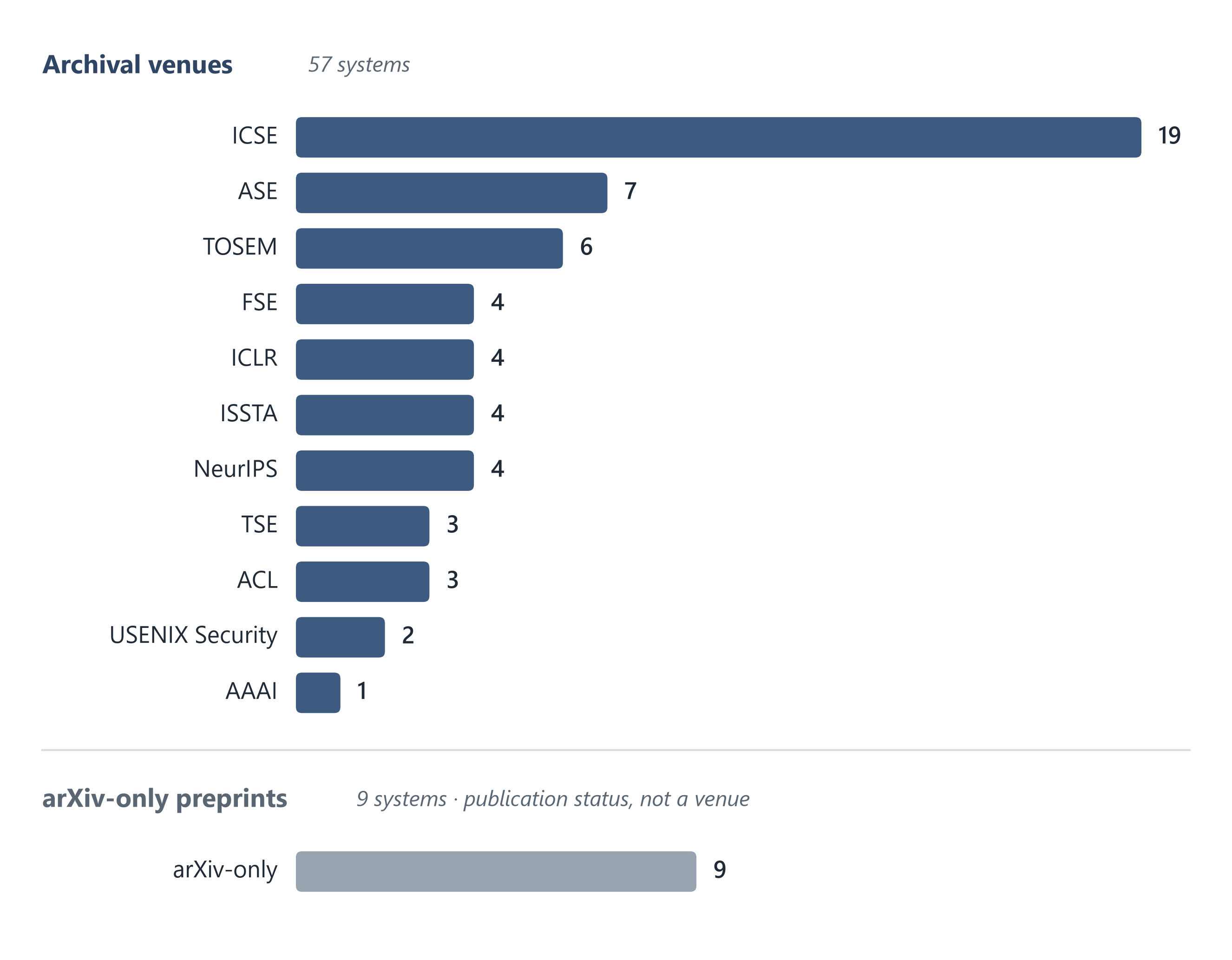}
    \caption{Publication status and archival venues of the 66 selected works.}
    \label{fig:venues}
\end{figure}

Figure~\ref{fig:venues} separates nine arXiv-only preprints from 57 archival publications. The arXiv bar reports publication status rather than a venue. The nine records and their retention rationales are enumerated in \path{artifact/version_status_audit.md}. Most archival systems appear in major archival software engineering venues and journals, with additional contributions from AI and security venues, indicating that LLM-based software repair is represented across multiple research communities within our criteria-bounded corpus.

\subsection{Temporal Distribution within the Criteria-Bounded Corpus}

Table~\ref{tab:apr-year-paradigm} summarizes how these 66 studies are distributed across years and paradigms. Within this representative corpus, 2023 contains many fine-tuning and prompting studies; 2024 adds procedural and agentic systems; 2025 has the largest number of systems across all paradigms; and the 2026 row covers papers in scope from January to April. These counts are descriptive corpus context rather than evidence of a complete field-wide trend.

\begin{table}[!htbp]
  \centering
  \caption{Annual distribution of LLM-based software repair applications by paradigm; 2026 covers January through April.}
  \label{tab:apr-year-paradigm}
  \setlength{\tabcolsep}{8pt}
  \renewcommand{\arraystretch}{1.1}
  \begin{tabular}{lccccc}
    \toprule
    \textbf{Year} & \textbf{Fine-Tuning} & \textbf{Prompting} & \textbf{Procedural} & \textbf{Agentic} & \textbf{Total} \\
    \midrule
    2022 & 0 & 2 & 0 & 0 & 2 \\
    2023 & 5 & 6 & 1 & 0 & 12 \\
    2024 & 7 & 1 & 5 & 3 & 16 \\
    2025 & 9 & 6 & 7 & 8 & 30 \\
    2026 & 0 & 2 & 2 & 2 & 6 \\
    \midrule
    Total & 21 & 17 & 15 & 13 & 66 \\
    \bottomrule
  \end{tabular}
\end{table}

\FloatBarrier

\begin{rqbox}{RQ1}
The taxonomy characterizes surveyed systems along multiple dimensions: parameter adaptation, runtime control authority, generation pattern, auxiliary evidence layer, and repair scope/deployment scenario. Its four primary paradigms separate fine-tuning, prompting, procedural pipelines, and agentic frameworks, while the auxiliary layer records retrieval, analysis, tests, human feedback, and domain knowledge without forcing hybrid systems into extra top-level buckets. Independent recoding further supports the reliability of this scheme: coders agreed on 65/66 primary-paradigm labels ($\kappa = 0.98$), 64/66 control-subtype labels ($\kappa = 0.97$), and all retrieval, analysis, and scenario tags.
\end{rqbox}

\section{Fine-Tuning Approaches}
\label{sec:finetuning}

Fine-tuning adapts a pre-trained LLM for software repair by updating its weights using bug-fix data. Research in this area now revolves around four adaptation strategies plus auxiliary-evidence variants. Fully supervised fine-tuning updates every parameter of the backbone and achieves strong accuracy across diverse benchmarks \cite{jiang2023impact,zhou2024out,jiang2024repaircat}. Parameter-efficient fine-tuning freezes most of the backbone and learns small adapter modules, which lowers memory cost while keeping competitive repair performance \cite{silva2025repairllama,yangmorepair,luo2025fine,li2024exploring,ruiz2025art}. Reinforcement learning fine-tuning trains the model with execution or security rewards so that it optimizes for functional correctness instead of log likelihood \cite{zhao2024repair,islam2024llm,wei2025swe,dai2025less,wen2025vul}. Knowledge distillation transfers bug-fixing skill from a large teacher or rule set to a smaller student, cutting inference latency without much accuracy loss \cite{jiang2023knod,wong2024distilrr,yang2024narrepair}. Some fine-tuned systems further inject static-analysis results, execution traces, retrieved patches, or template hints into the training input; we treat these signals as auxiliary evidence rather than as a separate fine-tuning subtype~\cite{bouzenia2023tracefixer,jin2023inferfix,chow2024pyty,huang2024template}.

\subsection{Full Fine-Tuning}

With the emergence of code-target language models based on transformers~\cite{vaswani2017attention} such as DeepSeek-Coder~\cite{guo2024deepseek,zhu2024deepseek}, CodeT5~\cite{wang2021codet5}, and CodeLlama~\cite{roziere2023code}, researchers have pursued full-model supervised fine-tuning on program repair scenarios. Recent work explores several variants of this strategy. Large-scale benchmark adaptation fine-tunes multiple backbone LLMs across standard datasets to quantify the effects of size and data~\cite{jiang2023impact}. Security-specific pipelines inject vulnerability patterns and CWE knowledge during full tuning to target secure patching~\cite{zhou2024out}. Domain-focused fine-tuning demonstrates that even 1B-parameter LLMs can learn high-yield repair heuristics when specialized data are available~\cite{jiang2024repaircat}. These efforts illustrate the breadth of full-parameter supervised fine-tuning for software repair.

\textbf{Huang~\etal} conducted an empirical study on fine-tuning LLMs (including CodeBERT~\cite{feng2020codebert}, GraphCodeBERT~\cite{guo_graphcodebert_2021}, PLBART~\cite{ahmad2021unified}, CodeT5~\cite{wang2021codet5}, and UniXcoder~\cite{guo2022unixcoder}) for program repair tasks~\cite{huang2023empirical}. Under their Defects4J protocol, the fine-tuned LLM variants repair more single-hunk and multi-hunk bugs than the earlier repair-tool baselines reported in the paper~\cite{just2014defects4j}. However, they reveal that LLMs usually fail to generate correct patches due to the input/output length limit of LLMs. On the other hand, the number of generated candidate patches is constrained by hardware resources, which often fail to meet the high computational demands of LLMs for massive candidate patches.

\textbf{Jiang~\etal} conducted a large-scale study that fully fine-tunes ten pre-trained code language models across four repair benchmarks to quantify the effect of supervised adaptation on repair performance \cite{jiang2023impact}. Under the protocol of Jiang et al., supervised fine-tuning substantially improved reported fix rates across the studied benchmarks, with the magnitude varying by model and dataset. The work also revealed heavy GPU demand and frequent overfitting to the suspected bug location, highlighting concerns about efficiency and generalization.

\textbf{VulMaster} fully fine-tunes a CodeT5~\cite{wang2021codet5} backbone on a security-focused corpus that blends complete vulnerable functions, abstract-syntax-tree structure, and CWE knowledge, then enlists ChatGPT at inference time to supply any missing contextual hints~\cite{zhou2024out}. On 1,754 real C/C++ projects with 5800 vulnerable functions, it doubles end-to-end effectiveness: exact-match rises from 10.2\% to 20.0\%, BLEU from 21.3\% to 29.3\%, and CodeBLEU from 32.5\% to 40.9\%. Strengths are its multi-source inputs and multi-LLM collaboration, but training demands $\sim$30 GB of GPU memory, and its gains taper on ultra-long files despite the enlarged context window.

\textbf{RepairCAT} is a fully fine-tuned 1B-parameter StarCoder~\cite{li2023starcoder,lozhkov2024starcoder}, trained on faults generated by AI code synthesizers to learn characteristic correction patterns~\cite{jiang2024repaircat}. Targeted training addresses a significant portion of such failures at a modest computational cost, demonstrating that small open models can benefit from full-model supervision in a specialized domain. Limited capacity and a narrow data distribution, however, curtail its effectiveness in addressing complex or conventional software defects.

Collectively, these methods demonstrate that full-parameter supervised fine-tuning can achieve strong repair accuracy while markedly reducing reliance on elaborate prompt engineering. Yet, practical challenges persist: high computational cost and, as large-scale benchmark studies indicate, a risk of catastrophic forgetting, dependence on the quality of auxiliary data, and susceptibility to overfitting when training data or evaluation suites are narrow. Tackling these issues remains essential for deploying large-scale, fine-tuned repair models in continuous integration environments.

\subsection{Parameter-Efficient Fine-Tuning}

Parameter-efficient fine-tuning (PEFT) alleviates the memory and compute burden of adapting LLMs for program repair by updating only a compact set of task-specific parameters. Current PEFT strategies for repair fall into two families: (i) \textbf{structural adapters}, including adapter layers~\cite{houlsby2019parameter}, prefix vectors~\cite{lester2021power}, and low-rank matrices such as LoRA~\cite{hu2021loralowrankadaptationlarge}; (ii) \textbf{quantization-aware adapters} typified by QLoRA, which merges 4-bit quantization with LoRA~\cite{dettmers2023qlora}. These designs shrink trainable parameters by up to three orders of magnitude, enable single-GPU training, and keep the frozen base model reusable for other tasks. Efficiency-oriented variants demonstrate promising effectiveness in program and vulnerability repair, broadening PEFT’s applicability to larger models and consumer-grade hardware~\cite{silva2025repairllama,yangmorepair,luo2025fine}. Nevertheless, attaining strong repair accuracy still requires judicious adapter placement, appropriate rank selection, and high-quality bug-fix data, because limited adapter capacity can miss semantic links that span multiple files~\cite{li2024exploring,ruiz2025art}.

\textbf{RepairLLaMA} augments a frozen CodeLlama-7B with rank-8 LoRA matrices (0.06\% of parameters) and fine-tunes them on 4,000 curated bug-fix pairs~\cite{silva2025repairllama}. Training on a single GPU fixes 144 Defects4J v2~\cite{just2014defects4j} bugs, 109 HumanEval-Java tasks~\cite{jiang2023impact}, and 20 GitBug-Java issues~\cite{silva2024gitbug}, more than the larger reported frozen-model baselines. Limited adapter capacity and reliance on static fault localization reduce accuracy for multi-hunk or cross-file defects.

\textbf{MORepair} applies QLoRA-style 4-bit quantization and rank-32 LoRA adapters, augmented with NEFTune noise, to multiple CodeLlama and other LLMs' backbones while jointly optimizing \emph{patch correctness} and \emph{repair rationale}~\cite{yangmorepair}. This configuration fixes the most bugs compared to standard fine-tuning on EvalRepair-Java and EvalRepair-C++, and in-depth case studies indicate the accuracy of both fixed code and generated rationale. However, performance still depends on rationale-annotated data, where the quality of rationale may influence the effectiveness of MORepair.

\textbf{Luo~\etal} apply LoRA adapters in a federated setting where each organization trains locally on proprietary code and a server aggregates updates, then evaluates on EvalRepair-Java~\cite{luo2025fine}. The protocol preserves privacy and matches centralized training within two percentage points while exchanging only megabytes of adapter weights per round. Convergence slows under highly heterogeneous client data, and a secure aggregation infrastructure is required.

\textbf{Li~\etal} benchmark LoRA, prefix-tuning, adapters, and (IA)$^{3}$~\cite{liu2022few} with CodeT5+ and StarCoder backbones~\cite{li2024exploring}. Under the authors' protocol, the (IA)$^{3}$ adapter resolves 58\% more bugs than the earlier LLM baselines reported in the paper while keeping memory usage low, but its gains plateau once adapter size or data volume exceeds moderate thresholds, so careful hyper-parameter tuning remains essential.

\textbf{Ruiz~\etal} compare full fine-tuning and LoRA on instruction-tuned code models under a total budget of ten patches per bug~\cite{ruiz2025art}. On HumanEval-Java, DeepSeek-Coder-Instruct~\cite{guo2024deepseek} with LoRA trained on 30K examples produces at least one plausible patch for 128 of 164 tasks (78.05\%) within 10 generations. Their experiments also show that iterative refinement benefits base models more consistently than fine-tuned models.

Overall, existing evidence suggests that PEFT can recover a substantial share of the repair accuracy reported for full fine-tuning in many settings, while using only a fraction of the computational and storage costs required for full fine-tuning. Effective practice requires balancing adapter capacity with data diversity, enriching supervision signals (for example, federated or rationale-aware training), and pairing lightweight models with search or analysis feedback to address multi-file and semantic defects.

\subsection{Knowledge Distillation}
\label{sec:distilling}

Knowledge distillation transfers bug-fixing skills from a high-capacity teacher or knowledge source to a smaller student, allowing inference costs to drop without compromising accuracy. Recent studies have applied classical teacher-student distillation (including soft-target and rationale distillation)~\cite{wong2024distilrr}, dataset distillation~\cite{yang2024narrepair}, and rule-guided distillation~\cite{jiang2023knod} to software repair, yielding strong empirical gains. These techniques either compress large teachers or replace them with structured knowledge, inject structural hints, restrict generation using structured rules, or enforce domain constraints while keeping the student lightweight. The following paragraphs review three representative methods that illustrate the breadth of distillation for program repair.  

\textbf{KNOD} distills formal syntax and semantic rules into a student decoder by aligning its token probabilities with a rule-derived teacher distribution~\cite{jiang2023knod}. The method repaired 72 Defects4J v1.2 bugs and 50 Defects4J v2.0 bugs, establishing new benchmark records. By restricting token choices to grammar-compliant options, KNOD greatly improves compile success and patch validity. Rule coverage is limited to statically checkable properties and offers less help for logical faults that do not violate syntax or types.

\textbf{DistiLRR} applies teacher-student distillation in which GPT-3.5 or GPT-4 produces bug fixes and natural-language rationales that supervise a CodeLlama-7B student~\cite{wong2024distilrr}. The distilled student nearly doubled the pass@1 success rate on low-resource languages such as Perl and Golang and matched the teacher on high-resource languages. Its main strength is the use of rich rationales, which give the student detailed guidance and boost repair in under-represented domains. The approach depends on high-quality synthetic rationales and yields smaller gains when the student already has ample native training data.

\textbf{NARRepair} first distills a CodeT5-large~\cite{wang2021codet5} teacher to build a cleaner student training set, then generates patches in parallel using action guidance and AST dependencies~\cite{yang2024narrepair}. It repairs 69, 41, and 23 bugs on Defects4J~\cite{just2014defects4j} v1.2, v2.0, and QuixBugs~\cite{lin2017quixbugs} while delivering 5.4-18.6 times faster inference than autoregressive baselines. The combination of dataset distillation and two-stage decoding bridges the usual quality gap of non-autoregressive generation without sacrificing speed. However, all evaluations are Java-only, so cross-language generality is unverified.

Distillation thus offers a practical route to simplify repair models while retaining high accuracy, and, in non-autoregressive settings, can markedly shorten decoding time. Classical teacher-student, dataset distillation, and rule-guided strategies each mitigate specific weaknesses of direct fine-tuning. Together, they show that transferring behavior, structural priors, or domain rules can deliver large gains without the runtime cost of very large models. Future work could blend these signals, including rationales and dataset curation, and extend distillation to reasoning traces, allowing lightweight students to tackle complex multi-file defects.

\subsection{Reinforcement Learning Fine-Tuning}

Recently, policy-gradient algorithms have been widely used in LLM-based research, including PPO~\cite{schulman2017proximal}, DAPO~\cite{yu2025dapo}, and Group Relative Policy Optimization GRPO~\cite{shao2024deepseekmath}. In this section, we also include preference-based tuning, such as DPO~\cite{rafailov2023direct}, which is not strictly an RL algorithm. Recent RL-based systems include RePair~\cite{zhao2024repair}, SecRepair~\cite{islam2024llm}, SWE-RL~\cite{wei2025swe}, and Vul-R2~\cite{wen2025vul}, while a preference-based alternative, AdaPatcher~\cite{dai2025less}, adopts Direct Preference Optimization (DPO) to refine its patch policy. In program-repair settings, reward design combines execution signals, such as compile-and-test success, with static cues like vulnerability removal or diff similarity to reduce sparsity~\cite{shojaeeexecution,ye2025process}. Some recent works adopt a two-stage pipeline that performs supervised fine-tuning to obtain a stable repair model and then applies RLFT for further alignment~\cite{dai2025less}. Compared with purely supervised fine-tuning, RLFT can push the model beyond the training distribution and align it directly with functional correctness; however, it incurs heavy computation because each gradient update may require running tests, rewards are sparse in repair tasks, and poorly shaped rewards can lead to over-fitting weak test oracles~\cite{dou2024stepcoder,liu2023rltf,ye2025process}.

\textbf{RePair} starts from a fully supervised StarCoder-15B model and adds an iterative RL loop mimicking human debugging~\cite{zhao2024repair}. A smaller critic model, trained to distinguish good from bad patches based on compile and test outcomes, supplies rewards to the actor via PPO. The actor may propose several successive fixes for one bug, receiving intermediate feedback at each step. This process enhances repair success on competitive-programming bugs to a level comparable to that of much larger proprietary LLMs. The trade-off is higher latency and the need to guard against infinite repair loops on hard-to-fix defects.

\textbf{SecRepair} targets vulnerability repair. It slices the code around each flaw, then fine-tunes a CodeGen2~\cite{nijkamp2023codegen2} actor with PPO, where the reward combines unit-test outcomes, semantic-similarity checks, and a security bonus that favors patches eliminating unsafe patterns~\cite{islam2024llm}. Trained on their built 36k-example InstructVul dataset and evaluated on zero- and N-day bugs from 6 open-source IoT operating systems, SecRepair fixes 58\% of vulnerable functions and generates concise commit messages that describe the flaw; the authors report up to 22 percentage points F1 improvement over a supervised baseline while adding 9\% runtime overhead. Strengths include security-aware rewards and explicit vulnerability descriptions; limitations are reliance on known flaw locations and increased token usage for descriptive comments.

\textbf{SWE-RL} fine-tunes a 70B-parameter Llama 3 backbone using policy-gradient RL on a large corpus of real software-evolution traces (GitHub issues, pull requests, and their ground-truth patches)~\cite{wei2025swe}. Its reward is a patch-level diff-similarity score, which eliminates costly execution while still offering fine-grained textual feedback. Training on authentic pull-request edits enables the model to capture developer reasoning, achieving a 41.0\% solve rate on SWE-bench Verified~\cite{jimenezswe}, strong reported open-source LLMs results under 100B, and transferring positively to five out-of-domain reasoning benchmarks. The approach scales to genuine project history at low cost, although its rule-based reward may undervalue semantically correct but stylistically different patches.

\textbf{AdaPatcher} adopts a strict two-stage pipeline: it first performs supervised fine-tuning on CodeLlama-Instruct-7B to train a Bug Locator that pinpoints faulty lines, then applies reinforcement learning, in the form of Direct Preference Optimization (DPO)~\cite{rafailov2023direct}, to a Program Modifier that refines the patch policy using preference pairs derived from unit-test outcomes and a ``fewer-changes'' reward. When evaluated on the ACPR benchmark, AdaPatcher achieves 62\% consistency and 67.57\% accuracy, higher than the purely supervised and prompt-based baselines reported in the same paper~\cite{dai2025less}. Its strength lies in marrying location-aware supervised fine-tuning with preference-based RLFT: the first stage supplies reliable edit anchors, and the second stage’s lightweight reward (diff similarity + compile-and-test bonus) lets the model learn to generate minimal yet correct edits without the high runtime of full Monte-Carlo execution.

\textbf{Vul-R2} treats vulnerability repair as a reasoning task, training a Qwen-based model that produces explanations and patches for C/C++ code through a two-stage pipeline: supervised fine-tuning for reasoning first, then reinforcement learning from verifiable rewards~\cite{wen2025vul}. In its Curriculum-based Verifiable Rewarded Training phase, an RLVR objective uses a critic LLM’s fix-or-not judgment plus ground-truth similarity and a simple format reward to score each patch. On the PrimeVul~\cite{ding2025vulnerability} and SVEN~\cite{he2023large} benchmarks, the authors report higher exact match and CodeBLEU over SFT and CoT baselines, and these gains coincide with longer, more reflective repair traces, giving unusually direct evidence that RLFT can improve reasoning rather than simply overfitting tests.

Existing RLFT studies confirm that execution- and security-based rewards can steer language models toward semantically correct patches, yet several obstacles remain. Many current methods still depend, to varying extents, on dynamic test execution, which can increase training costs when no proxy reward is available. Promising directions include hybrid reward functions that combine static cues with selective dynamic tests to reduce runtime, curriculum schedules to alleviate sparsity, and hierarchical planner-critic architectures that break down repair into tractable sub-goals. Resolving these challenges will be essential before RL-enhanced repair models can be reliably integrated into continuous integration pipelines.

\subsection{Auxiliary-Evidence Fine-Tuned Repair}

Some fine-tuned repair systems also augment the input with auxiliary evidence, such as static-analysis reports, runtime traces, inter-procedural dependency graphs, and historical patches, so the model can reason beyond the local snippet~\cite{bouzenia2023tracefixer,jin2023inferfix,chow2024pyty,huang2024template}. Approaches along this line inject compiler diagnostics, execution stack traces, or retrieved historical patches into the encoder input, guiding the model toward the root cause and constraining its edits~\cite{jin2023inferfix}. The main challenge is scalability: computing static or dynamic context for every training instance is costly, and long sequences may exceed LLMs' context limits~\cite{tian2023chatgpt}. Additionally, noisy or irrelevant signals can confuse the model, so effective selection and encoding of context remain open problems~\cite{parasaram2024fact,hollowayrole}.

\textbf{TraceFixer} demonstrates the value of dynamic feedback by feeding a pre-trained CodeT5 model with partial execution traces that capture the concrete program state where behavior diverges from expectations~\cite{bouzenia2023tracefixer}. The trace tokens are placed in a dedicated channel during fine-tuning, allowing the decoder to align state anomalies with the edits needed to repair them. This design repairs $13-20\%$ more bugs than a code-only baseline and excels on defects whose symptoms manifest only at runtime.

\textbf{InferFix} couples Meta’s Infer static analyzer with retrieval-augmented fine-tuning~\cite{jin2023inferfix}. For every alert, it retrieves analogous historical patches from a large corpus, concatenates the alert description, the buggy code, and the retrieved examples, and fine-tunes Codex-12B to generate patches that clear the analyzer warning. The additional context leads to 59 unique repairs that baseline LLMs miss and reduces false positives by filtering candidates that still trigger \textsc{Infer}.

\textbf{PyTy} uses the error message and location reported by \texttt{Pyre} as input to a model obtained by fine-tuning TFix, which had been trained to repair JavaScript linter warnings, on PyTyDefects~\cite{chow2024pyty}. On 281 held-out errors, its top-50 candidates remove the target error in 85.4\% of cases and exactly match developer fixes in 54.4\%, compared with 26.4\% exact match for GPT-4.

\textbf{NTR} (Neural Template Repair) formalizes template-based repair as a conditional text-to-text task where the selected template identifier is injected as a control token~\cite{huang2024template}. A lightweight classifier first ranks candidate templates, then a billion-parameter decoder is fine-tuned to translate the buggy code and template hint into a concrete patch. An \textit{OtherTemplate} token lets the model learn fixes beyond the known catalogs. NTR with CodeLlama-70B correctly repairs 139 bugs on Defects4J v1.2 and 136 of 163 tasks on HumanEval-Java under a budget of up to 100 candidate patches per task.

Overall, these recent studies confirm that richer semantic, dynamic, historical, and inter-procedural dependency graph contexts can materially improve LLM-based repair. Yet the cost of gathering and encoding such information remains significant, and ever-longer inputs challenge current context windows. Future work will need more selective retrieval, adaptive compression, and context-aware parameter-efficient fine-tuning strategies to retain the benefits of additional signals without compromising scalability or robustness.

\begin{overviewbox}
\textbf{Key finding.} Fine-tuning improves reported repair accuracy when enough high-quality bug-fix data or reward signals are available, but it trades generality and training cost for specialization. PEFT, distillation, and auxiliary-evidence training reduce this cost, while RL/preference objectives remain sensitive to reward design and test availability.
\end{overviewbox}

\section{Prompting Approaches}
\label{sec:prompting}

Prompt-based software repair treats the LLM as a frozen component that returns a complete patch after a single query, while every step of retrieval, example curation, or validation runs outside the model. Two core prompting settings and two recurring auxiliary-evidence patterns emerge. Zero-shot prompting supplies only the buggy code and a brief fix instruction, counting on the LLM’s prior knowledge to generate a patch~\cite{xia2022less,prenner2022can,tian2023chatgpt}. Few-shot prompting precedes the defect with a handful of similar bug-fix pairs, enabling the model to imitate their edit pattern~\cite{xia2023automated,gao2023makes,ahmed2023majority,nashid2023retrieval}. Retrieval-augmented prompting attaches code fragments, templates, or knowledge-graph facts fetched from external resources, grounding the patch in concrete project or domain information~\cite{chen2024large,ehsani2025hierarchical,ouyang2025knowledge,shao2026pailgen}. Analysis-enriched prompting injects diagnostics derived from static or dynamic analysis, such as failing-test logs or semantics-aware slices, to guide the model toward the root cause~\cite{xu2025aligning,nong2025appatch,haque2025traceprompt,zhang2025acfix,zhuo2026apimisuse}. These approaches share the same single-turn interaction but trade off preparation effort, token budget, and repair accuracy in different ways.

\subsection{Zero-shot Prompting}

Zero-shot prompting keeps a pre-trained model frozen and asks it to generate a patch in a single turn, given only the buggy code and a terse instruction such as ``\texttt{fix the bug}''. Because the approach omits fine-tuning, retrieval, and multi-turn dialogue, deployment is rapid and incurs no additional training cost. AlphaRepair~\cite{xia2022less} was the first study to demonstrate that LLMs can be competitive with established repair techniques under zero-shot repair. Empirical studies such as Prenner~\etal~\cite{prenner2022can} and Fan~\etal~\cite{fan2023automated} on Codex and Tian~\etal~\cite{tian2023chatgpt} on ChatGPT show that careful wording plus a sufficiently large model can already be competitive with several traditional repair pipelines under their reported protocols. At the same time, they expose strong prompt sensitivity, but their robustness beyond single-hunk functions remains unclear.

\textbf{AlphaRepair}~\cite{xia2022less} masks the tokens of the buggy lines and leverages CodeBERT to infill the correct code, thereby predicting the fixed codes directly instead of learning edit scripts. On Defects4J-v1.2/2.0, the authors report up to 3.3 times more repaired bugs than the baselines in their study, and they also report competitive QuixBugs results under their protocol, demonstrating that a frozen pre-trained LLM can be effective without further training.

\textbf{Prenner~\etal} evaluated Codex (12B)~\cite{chen2021evaluatinglargelanguagemodels} on all 40 QuixBugs~\cite{lin2017quixbugs} bugs in Python and Java~\cite{prenner2022can}. They tried five one-shot prompt variants that differed only in the static text preceding the buggy function. With temperature 0 and \texttt{max\_tokens}=1024, Codex generated ten candidates per bug; a patch was accepted if QuixBugs’ reference tests all passed. The highest-repair prompt variant in the authors' study repaired 23 Python and 14 Java bugs, which they compare against three specialized neural repair systems despite Codex never having seen any examples. The study also showed that adding a simple ``\#\#\# bug is here'' hint sometimes hurts accuracy, illustrating how small prompt edits can swing outcomes.

\textbf{Fan~\etal} conducted a large-scale evaluation of \emph{zero-shot} Codex on competitive-programming tasks and then studied how software repair can salvage its failures~\cite{fan2023automated}. Their prompt contained only the LeetCode problem description and the target function’s signature; no exemplar fixes were provided. By generating 50 solutions per task, Codex solved 46 of the 113 tasks on their new \textsc{LMDefects} benchmark, showing that LLM with naive zero-shot prompting already covers a substantial fraction of real bugs and that lightweight repair can extend this reach.

\textbf{Tian~\etal} explored ChatGPT (GPT-3.5) as a programming assistant across code generation, summarization, and program repair~\cite{tian2023chatgpt}. For repair, they built a benchmark of 1,783 incorrect student assignments covering five classic problems. A single-shot prompt that contained only the buggy function (or that function plus its textual assignment description) asked ChatGPT for a ``corrected version''. Without any retries or analysis feedback, ChatGPT fixed 84\% of the tasks at TOP-5 and averaged 60\% correct among its five completions, rivaling the Refactory~\cite{hu2019re} semantic repair tool. Notably, providing the verbose assignment description reduced accuracy, confirming that concise prompts better focus the model’s attention.

In summary, single-turn zero-shot prompting already addresses a non-trivial portion of real bugs, ranging from small algorithmic errors to student-level assignments, and several papers report higher repair counts for larger models under their own protocols. Nevertheless, the approach remains prompt-sensitive and fragile, and its reliability with larger, multi-hunk patches or when extra but imprecise context is supplied remains uncertain.

\subsection{Few-shot Prompting}

Few-shot prompting keeps the language model frozen and places a small set of bug-fix exemplars before the target defect so that the model can infer an edit pattern in a single decoding pass. Compared with zero-shot requests, these demonstrations often stabilize generation and improve accuracy. However, they must fit within the context window. As shown by work on basic few-shot~\cite{xia2023automated}, similarity-based selection~\cite{gao2023makes}, explanation-rich exemplars~\cite{ahmed2023majority}, and retrieval-driven prompting such as CEDAR~\cite{nashid2023retrieval}, their composition requires careful curation.

\textbf{Xia~\etal} performed the first large-scale study that applies nine modern LLMs to repair without fine-tuning~\cite{xia2023automated}. They evaluated three prompt settings—complete function generation, code infilling with prefix and suffix, and single-line edits, over 1,094 bugs drawn from Defects4J 1.2/2.0 (Java)~\cite{just2014defects4j}, QuixBugs (Java \& Python)~\cite{lin2017quixbugs}, and ManyBugs (C)~\cite{le2015manybugs}. For few-shot generation, they prefixed two exemplar bug-fix pairs ahead of the target defect and requested up to 200 samples per bug. Under this setting, Codex repaired sixty-three single-function Java bugs on \textsc{Defects4J}, more than the earlier repair-tool baseline reported in the paper. The study further reveals a clear scaling trend (larger models fix more bugs), highlights the value of including suffix context during infilling, and demonstrates that simple entropy scores enable frozen LLMs to rank patches effectively. These findings suggest that even a minimal number of few-shot prompts, when paired with sufficient sampling, can rival specialized learning-based systems, but prompt design must balance exemplar relevance, context length, and sample budget.

\textbf{Gao~\etal} designed more than two hundred Codex prompts for Java repair~\cite{gao2023makes}. Each prompt began with exactly four bug-fix pairs that were retrieved by lexical similarity and copied verbatim, in the same order, followed by the new buggy snippet and the instruction for producing a fixed version. Under this setting, exact-match accuracy more than doubled relative to random exemplars and tripled relative to zero-shot. The authors concluded that prompts should prioritize highly similar examples, keep their order stable, and stop adding demonstrations once the prompt nears the context limit; otherwise, gains vanish when no close fixes exist.

\textbf{Ahmed~\etal} prefixed eight bug-fix demonstrations, each followed by its original commit message that briefly explained the change~\cite{ahmed2023majority}. After these exemplars, they placed the buggy JavaScript function and asked the frozen \texttt{code-davinci-002} model to return a corrected version. They sampled 30 completions and selected the patch that appeared most often. The commit messages provided lightweight reasoning traces that helped the model generalize, and majority voting raised repair accuracy by roughly ten percentage points. This approach increases token usage and assumes that informative commit logs are available.

\textbf{CEDAR} automatically retrieves exemplar pairs from a large corpus using embedding or token-frequency similarity and inserts them into the prompt without any model fine-tuning~\cite{nashid2023retrieval}. On Java and Python repair tasks, four retrieved demonstrations yielded about 76.55\% exact-match accuracy, which the authors report as higher than task-specific baselines and competitive with fully fine-tuned models, showing that systematic retrieval can rival manual curation while keeping the model frozen.

These studies confirm that well-chosen exemplars, whether similarity-ranked, retrieval-selected, or explanation-enriched, can guide frozen LLMs to competitive one-step repairs. Open challenges include finding suitable examples for projects without a history, compressing prompts when defects span multiple lines, and controlling costs when majority-voting or other multi-sample strategies increase token usage.

\subsection{Retrieval-Augmented Generation Enhanced Prompting}

Retrieval-enriched prompting, a repair-specific use of retrieval-augmented generation (RAG), enriches a single-turn repair prompt with external artifacts fetched on demand. By retrieving repository structure information~\cite{chen2024large}, layered knowledge injection~\cite{ehsani2025hierarchical}, or domain-specific API knowledge encoded in a graph~\cite{ouyang2025knowledge}, a frozen language model can ground its patch in concrete evidence instead of relying solely on parametric memory. Compared with plain zero- or few-shot prompts, this design can improve factual grounding and supply missing context while deployment remains as light as a one-shot generation.

\textbf{RLCE} targets repository-level bugs that span multiple files. A static parser builds a lightweight project map and then retrieves only those out-of-file definitions and usages that interact with the faulty code; the selected fragments are appended to the prompt that asks GPT-3.5 or GPT-4 to produce the fixed version~\cite{chen2024large}. This single-shot context expansion raises GPT-3.5’s top-1 fix rate on their built 124-bug RepoBugs suite from 22.6\% to 56.0\% and lifts GPT-4 from 41.1\% to 81.5\%, more than doubling accuracy without extra inference rounds or model updates.

\textbf{Ehsani~\etal} propose a layered knowledge injection framework for LLM-based software repair that incrementally augments prompts with Bug Knowledge, Repository Knowledge, and Project Knowledge automatically extracted from the buggy function, co-changing files, Git history, documentation, and previously resolved issues~\cite{ehsani2025hierarchical}. On the BugsInPy~\cite{widyasari2020bugsinpy} benchmark of 314 Python bugs, this framework with Llama 3.3 raises the fix rate to 79\% by correctly repairing 250 bugs, a 23\% improvement over prior work, with 65\% fixed using only the Bug Knowledge Layer, a further 9\% gain from adding the Repository Knowledge Layer, and an additional 5\% gain from the Project Knowledge Layer. They find that retrieved context should be injected hierarchically and in a bug-type-aware way, rather than being concatenated all at once, because different bug types benefit from different kinds and depths of retrieved repository and project knowledge for effective repair.

\textbf{DSrepair} retrieves structured API knowledge from a purpose-built Data-Science Knowledge Graph and injects the retrieved triples plus the localized buggy snippet into a single repair prompt for GPT-3.5, GPT-4, and two open-source code LLMs, DeepSeek-Coder~\cite{guo2024deepseek} and Codestral~\cite{mistralai2024codestral}, without any fine-tuning~\cite{ouyang2025knowledge}. On the DS-1000 benchmark~\cite{lai2023ds}, the authors report 44.4\%, 14.2\%, 20.6\%, and 32.1\% more fixed data-science bugs than their baseline configurations while cutting prompt tokens by up to 34\%, showing that knowledge-graph evidence can improve one-shot repair when the retrieved facts are well aligned with the task.

\textbf{PailGen} applies retrieval- and pattern-aware prompting to vulnerability patch generation~\cite{shao2026pailgen}. It first retrieves vulnerability-fix examples related to the target vulnerable function, mines recurring fix patterns from these examples, and injects both the retrieved patches and the abstracted edit pattern into a single in-context repair prompt. With DeepSeek-Coder-V2, PailGen reaches 23.23\% exact match on Big-Vul\_CVEfixes, compared with 5.07\% for the corresponding baseline prompt. The result shows that retrieval is most useful when it is not treated as a generic few-shot context, but is converted into a fixed-pattern evidence that constrains the expected security edit.

These studies illustrate the versatility of retrieval-enriched prompting: knowledge-graph retrieval supports domain-specific API fixes, repository-aware retrieval supports cross-file repair, and template-guided retrieval accelerates security patching, all while preserving the simplicity of a one-shot prompt and avoiding procedural or agentic orchestration.

\subsection{Program-Analysis Evidence in Prompting}

Program-analysis-supported prompting feeds a language model with artifacts such as compiler diagnostics, fault-localization scores, data-flow constraints, failing-test logs, or execution traces. These artifacts typically appear in three forms: failing-test evidence that exposes runtime failures, semantics-aware slices that prune unrelated code, and concise execution traces that embed dynamic behavior. Recent systems illustrate each form in turn: D4C~\cite{xu2025aligning} injects failing-test evidence, Appatch~\cite{nong2025appatch} applies semantics-aware slices, and TracePrompt~\cite{haque2025traceprompt} contributes concise execution traces. Each patch is grounded in concrete evidence. Compared to plain zero-shot prompts, these systems can improve repair accuracy and keep edits focused on fault-relevant code, while ensuring that extra context remains within the model window.

\textbf{D4C} aligns the prompt with decoder-only models’ next-token objective and lets GPT-4 complete the whole function while embedding the failing test case and its output~\cite{xu2025aligning}. Without any fault-localization hints, D4C fixes 180 of 437 single-function bugs in Defects4J under the authors' protocol, which they report as 10\% more than a prior localized baseline while using ten samples per bug. The approach relies on dynamic evidence, so its gains diminish when tests give little diagnostic detail.

\textbf{APPATCH} applies semantics-aware scoping and adaptive exemplar prompting to vulnerability repair~\cite{nong2025appatch}. It slices the program to include only code relevant to the vulnerability, mines an exemplar database, and selects examples on the fly to guide patch generation. Across 97 zero-day vulnerabilities and 20 vulnerabilities from ExtractFix, APPATCH reaches up to 36.46\% and 73.86\% F1, respectively, compared with 28.41\% and 68.54\% for the best prompting baseline. Strengths include reduced context, tailored exemplars, and ensemble validation across models; limitations are the need for a known vulnerability location and extra token overhead from exemplar insertion.

\textbf{TracePrompt} augments the buggy code and failing test with a structured execution trace produced by PySnooper~\cite{rachum2019pysnooper}, then issues one repair request to GPT-4~\cite{haque2025traceprompt}. Across Refactory~\cite{hu2019re}, RunBugRun~\cite{prenner2023runbugrun}, and HumanEval-Java~\cite{jiang2023impact}, trace-based prompts improve GPT-4’s correct-program accuracy by up to 6\% over error-only prompts. However, the benefit vanishes for GPT-3.5 and for traces that exceed 200 lines, showing that runtime context helps powerful models only when it is concise.

\textbf{ACFix} targets access-control vulnerabilities in smart contracts by mining common role-based access-control practices and injecting contract-specific access-control context into a GPT-4 repair prompt~\cite{zhang2025acfix}. Its offline phase builds a role-permission taxonomy from on-chain contracts, while its online phase constructs an access-control graph for the subject contract, extracts relevant role and permission evidence, and validates the generated patch with static and semantic checks. On 118 real-world access-control vulnerabilities, ACFix repairs 94.92\% of cases, compared with 52.54\% for a direct GPT-4 baseline. This makes it a representative analysis-enriched prompting system: the repair remains a prompt-based generation step, but the prompt is grounded in contract-specific access-control analysis rather than only in the vulnerable code.

\textbf{Dr.Fix} repairs API misuses in LLM-generated code by making the misuse category explicit in the repair prompt~\cite{zhuo2026apimisuse}. The system first detects whether a generated snippet contains an API misuse, classifies the misuse into categories such as intent mismatch, hallucination, redundancy, or missing item, and then asks an instruction-tuned LLM to repair the masked API element with this diagnostic label and a demonstration. On the Java method-misuse repair task, GPT-4o with Dr.Fix reaches 96\% exact match, improving over the 52\% baseline configuration. This setting differs from conventional benchmark repair in that the faulty artifact is generated code and correctness is measured at the API-element level, yet it remains a software repair task in which analysis-derived misuse labels guide an LLM patch.

These studies reveal complementary analysis-enriched strategies: semantics-aware slices narrow the search space, failing-test evidence enables direct debugging without prior localization, compact execution traces inject runtime semantics, and domain taxonomies or mined security practices guide the model toward legally or semantically appropriate edits. All approaches preserve the simplicity of one-shot prompting and avoid the need for parameter fine-tuning. However, each must balance the need for extra context against the limited tokens and potential prompt bias. Future work should explore adaptive artifact selection and summarization to keep this balance in large-scale settings.

\begin{overviewbox}
\textbf{Key finding.} Prompt-based repair benefits most from focused auxiliary context: similar fixes, relevant repository/API snippets, failing tests, or compact traces. Irrelevant or verbose context can hurt performance, so the main open problem is adaptive context selection under prompt-length and cost limits.
\end{overviewbox}

\section{Procedural Approaches}
\label{sec:procedural}

Procedural approaches treat software repair as a fixed pipeline where the language model runs only at scripted invocation points and never alters its weights or the surrounding logic. Every choice about search scope, context retrieval, validation order, and iteration budget is hard-coded, which yields reproducible outcomes and predictable costs. Within this umbrella, systems differ by the dominant feedback or evidence source. Test-feedback-loop pipelines alternate generation with unit tests and take each failure as an oracle \cite{xia2024automated,yin2024thinkrepair,tang2024code,kong2025contrastrepair}. Human-in-the-loop pipelines thread developer insight into the loop through hints, plans, or reviews that shape the next prompt \cite{yang2024cref,takerngsaksiri2025human,zhao2024enhancing}. Retrieval-enriched scripted loops insert concise evidence fetched by static rules between calls \cite{xia2025demystifying,zhang2025patch,yang2025enhancing}. Analysis-enriched scripted loops invoke compilers, debuggers, or security analyzers to feed diagnostics or traces that steer regeneration toward the root cause \cite{kim2025logs,xiao2025predicatefixrepairingstaticanalysis,fakih2025llm4cve,wei2023copiloting,du2026environmentcrash}. These systems differ mainly in the source of the feedback they inject and in the auxiliary tools that collect it. At the same time, they all rely on scripted control to keep the repair deterministic.

\subsection{Test-Feedback Loop Pipelines}
\label{subsec:titl}
Test-feedback-loop pipelines alternate patch generation with automatic test execution under a designer-specified loop. Lightweight conversational loops, exemplified by ChatRepair~\cite{xia2024automated}, feed failing assertions back to the LLM and regenerate patches in the loop. Reasoning-enhanced variants such as ThinkRepair~\cite{yin2024thinkrepair} cache a chain of thought and partial passes. Bandit-guided search, represented by REx~\cite{tang2024code}, allocates queries based on test rewards, whereas contrast-aware schemes, such as ContrastRepair~\cite{kong2025contrastrepair}, pair failing and passing tests to isolate the faulty logic. Because every step and data flow is predetermined by the script rather than the model, these pipelines differ from fully agentic frameworks while still turning the test suite into an iterative oracle that steers the LLM toward a correct fix.

\textbf{ChatRepair} prompts GPT-3.5 or GPT-4 to propose a patch, executes the full test suite, then feeds the exact failing assertion messages plus the buggy lines back to the model for an average of 21.86 regeneration rounds~\cite{xia2024automated}. Its conversational prompts are lightweight and cost-efficient, letting ChatGPT fix 162/337 Defects4J bugs at roughly \$0.42 per bug while requiring no auxiliary analysis. Strengths include minimal engineering effort and strong empirical performance on real Java defects. Weaknesses include prompt sensitivity and limited scalability to multi-hunk or cross-file bugs, as failure messages alone may not pinpoint the root cause. All loop conditions, timeouts, and data exchanges are hard-wired; the LLM only supplies patch text and never alters the surrounding script.

\textbf{ThinkRepair} first asks a frozen ChatGPT to write an explicit chain-of-thought that diagnoses the bug, stores any partially correct patches that pass some tests, then iteratively re-prompts the model with those traces plus the remaining failing tests~\cite{yin2024thinkrepair}. This two-phase Collect-Fix loop converts the test suite into a knowledge source, reusing previous reasoning to accelerate convergence. Under the authors' Defects4J protocol, the explicit reasoning and knowledge pool increases the number of fixed bugs to 98 compared with the conversational baselines reported in the same study. Its drawback is longer prompts and higher token consumption, which may stress context limits on very large projects.

\textbf{REx} (Refine-Explore-Exploit) frames iterative program repair as a Thompson-sampled multi-armed bandit~\cite{tang2024code}. It views every partial patch as an arm whose reward is the proportion of tests currently passing. Starting from a pool of seed fixes, the algorithm samples from each arm’s Beta posterior, chooses the most promising candidate, and prompts GPT-4 with its remaining failing tests. This adaptive explore-exploit policy reduces LLM calls by roughly 2- to 5-fold on challenging benchmarks compared to breadth-first, greedy, or fixed-width schedules. On the Nonlinear Loop Invariant dataset, REx solves 73.7\% of tasks within 300 GPT-4 queries, compared with the 60.5\% reported specialized-solver baseline. Performance still depends on a well-calibrated Beta prior; if the prior is off, the search can miss rare but vital repair paths.

\textbf{ContrastRepair} augments conversation-based software repair by supplying ChatGPT with a contrastive pair of nearly identical test cases (one failing and one passing) to focus the model on the specific code change that triggers the bug~\cite{kong2025contrastrepair}. It repairs 143 of 337 bugs on Defects4J~\cite{just2014defects4j}, and the authors report higher QuixBugs~\cite{lin2017quixbugs} and HumanEval-Java~\cite{jiang2023impact} results than their listed baselines under the same study protocol. The authors demonstrate that selecting the passing test with the smallest Damerau-Levenshtein distance~\cite{damerau1964technique,levenshtein1966binary} from the failing test yields richer feedback and reduces the number of conversation rounds required for a correct patch. However, this method relies on the presence of at least one closely matching passing test and extra retrieval time, so its benefit diminishes when no closely matching passing case exists, when generating one by mutation exceeds the time budget, or when the synthetic passing test hides the real fault.

These representatives illustrate two consistent patterns. First, richer feedback yields higher fix rates: moving from raw assertion messages to contrastive test pairs, chain-of-thought traces, or adaptive candidate selection progressively boosts success without altering model weights. Second, feedback granularity trades off against resource usage. Lightweight conversations scale to hundreds of bugs at minimal cost. In contrast, trace-driven, contrast-driven, or search-intensive loops achieve greater accuracy at the expense of longer prompts, more LLM calls, and added engineering complexity. A promising future direction is an adaptive tiered pipeline that escalates from inexpensive cues to detailed analysis only when simpler signals fail, balancing efficiency with effectiveness across diverse defect types.

\subsection{Human-in-the-Loop Pipelines}
\label{subsec:hitl}

Earlier non-LLM studies have underscored the necessity of a human in the loop for software repair~\cite{geethal2023human}. Human-in-the-loop pipelines embed frozen LLMs into a structured workflow that alternates between LLM inference and human insight. Recent work spans interactive tutoring for novice programmers~\cite{yang2024cref}, planner-and-coder collaborations on industrial issues~\cite{takerngsaksiri2025human}, and reviewer-driven patch refinement~\cite{zhao2024enhancing}. For each human-in-the-loop approach, progress occurs when the LLM revises its patch after receiving new human feedback, such as tutor hints, input labels, design rationales, or reviewer advice.

\textbf{CREF} partners a programming tutor with a frozen GPT-4 in an interactive repair dialogue~\cite{yang2024cref}. Each cycle shows the model the buggy function, failing tests, and the tutor’s natural-language hint. The tutor then reviews the candidate's patch, supplying a fresh hint or approval that becomes part of the next prompt. Because new human guidance is injected at every turn, the dialogue converges quickly and achieves an AVG-5 score of 76.6\% on the 1,239-bug \textsc{TutorCode}~\cite{TutorCode} benchmark, 17-25\% above no-tutor baselines.

\textbf{HULA} is a human-in-the-loop LLM agent framework, where an AI Planner and an AI Coding Agent work with a human engineer to localize files, draft a coding plan, and generate code patches for each issue~\cite{takerngsaksiri2025human}. Experimental results on SWE-bench, the planner reached 86\% file-recall and the coder attained 45\% LLM-based code-similarity, while in an online study with 663 real Jira issues 82\% of plans were approved and 56 pull requests were merged. The authors find that detailed issue descriptions and interactive human feedback are critical, since practitioner agreement rises when engineers enrich descriptions and review agent output. Nevertheless, HULA still struggles to guarantee functional correctness beyond unit tests and requires substantial context preparation.

\textbf{DRCodePilot} embeds frozen GPT-4 in a dialogue-style loop that repeatedly injects developer design rationales and reviewer-style feedback~\cite{zhao2024enhancing}. Each cycle starts with DRMiner~\cite{zhao2024novel}-extracted solution information, letting GPT-4 locate the buggy segment and draft a patch; a tuned CodeT5P~\cite{wang2023codet5+} then supplies a reference patch, while an identifier recommender offers renaming hints. GPT-4 reflects on this mixed human feedback and produces a refined fix. On a new 938-issue Flink and Solr benchmark, it yields 109 and 18 exact-match patches, respectively, 4.7x and 3.6x more than GPT-4 zero-shot, with CodeBLEU gains of 5.4\% and 3.9\%.

These studies illustrate complementary avenues for leveraging human insight into procedural repair loops, ranging from issue-tracker rationales to iterative tutor coaching, developer-supplied documents and hints, reviewer comments, and expert Socratic questioning. Across these settings, human feedback injected between LLM generations transforms a frozen model into an effective repair partner, while minimizing computational cost and workflow complexity.

\subsection{Retrieval-Enriched Scripted Pipelines}

Retrieval-enriched scripted loops keep the LLM frozen and insert a deterministic retrieval stage between successive generations, enriching each new prompt with fresh, highly targeted evidence. The control flow is scripted by the designer rather than chosen by the model, placing these methods squarely in the procedural paradigm. Within this paradigm, hierarchical repository pruning exemplified by Agentless~\cite{xia2025demystifying}, multi-stage retrieving as used in PATCH~\cite{zhang2025patch}, and knowledge-graph expansion adopted in KGCompass~\cite{yang2025enhancing} all show that precise retrieval can raise repair accuracy while containing token cost.

\textbf{Agentless} follows a three-phase loop that first asks GPT-4o to rank suspicious files, then shows only their signatures so the model can pick concrete edit lines, and finally reveals full code for patch generation, followed by test validation~\cite{xia2025demystifying}. Hierarchical retrieval provides exactly the required context at each stage and avoids exposing the entire repository up front. On SWE-bench Lite, the initial pipeline with GPT-4o repaired 32.0\% of issues at about \$0.70 per bug under the authors' protocol, with lower token use than several tool-driven autonomous-agent baselines reported in that paper. While promising, the repository traversal can be inefficient or imprecise on very large projects, and the fixed three rounds may under-explore complex defects, suggesting adaptive depth or early-exit heuristics.


PATCH is a stage-wise framework that augments the buggy code snippet with class and repository-level dependence context, commit message programmer intent, and retrieved bug-fixing demonstrations, then simulates tester, developer, and reviewer roles with three ChatGPT agents across bug reporting, bug diagnosis, patch generation, and patch verification~\cite{zhang2025patch}. On the BFP~\cite{tufano2019empirical} benchmark, PATCH reaches pass@1 33.97\%, pass@3 37.08\%, and pass@5 39.81\% with a Levenshtein distance of 21.44, compared with GPT-4 at 19.96\%, 24.42\%, 26.00\% and 26.07\% respectively. For retrieval-enriched procedural repair, the main lesson is that retrieval of dependence context, commit messages, and similar bug fixing pairs is most effective when combined with an explicit multi-role, multi-stage interaction loop that lets the LLM iteratively explain, reason, and review patches instead of treating repair as a single pass generation task.

\textbf{KGCompass} converts a repository and its issue tracker into a heterogeneous knowledge graph, then, before each regeneration, expands from the buggy node to fetch the twenty most relevant code blocks along dependency edges~\cite{yang2025enhancing}. These snippets and the current failing assertions are concatenated into the next repair-model prompt, providing high-signal structural context.
KGCompass reports SWE-bench Lite pass@1 of 58.30\% while achieving 56.0\% function-level localization accuracy at roughly \$0.20 per bug. Constructing and updating the knowledge graph incurs pre-processing overhead and can miss dynamic links, such as reflection. Incremental graph maintenance and lightweight dynamic traces are promising extensions.

These studies confirm that designer-defined retrieval, hierarchical pruning, forum mining, search-tree feedback, or graph expansion delivers concise, high-signal context that steers a frozen model toward correct edits while keeping low costs. Remaining challenges include pre-processing overhead, coverage of dynamic behavior, and balancing retrieval granularity against token economy; addressing these issues will further narrow the gap between retrieval-enriched procedural pipelines and fully autonomous repair agents.

\subsection{Analysis-Enriched Scripted Pipelines}
\label{subsec:analysisloop}

Analysis-enriched scripted loops interleave an automated program analysis pass with LLM repair steps. Evidence can stem from sanitizer logs analyzed by SAN2PATCH~\cite{kim2025logs}, from semantics-based token pruning in Repilot~\cite{wei2023copiloting}, or from static and hybrid security analyzers such as CodeQL~\cite{de2007ql}, adopted by PredicateFix~\cite{xiao2025predicatefixrepairingstaticanalysis}, and from exploit validators embedded in LLM4CVE~\cite{fakih2025llm4cve}. A fixed script orchestrates the analysis stages and regeneration steps, keeping these pipelines within the procedural paradigm. The analysis signals guide regeneration without relying solely on unit tests.

\textbf{Repilot} leverages a frozen LLM with a semantics-based Completion Engine—concretely, the incremental parser and code-completion API of \emph{Eclipse JDT Language Server}~\cite{EclipseJDTLS2023,UniverseFlyJDTLS2023}, which inspects the partial AST, enumerates type-, scope-, and import-consistent continuations, prunes invalid tokens, and actively completes rare identifiers during autoregressive decoding~\cite{wei2023copiloting}. The workflow iterates token by token: the LLM supplies a probability list for the next token, the engine eliminates options statically infeasible, and occasionally inserts a full identifier when only one continuation type-checks, after which generation resumes. Evaluated on single-hunk Java bugs from \textsc{Defects4J} 1.2/2.0, the authors report that Repilot fixes 66 and 50 bugs, respectively, 27\% and 47\% more than prior strong repair tools in their study, and sustains a 59\% compilation rate at 5000 samples per bug. Because static analysis is consulted for every partial program, Repilot delivers markedly higher validity and unique fixes but incurs roughly 7\% overhead for CodeT5-large~\cite{wang2021codet5} and negligible overhead for InCoder-6.7B~\cite{friedincoder}. The authors argue that such fine-grained pruning can be applied with different model choices and generalized to other generation tasks. However, scalability to multi-file projects and languages beyond Java remains an open question.


SAN2PATCH is an automated vulnerability repair system that uses only sanitizer logs and source code, running a four-stage Tree of Thought LLM pipeline (Vulnerability Comprehension, Where-To-Fix, How-To-Fix, Candidate Generation) with AST-based code context retrieval and strict patch validation to synthesize and verify patches~\cite{kim2025logs}. On the VulnLoc~\cite{shen2021localizing} benchmark, SAN2PATCH with GPT-4o correctly repairs 31 of 39 vulnerabilities, achieving a 79.5\% success rate. For analysis-enriched vulnerability repair, the key insight is that careful prompt structuring and multi-stage LLM reasoning over lightweight runtime artifacts like sanitizer logs can replace heavy program analysis and manual localization while still producing functionally correct patches at scale.

\textbf{PredicateFix} targets security alerts raised by CodeQL~\cite{de2007ql} and GoInsight. Their analyzer inspects the rule predicates behind an alert, derives a bridging predicate that links the defect to missing sanitization, and then retrieves key examples that satisfy that predicate~\cite{xiao2025predicatefixrepairingstaticanalysis}. These examples, along with the original alert, are incorporated into a retrieval-augmented prompt that guides GPT-4 or other models to generate a patch. After each proposal, the analyzer reruns; unresolved alerts trigger another round while already-fixed code is frozen. Across 6,027 CVE-based alerts, the pipeline raised the number of correct repairs by $27.1\%$-$72.5\%$ relative to plain retrieval-augmented baselines.

\textbf{LLM4CVE} presents an iterative vulnerability-repair loop for real-world C and C++ functions~\cite{fakih2025llm4cve}. The pipeline first asks an LLM for a candidate fix, then applies a combination of static heuristics, compilation checks, and custom proof-of-concept exploits to judge whether the vulnerability is eliminated. Failure feedback, including compiler errors and exploit traces, is summarized and fed back into the next prompt. Using Llama-3-70B~\cite{grattafiori2024llama}, three feedback rounds improved ground-truth similarity by $20\%$ and earned a human quality score of $8.51/10$.

\textbf{IntDiagSolver} targets crash bugs whose fixes require code or environment diagnosis rather than only editing a faulty statement~\cite{du2026environmentcrash}. Its workflow is organized as a fixed loop: the LLM first asks targeted questions to complete missing contextual information, then generates a candidate solution, and finally the solution is checked by execution validation before another refinement round is triggered. With DeepSeek-R1 on 30 environment-related crash bugs, IntDiagSolver repairs 26 cases, corresponding to 86.7\% repair accuracy, compared with 43.3\% under the basic prompt. Its loop structure, validation step, and information-collection policy are scripted, while the injected evidence comes from crash context, dependency versions, configuration information, and execution outcomes.

Continuous analyzer feedback filters hallucinations, improves localization, and reduces overfitting to weak oracles. This benefit comes at the cost of longer running times and larger prompts, especially when many alerts remain. Future work should further cut overhead by reusing partial analysis results.

\begin{overviewbox}
\textbf{Key finding.} Procedural workflows improve predictability by fixing the retrieval, testing, and feedback loop in advance. Their main advantage is controllable cost and reproducibility; their main limitation is that deeper analysis must be triggered selectively to avoid unnecessary overhead.
\end{overviewbox}

\section{Agentic Approaches}
\label{sec:agentic}
Agentic frameworks delegate the choice of each repair step to the language model, rather than prescribing a fixed sequence in code. The key distinction is straightforward: when handwritten logic selects the next action, the approach is procedural, whereas a model-driven choice makes it agentic. Within this category, autonomy rises along three tiers. Tool-augmented agents place one model at the center and let LLMs call repository utilities, build or test commands, and even perform graphical actions, while the outer loop stays static \cite{yang2024swe,yang2025swem,wang2024openhands,bouzenia2025repairagent,zhang2024autocoderover,luo2025unlocking,liu2025agent}. LLM-gated review systems detach generation from evaluation by adding a critic LLM whose judgments gate which candidates or search branches are pursued next \cite{he2025code,zhou2024leveraging,li2024cleanvul,yadavally2025large}. Self-controlled systems hand one or more LLMs full authority over planning, delegation, and termination, enabling them to tackle function-level or multi-file bugs at the cost of heavier coordination \cite{tao2024magis,antoniades2025swe,orwall2024moatless,su2025learn}. The remainder of this section examines these three Agentic subtypes in depth.

\subsection{Tool-Augmented Agents}
\label{subsec:toolaug}
Tool-augmented agents place a frozen LLM at the center of a fixed perceive-think-act loop and expose a curated toolbox to that loop. Many systems in this family instantiate a ReAct-style interaction pattern~\cite{yao2023react}, in which reasoning steps and tool actions are interleaved, and observations from external tools condition subsequent repair decisions. In software repair, this template appears as repository shell/edit/test loops, browser or screenshot actions for multimodal repair, translation-and-compilation loops for low-resource languages, and debugger-driven vulnerability repair. We still group these systems by runtime control authority and tool/validation design, because ReAct describes the interaction pattern rather than the repair objective, benchmark scope, or validation mechanism. Systems that follow this pattern include SWE-Agent~\cite{yang2024swe}, SWE-Agent M~\cite{yang2025swem}, OpenHands~\cite{wang2024openhands}, RepairAgent~\cite{bouzenia2025repairagent}, AutoCodeRover~\cite{zhang2024autocoderover}, LANTERN~\cite{luo2025unlocking}, and VulDebugger~\cite{liu2025agent}. Their reported results span different benchmarks and repair scopes, so we treat them as protocol-specific evidence for tool-mediated repair rather than as a single accuracy ranking.

\textbf{SWE-Agent}~\cite{yang2024swe} uses GPT-4 Turbo in a ReAct~\cite{yao2023react} loop that interleaves \texttt{git\_grep}, targeted file edits, and unit-test runs to drive repair without any predefined call order. In large-scale experiments on SWE-bench~\cite{jimenezswe} it fixed 12.47\% of 2294 real issues under the authors' protocol while requiring no fine-tuning. The agent’s chief advantage is adaptability, as the model can dynamically choose the tool that seems most informative at each stage. Its drawback is cost: every reasoning step consumes new tokens and runtime, so the repair process is computationally expensive. Moreover, SWE-Agent cannot invent brand-new actions at run time; it can only select from the predefined toolkit registered by the authors.

\textbf{SWE-Agent M} extends the JavaScript-adapted SWE-Agent interface with web-browsing, screenshot-capture, and image-viewing tools for inspecting rendered interfaces~\cite{yang2025swem}. On the 517-task SWE-bench Multimodal test split, SWE-Agent M with GPT-4o resolves 12.2\% of tasks, compared with 12.0\% for SWE-Agent Base and 9.2\% for SWE-Agent JS.

\textbf{OpenHands} is an open platform for developing generalist and specialist agents that write and edit code, execute Bash commands and Python code in a Docker sandbox, browse the web, and delegate tasks to other agents~\cite{wang2024openhands}. Its CodeActAgent v1.8 with Claude 3.5 Sonnet resolves 26.0\% of the 300 SWE-bench Lite tasks without hint text. Integrating and orchestrating many external actions adds engineering overhead, and the system still depends on the model's judgment to apply each action correctly.

\textbf{AutoCodeRover} views each failing test as a sub-goal~\cite{zhang2024autocoderover}. It navigates the abstract syntax tree to focus the LLM on suspicious functions, applies spectrum-based fault localization when tests exist, and retests immediately after every patch. The agent repaired about 19\% of SWE-bench Lite issues at a mean cost of \$0.43 per bug. Structured context retrieval enhances search precision and minimizes tokens. Effectiveness drops when tests are weak or the project lacks a clear modular structure.

\textbf{RepairAgent} wraps an LLM inside a finite-state controller that exposes tools for reading files, running tests, compiling code, and editing sources~\cite{bouzenia2025repairagent}. The model chooses actions, observes outputs, and loops until tests pass or progress stalls, reaching 164 fixes on the Defects4J v1.2/2.0 benchmark. Tool freedom allows the agent to mimic a developer and achieve high accuracy on this suite. Long exploratory sessions consume many tokens, and careful guardrails are required to avoid infinite loops.

\textbf{LANTERN} tackles low-resource languages by translating buggy code into a language where the LLM is stronger, repairing it there, and translating the patch back~\cite{luo2025unlocking}. A decision module picks the intermediate language, and iterative testing guides further rounds of translation and refinement. On the xCodeEval~\cite{khan2024xcodeeval} benchmark, it raised Rust fix rates by over 22\% and improved other under-represented languages as well. Cross-language transfer leverages the strengths of LLMs beyond the source language. Accurate round-trip translation is challenging, and the multi-stage process introduces latency and the risk of semantic drift.

\textbf{VulDebugger} embodies a ``debug like a human'' philosophy for security repair~\cite{liu2025agent}. Starting from a proof-of-concept crash, the agent launches the program under a debugger, compares the \emph{actual} runtime state with \emph{expected} crash-free constraints, and iteratively edits code until the constraints hold. Across 50 real-world C/C++ projects, VulDebugger reports pass@1 of 60.00\% (30/50), which the authors compare favorably with prior security-agent repair results under their protocol. Dynamic state inspection pinpoints root causes that static retrieval misses, but the heavy reliance on platform-specific debuggers complicates deployment and slows throughput.

In summary, tool-augmented agents demonstrate that granting an LLM selective tool access can improve reported fix rates over one-shot prompting or fixed scripts in the evaluated settings. However, every additional reasoning step and external call still follow the same static loop. Future work should consider caching tool outputs, incorporating lightweight heuristics, or adaptively adjusting the reasoning budget to maintain flexibility while reducing the efficiency gap with purely scripted pipelines.

\subsection{LLM-Gated Review}
\label{subsec:llmjudge}

LLMs can serve not only as code generators but also as judges that evaluate candidate patches. This discriminative role, highlighted by He~\etal~\cite{he2025code}, positions them as scalable proxies for human reviewers when assessing whether a change fixes a defect or eliminates a vulnerability. In our taxonomy, a critic is agentic only when its output gates acceptance, branching, regeneration, or stopping; an offline score alone is auxiliary evaluation. Concrete judge modules now appear in 4 emerging lines of work: differential patch testing for benchmark validation~\cite{wang2025solved}, tree-search-guided patch evaluation in iterative repair loops~\cite{hu2025tsaprtreesearchframework}, dual abstention-and-validation funnels~\cite{cambronero2025abstain}, and patch validation~\cite{ruan2025specrover}.

\textbf{TSAPR} instantiates a tree-search APR framework in which an LLM collaborates with Monte Carlo Tree Search to iteratively select, refine, and evaluate candidate patches instead of following a single-path trial-and-error loop~\cite{hu2025tsaprtreesearchframework}. In this framework, the LLM plays a dual role as generator and judge: it produces new patches with chain-of-thought reasoning, then reuses the same model as a gating judge to score failing candidates based on buggy context, test results, execution feedback, and self-reflection traces, while falling back to a Test-as-Judge policy when test suites provide dense coverage. Empirically, TSAPR repairs 201 out of 835 bugs on Defects4J~\cite{just2014defects4j} under a pass@32 budget and 27 out of 79 vulnerabilities on VUL4J~\cite{bui2022vul4j}; the authors report these results against APR and AVR baselines with different patch budgets. These results suggest that combining global search with self-evaluating LLM critics can turn sparse test outcomes into informative rewards, strengthening LLM-gated review for complex multi-location and security-sensitive repairs.

\textbf{Abstain and Validate} introduces a dual LLM policy layer on top of an industrial agentic APR system that learns when to attempt a repair and when to surface a candidate patch to developers, aiming to reduce noise and wasted compute at the repository level during program repair~\cite{cambronero2025abstain}. As gating components, the bug abstention model reads the bug report and predicts the success probability of the underlying agent, deciding to skip low probability bugs. In contrast, the patch validation model generates a textual fix specification and then judges each candidate patch and its trajectory, assigning a correctness score that filters patches even after build and regression tests pass. On 2 additional sets of 198 null pointer exceptions and 50 sanitizer bugs from Google’s codebase, the dual policies raise filtered fail-to-pass@1 on human bugs from 11.29\% to 53\% and improve filtered accept@1 by up to 24pp on NPE bugs. These results show that wrapping an agentic repair loop with front-end abstention and back-end judge validation can sharply increase the precision and perceived reliability of APR in production.

\textbf{SpecRover} builds on AutoCodeRover by coupling iterative specification inference with a multi-agent LLM workflow that searches the codebase, synthesizes function-level intent summaries, and uses them to drive patch generation for GitHub issues on SWE-bench~\cite{ruan2025specrover}. In its review stage, a dedicated LLM reviewer agent acts as a gating reviewer, executing the reproducer test on the original and patched programs, inspecting inferred specifications, tests, and issue descriptions, and then either rejecting the patch with natural-language feedback or approving it and triggering regression testing. Empirically, SpecRover solves 19.3\% of issues on the full SWE-bench and 31.0\% on SWE-bench Lite at an average cost of \$0.65 per issue. These results suggest that positioning an LLM as a high-level reviewer that reconciles specifications, tests, and natural language intent can substantially strengthen LLM-gated review, turning it into a calibrated confidence filter and explanation generator rather than a standalone oracle.

LLM-gated review components recast patch evaluation as a learned decision problem, letting repair systems trade broader exploration for higher precision at acceptance time. Across tree search, abstention and validation policies, and specification-guided review, they already improve solve rates, suppress noisy patches that merely satisfy weak tests, and provide textual justifications that help developers understand decisions. Remaining challenges include calibration, robustness to spurious cues, and alignment with stronger oracles, so future work should more closely couple judges with static analysis and regression testing, learn explicit abstention behavior, and treat these critics as conservative gatekeepers rather than standalone authorities.

\subsection{Self-Controlled System}
\label{subsec:selfctrl}

Self-controlled systems remove the handwritten skeleton altogether. One or more LLMs generate or revise the global plan at run time, spawn new sub-tasks when needed, and decide when to terminate, so the model, not the script, controls the entire outer loop. Representative examples are MAGIS~\cite{tao2024magis}, which distributes work to four specialized roles; SWE-Search~\cite{antoniades2025swe,orwall2024moatless}, which couples LLMs with Monte Carlo tree search to debate and apply patches; and Learn-by-Interact~\cite{su2025learn}, which lets a single model explore an environment, refine its strategy, and distill the trajectory into reusable instructions. With full control, these agents can tackle multi-file or cross-module defects but incur higher coordination costs and safety risks.

\textbf{MAGIS} models a software team with four role-specific agents: a Manager plans high-level steps, a Repository Custodian pinpoints relevant files, Developer agents propose edits, and a QA agent validates patches~\cite{tao2024magis}. The agents communicate through shared memory, iteratively refine the plan, and trigger new sub-tasks until all tests pass or a time limit expires. On SWE-bench Lite, the original paper reports 25.33\% pass@1 under the authors' protocol. Division of labor gives MAGIS repository coverage and internal quality checks, while the coordination overhead remains high and many hard bugs are unresolved.

\textbf{SWE-Search} combines Monte Carlo Tree Search with a multi-agent control loop in which GPT-4o iteratively searches the repository, plans edits, and debates candidate patches before applying a fix~\cite{antoniades2025swe,orwall2024moatless}. On SWE-bench Lite, the original paper reports a 31\% solve rate with GPT-4o and a 23\% relative gain across five LLMs under the authors' protocol snapshot, while Claude 3.5 Sonnet reaches 39\% fixes at 2.7 issues per dollar. However, the search depth scales with inference-time computation and has been noted as ``probably pretty expensive,'' pointing to future work on cost-aware depth control and quicker repository traversal.

\textbf{Learn-by-Interact} is a self-controlled agentic framework that lets an LLM explore a new software or web environment, plan its action sequence, execute those actions, and retrospectively distill the trajectory into reusable instructions, so the LLM, rather than a fixed script, decides every step of the outer loop~\cite{su2025learn}. The authors fine-tune a Codestral-22B backbone on synthesized trajectories and also provide a training-free variant where Claude 3.5 Sonnet retrieves the distilled instructions at run time. On SWE-bench, the original paper reports 60.0\% pass@1 for Claude 3.5 Sonnet and up to 12.2pp improvement over the baseline under this protocol snapshot. The approach depends on accurate and up-to-date environment documentation; when documentation is sparse or changing, the data-synthesis phase can yield low-quality trajectories and incur substantial computational overhead.

Self-controlled agents show three recurring lessons in the surveyed public-benchmark studies. First, explicit role separation or self-generated sub-goals narrows the search space without sacrificing autonomy. Second, stable feedback signals, such as full test suites or critical debates, prevent blind exploration. Third, careful cost management remains crucial because token budgets, tool latency, and coordination overhead can offset gains in accuracy. Continued progress on adaptive stopping rules, incremental context caching, and lightweight self-reflection is essential before these fully autonomous agents can meet the throughput and reliability demands of continuous integration.

\begin{overviewbox}
\textbf{Key finding.} Agentic repair expands repository-level autonomy through tool use, planning, and iterative feedback, but its gains are limited by token cost, tool latency, context drift, and unstable control loops. Practical systems therefore need stopping rules, context caching, and lightweight safety checks.
\end{overviewbox}

Having surveyed fine-tuning, prompting, procedural pipelines, and agentic frameworks, we now synthesize how the primary taxonomy paradigms and auxiliary evidence layers jointly organize the design space and answer RQ2.

\begin{table}[h]
\centering
\caption{Distribution of the 66 selected systems across primary taxonomy paradigms and deployment scenarios. Counts are derived from the released scenario-assignment audit file.}
\label{tab:scenario-evidence}
\resizebox{\textwidth}{!}{%
\begin{tabular}{lrrrrr}
\toprule
\textbf{Deployment scenario} & \textbf{Fine-tuning} & \textbf{Prompting} & \textbf{Procedural} & \textbf{Agentic} & \textbf{Total} \\
\midrule
Localized benchmark repair & 17 & 13 & 6 & 3 & 39 \\
Repository-level issue resolution & 1 & 1 & 3 & 8 & 13 \\
Vulnerability repair & 3 & 3 & 3 & 1 & 10 \\
Industrial / practitioner workflow & 0 & 0 & 2 & 1 & 3 \\
Educational tutoring & 0 & 0 & 1 & 0 & 1 \\
\midrule
Total & 21 & 17 & 15 & 13 & 66 \\
\bottomrule
\end{tabular}}
\end{table}

Across the 66 systems, 15 of 29 retrieval-tagged systems target repository-level or industrial settings, compared with 1 of 37 systems without retrieval (odds ratio $38.57$; Fisher's exact $p=3.4\times10^{-6}$). This is the strongest of the 21 exploratory tag-outcome association tests we ran (three auxiliary-evidence tags crossed with seven coded outcomes) and the only one that remains significant after Benjamini--Hochberg correction across the full family ($q=7.2\times10^{-5}$). This descriptive association is non-causal: paradigm, scenario, task, and publication year are confounded. It supports treating context acquisition as an orthogonal design dimension, not a claim that retrieval or any paradigm universally improves repair. System-level counts, the complete test family, and definitions are released in \path{artifact/auxiliary_evidence_value_by_system.csv}, \path{artifact/auxiliary_evidence_value_tests.csv}, and \path{artifact/auxiliary_evidence_value_audit.md}.

\begin{rqbox}{RQ2}
The LLM-based software repair taxonomy is organized by four primary paradigms derived from parameter adaptation and runtime control, while auxiliary evidence sources such as retrieval, analysis, tests, human feedback, and domain knowledge cut across those paradigms. Table~\ref{tab:scenario-evidence} shows that localized benchmark repair accounts for the largest share of the corpus and is especially common among fine-tuning and prompting systems. Repository-level issue resolution, in contrast, is concentrated in procedural and agentic workflows, where systems can search files, retrieve project context, run tools, and validate patches across multiple steps. Vulnerability repair is more evenly distributed across fine-tuning, prompting, and procedural systems, reflecting its mixture of domain-specific training data, specialized context, and analysis-guided pipelines. These counts support a protocol-conditioned interpretation of paradigm choice rather than a claim that one paradigm is universally preferable.
\end{rqbox}

\section{Benchmark Protocol Profiles and Reported Metrics}
\label{sec:bench-trends}
This section consolidates published pass@$k$ and task-specific repair metrics for widely used software repair benchmarks as protocol snapshots rather than as a live leaderboard. The unit of analysis is a reported repair-system result, not the benchmark dataset alone. We report each system's published metric and record the evaluation assumptions alongside the score. This maintains fidelity to the original evaluation setups and makes explicit differences in fault localization, scope, model choice, and candidate budget.

Shared benchmark names alone do not determine whether published repair scores are comparable. We therefore ask a narrower question: when can published repair-system results from different systems be compared without changing the meaning of the score? We analyze benchmark comparability at two levels. First, we identify narrow, protocol-aligned windows in which bounded comparison is defensible. Second, we summarize results in benchmark-family groups, so systems evaluated under different metrics or assumptions still inform the comparability analysis without being forced into an invalid ranking.

We use pass@$k$ as a unified candidate-budget label: up to $k$ candidate patches are sampled or considered, and a task is counted as solved if any candidate passes the test suite. When a wall-clock or token budget is enforced, we state it explicitly, for example, pass@$k$ within 5 hours. Metrics such as exact match, F1, CodeBLEU similarity, correct repairs, success rate, repair accuracy, avg@5, merged pull requests, and accept@1 are kept as task-specific metrics rather than being converted to pass@$k$.

We construct a protocol-aligned window only when reported rows share the same benchmark variant and metric, and we mark a narrower model-controlled window only when the base-model family and key repair-scope assumptions are also comparable. Differences that remain within a window, such as the tool interface, search budget, retrieval policy, validation depth, training data, or the exact model snapshot, are listed as confounders rather than hidden. Rows that fail these criteria stay in benchmark-family snapshot groups, so the analysis covers the full corpus without converting heterogeneous results into a leaderboard.

\begin{table}[t]
\centering
\caption{Protocol-aligned comparison windows identified from reported repair results. We interpret results only within these windows; other rows remain protocol snapshots. Rows within a window are grouped by control profile rather than ranked by score.}
\label{tab:protocol-aligned-windows}
\resizebox{\textwidth}{!}{
\begin{tabular}{p{3.2cm} p{4.8cm} p{3.2cm} p{4.2cm} p{4.5cm}}
\toprule
\textbf{Window} & \textbf{Systems and scores} & \textbf{Controlled factors} & \textbf{Remaining confounders} & \textbf{Supported interpretation} \\
\midrule
SWE-bench Lite pass@1 &
\begin{tabular}[t]{@{}l@{}}
KGCompass 58.30\% \\
Agentless 32.00\% \\
SWE-Agent 18.00\% \\
AutoCodeRover 19.00\% \\
OpenHands 26.00\% \\
SpecRover 31.00\% \\
MAGIS 25.33\% \\
SWE-Search 31.00\%
\end{tabular} &
Same benchmark variant and pass@1 metric & Base model, cost budget, search depth, retrieval policy, validation strategy, and publication snapshot & Whole-system repository-repair stacks are comparable as protocol snapshots; results should not be read as isolated paradigm effects. \\
\midrule
EvalRepair-Java pass@10, CodeLlama 13B, no fault-location prompt &
\begin{tabular}[t]{@{}l@{}}
MORepair 77.90\% \\
Luo et al. 76.88\%
\end{tabular} &
Same augmented benchmark, pass@10 metric, base model, and no-FL-prompt setting & Training corpus, adaptation objective, and federated versus multi-objective fine-tuning setup & Supports a narrow comparison of adaptation strategies under an augmented Java repair oracle. \\
\midrule
HumanEval-Java pass@10, adapter-tuning-oriented systems &
\begin{tabular}[t]{@{}l@{}}
Ruiz et al. 78.05\% \\
Li et al. 68.10\% \\
RepairLLaMA 67.28\%
\end{tabular} &
Same benchmark and pass@10 metric & Base model, training data, adaptation objective, and localization reporting differ & Supports a bounded adapter-tuning-oriented discussion, but not a model-controlled ranking. \\
\bottomrule
\end{tabular}}
\end{table}

\begin{table}[t]
\centering
\caption{Benchmark-family protocol groups. These groups summarize how the surveyed systems contribute to the comparability analysis while preserving protocol differences.}
\label{tab:benchmark-coverage-groups}
\resizebox{\textwidth}{!}{
\begin{tabular}{p{3.4cm} p{0.8cm} p{4.3cm} p{4.8cm} p{4.8cm}}
\toprule
\textbf{Benchmark group} & \textbf{\#} & \textbf{Shared basis} & \textbf{Non-comparable dimensions} & \textbf{Safe interpretation} \\
\midrule
Defects4J, Perfect-FL or function-level setting & 6 & Same benchmark family; all rows use localized Java APR settings with Perfect-FL or comparable localized inputs & Candidate budgets differ (pass@$k$, pass@1, pass@5, pass@10, pass@200, pass@1000), as do denominators, benchmark versions, and base models & Shows how reported Defects4J scores depend on localization hints and sampling budget, not a direct ranking. \\
Defects4J, single-function setting & 3 & Same benchmark family and single-function scope & pass@10/pass@25/pass@500 budgets and denominators differ & Supports comparison of narrowed-scope repair protocols and test-feedback budgets. \\
Defects4J, tool/search systems without explicit Perfect-FL control & 2 & Same broad benchmark family and GPT-3.5-class repair systems & pass@1 versus a 32-patch search budget, tool autonomy, search policy, and FL assumptions differ & Provides end-to-end or search-heavy protocol evidence without claiming score comparability. \\
Defects4J, single-hunk setting & 1 & Same benchmark family but singleton scope & No surveyed peer uses the same single-hunk protocol & Records subset specialization. \\
HumanEval-Java pass@10 & 3 & Same benchmark and pass@10 metric & Base model, training data, PEFT objective, and localization reporting differ & Supports bounded comparison of adapter-tuning-oriented repair systems. \\
HumanEval-Java, other metrics & 3 & Same benchmark & pass@100, pass@40, and accuracy are different metrics & Shows metric heterogeneity within the same benchmark. \\
EvalRepair-Java pass@10 & 2 & Same HumanEval-Java-derived augmented benchmark, pass@10 metric, CodeLlama 13B base model, and no-FL-prompt setting & Training data, adaptation objective, and privacy/federation settings differ & Useful for showing benchmark evolution and augmented-oracle practice, but kept separate from HumanEval-Java pass@10 comparisons. \\
SWE-bench Lite pass@1 & 8 & Same benchmark variant and pass@1 metric & Base model, model snapshot, cost budget, search depth, retrieval policy, and validation depth differ & Most populated protocol-aligned whole-stack comparison window among the surveyed systems. \\
Other SWE-bench variants & 3 & Same repository-level benchmark family & Verified, Multimodal, full-benchmark, and environment-conditioned variants differ & Contributes to repository-level protocol analysis but remains separate from Lite pass@1. \\
Vulnerability, API-misuse, and crash repair & 11 & Security or robustness-oriented repair tasks & Datasets and metrics are disjoint, including exact match/EM, F1, CodeBLEU similarity, correct repairs, success rate, and repair accuracy & Reveals the lack of a consolidated benchmark for security/API/crash repair. \\
Other task-specific or singleton benchmarks & 24 & Systems outside the shared families above & Rows span non-overlapping or singleton benchmarks, including APR Competition, QuixBugs, BugsInPy, RepoBugs, DS-1000, TutorCode, Flink, and Google accept@1 & Provides coverage and protocol-diversity evidence, not cross-paper ranking. \\
\bottomrule
\end{tabular}}
\end{table}

\begin{table}[t]
\centering
\caption{Distribution of reported pass@$k$ and related metric families by publication year.}
\label{tab:passk-reporting}
\resizebox{\textwidth}{!}{
\begin{tabular}{l r r r r r r r}
\toprule
\textbf{Year} & \textbf{Total} & \textbf{pass@1} & \textbf{pass@5} & \textbf{pass@10} & \textbf{Other pass@$k$} & \textbf{Single-candidate non-pass@$k$} & \textbf{Non-pass@$k$} \\
\midrule
2022 & 2 & 1 & 0 & 0 & 1 & 0 & 0 \\
2023 & 12 & 1 & 3 & 2 & 4 & 0 & 2 \\
2024 & 16 & 5 & 1 & 1 & 4 & 0 & 5 \\
2025 & 30 & 13 & 3 & 5 & 3 & 0 & 6 \\
2026 & 6 & 1 & 0 & 1 & 0 & 1 & 3 \\
\midrule
Total & 66 & 21 & 7 & 9 & 12 & 1 & 16 \\
\bottomrule
\end{tabular}}
\end{table}

Table~\ref{tab:passk-reporting} shows that pass@1 is the most common single pass@$k$ value in the surveyed systems, appearing in 21 of 66 rows. Its use increases in the 2024--2025 portion of the corpus: 1 of 12 systems in 2023, 5 of 16 systems in 2024, and 13 of 30 systems in 2025 report pass@1. We keep accept@1-style single-candidate metrics separate from pass@$k$ because they do not denote the same sampling protocol. The pattern is not uniform: the 11 security/API/crash-repair rows mix pass@1/pass@5 with exact match/EM, F1, CodeBLEU similarity, correct repairs, success rate, and repair accuracy rather than one shared reporting convention.

\begin{table}[h]
\centering
\caption{Defects4J protocol snapshot.}
\label{tab:defects4j-trend}
\resizebox{\textwidth}{!}{
\begin{tabular}{l l l l l l l}
\toprule
Year & Paradigm & System & Base model & Metric & Notes & Ref. \\
\midrule
2022 & Prompting (zero-shot) & AlphaRepair & CodeBERT & pass@$k$ within 5 hours: 74/109 & Perfect fault localization. & \cite{xia2022less} \\
2023 & Prompting (few-shot) & Xia \textit{et al.} & Codex 12B & pass@200: 99/255 & Perfect fault localization. & \cite{xia2023automated} \\
2023 & Full fine-tuning & Huang \textit{et al.} & CodeT5 & pass@5: 82/107 & Perfect fault localization. & \cite{huang2023empirical} \\
2023 & Full fine-tuning & Jiang \textit{et al.} & InCoder 7B & pass@10: 31/138 & Perfect fault localization. & \cite{jiang2023impact} \\
2023 & Scripted Tool Loop + analysis tag & Repilot & InCoder 7B & pass@5000: 116/273 & Single-hunk Java bugs. & \cite{wei2023copiloting} \\
2023 & Knowledge distillation & KNOD & Not specified & pass@1000: 122/837 & Perfect fault localization. & \cite{jiang2023knod} \\
2025 & Prompting + analysis tag & D4C & GPT-4 & pass@10: 180/437 & Single-function setting, no FL hints. & \cite{xu2025aligning} \\
2024 & Test-Feedback Loop & ChatRepair & GPT-3.5 & pass@500: 162/337 & Single-function setting. & \cite{xia2024automated} \\
2024 & Tool-augmented agent & RepairAgent & GPT-3.5 & pass@1: 164/835 & Repository-level execution with tools. & \cite{bouzenia2025repairagent} \\
2024 & Distillation (non-AR student) & NARRepair & CodeT5 & pass@1: 110/127 & Perfect fault localization. & \cite{yang2024narrepair} \\
2025 & LLM-Gated Review & TSAPR & GPT-3.5 & pass@32: 24.07\% (201/835) & Judge-guided search; FL assumption not explicit. & \cite{hu2025tsaprtreesearchframework} \\
\bottomrule
\end{tabular}}
\end{table}

On Defects4J (Table~\ref{tab:defects4j-trend}), the main signal is protocol diversity within the same benchmark family rather than a clean temporal curve. Raw solved-bug counts are hard to compare because pass@$k$ ranges from 1 to 5000, several rows assume perfect localization, and some systems evaluate only on selected subsets, single-hunk bugs, or single-function variants. Table~\ref{tab:benchmark-coverage-groups} therefore separates Defects4J rows into localized/Perfect-FL, single-function, tool/search, and single-hunk groups; these rows support analysis of protocol differences rather than a global score ordering.

\begin{table}[h]
\centering
\caption{HumanEval-Java and EvalRepair-Java protocol snapshots.}
\label{tab:hejava-trend}
\resizebox{\textwidth}{!}{
\begin{tabular}{l l l l l l l}
\toprule
Year & Paradigm & System & Base model & Metric & Notes & Ref. \\
\midrule
\multicolumn{7}{c}{\textit{Group H1: HumanEval-Java pass@10}} \\
2025 & Adapter Tuning (iterative) & Ruiz \textit{et al.} & DeepSeek-Coder 6.7B & pass@10: 78\% & LoRA fine-tuning; iterative repair; source-reported percentage. & \cite{ruiz2025art} \\
2024 & Adapter Tuning & Li \textit{et al.} & DeepSeek-Coder 6.7B & pass@10: 68.10\% & Perfect fault localization. & \cite{li2024exploring} \\
2025 & Adapter Tuning & RepairLLaMA & CodeLlama 7B & pass@10: 67.28\% & Code-region localization. & \cite{silva2025repairllama} \\
\midrule
\multicolumn{7}{c}{\textit{Group H2: EvalRepair-Java pass@10, CodeLlama 13B}} \\
2025 & Adapter Tuning (multi-objective) & MORepair & CodeLlama 13B & pass@10: 77.90\% & EvalPlus-augmented tests; no fault-location prompt. & \cite{yangmorepair} \\
2025 & Adapter Tuning (federated) & Luo \textit{et al.} & CodeLlama 13B & pass@10: 76.88\% & EvalPlus-augmented tests; no fault-location prompt. & \cite{luo2025fine} \\
\midrule
\multicolumn{7}{c}{\textit{Group H3: HumanEval-Java with different metrics}} \\
2025 & Full FT + evidence & NTR & CodeLlama 70B & pass@100: 90.6\% & Template guidance; perfect FL. & \cite{huang2024template} \\
2025 & Test-Feedback Loop & ContrastRepair & GPT-3.5 & pass@40: 84.05\% & Contrastive test pairs. & \cite{kong2025contrastrepair} \\
2025 & Prompting + analysis tag & TracePrompt & GPT-4 & accuracy: 71.30\% & Concise execution traces. & \cite{haque2025traceprompt} \\
\bottomrule
\end{tabular}}
\end{table}

On HumanEval-Java and EvalRepair-Java (Table~\ref{tab:hejava-trend}), the most comparable rows are grouped contiguously rather than ranked as a single list. Group H1 controls benchmark and pass@10 but still differs in model and localization reporting. Group H2 is the narrowest model-controlled Java repair window in the corpus: both rows use EvalRepair-Java, CodeLlama 13B, pass@10 reporting, and no fault-location prompt. Group H3 keeps HumanEval-Java rows with pass@100, pass@40, and accuracy as protocol snapshots rather than ranking them against pass@10 rows.

\begin{table}[h]
\centering
\caption{SWE-bench protocol snapshot. Verified and Lite variants are reported separately.}
\label{tab:swebench-trend}
\resizebox{\textwidth}{!}{
\begin{tabular}{l l l l l l l l}
\toprule
Year & Variant & Paradigm & System & Base model & Metric & Notes & Ref. \\
\midrule
2024 & Lite & Tool-augmented agent & SWE-Agent & GPT-4 Turbo & pass@1: 18.00\% & Repository-level issue repair. & \cite{yang2024swe} \\
2025 & Lite & Tool-augmented agent & OpenHands & Claude 3.5 Sonnet & pass@1: 26.00\% & Generalist CodeAct-style agent. & \cite{wang2024openhands} \\
2024 & Lite & Tool-augmented agent & AutoCodeRover & GPT-4 & pass@1: 19.00\% & AST-guided search and test feedback. & \cite{zhang2024autocoderover} \\
2024 & Lite & Scripted Tool Loop + retrieval tag & Agentless & GPT-4o & pass@1: 32.00\% & Three-stage localization, repair, validation. & \cite{xia2025demystifying} \\
2025 & Lite & Scripted Tool Loop + retrieval tag & KGCompass & Claude Sonnet 4 & pass@1: 58.30\% & Repository-aware knowledge graph. & \cite{yang2025enhancing} \\
2025 & Lite & LLM-Gated Review & SpecRover & Claude 3.5 Sonnet + GPT-4o & pass@1: 31.00\% & Specification-guided patch review. & \cite{ruan2025specrover} \\
2024 & Lite & Self-controlled agent & MAGIS & GPT-4 & pass@1: 25.33\% & Multi-agent role decomposition. & \cite{tao2024magis} \\
2025 & Lite & Self-controlled agent & SWE-Search & GPT-4o & pass@1: 31.00\% & Search-based repository exploration. & \cite{antoniades2025swe} \\
\midrule
2025 & Verified & RLFT + procedural & SWE-RL & Llama 3 70B & pass@1: 41.00\% & Verified singleton in this corpus; uses an Agentless-style pipeline. & \cite{wei2025swe} \\
2025 & Multimodal & Tool-augmented agent & SWE-Agent M & GPT-4o & pass@1: 12.2\% & Visual/front-end issue variant. & \cite{yang2025swem} \\
2025 & Full/conditioned & Self-controlled agent & Learn-by-Interact & Claude 3.5 Sonnet & pass@1: 60.00\% & Environment documentation and synthesized trajectories. & \cite{su2025learn} \\
\bottomrule
\end{tabular}}
\end{table}

\begin{table}[h]
\centering
\caption{SWE-bench Lite pass@1 comparison window. These rows share the benchmark variant and metric, so they support a bounded whole-stack comparison, but not an isolated comparison of algorithms or paradigms. Rows are grouped by control profile, not ranked by score.}
\label{tab:swebench-lite-pass1-window}
\resizebox{\textwidth}{!}{
\begin{tabular}{l l l l l}
\toprule
System & Paradigm & Base model & Reported pass@1 & Main protocol distinction \\
\midrule
KGCompass & Scripted Tool Loop + retrieval tag & Claude Sonnet 4 & 58.30\% & Knowledge-graph-guided repository retrieval. \\
Agentless & Scripted Tool Loop + retrieval tag & GPT-4o & 32.00\% & Staged localization, repair, and validation. \\
SWE-Agent & Tool-augmented CLI agent & GPT-4 Turbo & 18.00\% & Agent-computer interface for repository repair. \\
AutoCodeRover & Tool-augmented repository agent & GPT-4 & 19.00\% & AST-guided search and test feedback. \\
OpenHands & Tool-augmented generalist agent & Claude 3.5 Sonnet & 26.00\% & Generalist CodeAct-style action interface. \\
SpecRover & LLM-Gated Review/specification-guided agent & Claude 3.5 Sonnet + GPT-4o & 31.00\% & Specification-guided patch review. \\
MAGIS & Multi-agent self-controlled system & GPT-4 & 25.33\% & Multi-agent role decomposition. \\
SWE-Search & Self-controlled search agent & GPT-4o & 31.00\% & Search-based repository exploration. \\
\bottomrule
\end{tabular}}
\end{table}

\begin{table}[h]
\centering
\caption{SWE-bench Verified pass@1 results available in surveyed-paper contexts. Rows cited to SWE-RL are comparison baselines reported by SWE-RL rather than the original systems' own SWE-bench Verified reports.}
\label{tab:swebench-verified-context}
\resizebox{\textwidth}{!}{
\begin{tabular}{l l l l}
\toprule
System & Model & Reported pass@1 & Ref. \\
\midrule
OpenHands & Claude 3.5 Sonnet & 53.00\% & \cite{wei2025swe} \\
Agentless & Claude 3.5 Sonnet & 50.80\% & \cite{wei2025swe} \\
AutoCodeRover & Claude 3.5 Sonnet & 46.20\% & \cite{wei2025swe} \\
SWE-RL & Llama3-SWE-RL-70B with Agentless Mini & 41.00\% & \cite{wei2025swe} \\
Agentless & GPT-4o & 38.80\% & \cite{xia2025demystifying} \\
SWE-Agent & Claude 3.5 Sonnet & 33.60\% & \cite{wei2025swe} \\
SWE-Agent & GPT-4o & 23.20\% & \cite{wei2025swe} \\
\bottomrule
\end{tabular}}
\end{table}

On SWE-bench (Tables~\ref{tab:swebench-trend}, \ref{tab:swebench-lite-pass1-window}, and~\ref{tab:swebench-verified-context}), the most populated comparison window among the surveyed results is SWE-bench Lite pass@1. Within that window, scores reported by the original papers range from 18.00\% to 58.30\%, but they should be interpreted as whole-system stacks under the respective authors' protocol rather than isolated paradigm effects because base models, search depth, cost budgets, and validation policies differ. SWE-bench Verified adds a human-checked, repository-level benchmark, but it is currently reported less uniformly across the surveyed systems. Table~\ref{tab:swebench-verified-context} therefore records Verified pass@1 results available in surveyed-paper contexts, including comparison baselines reported by SWE-RL, and we use it only as repository-level context rather than as an additional protocol-aligned comparison window.

Heterogeneous metrics and sampling budgets further obscure whether improvements arise from model choice, more generous pass@$k$, more extensive retrieval, or additional validation. Table~\ref{tab:protocol-aligned-windows} makes this distinction explicit: only a few windows support bounded comparison, while the remaining rows are used to diagnose protocol fragmentation and reporting gaps.

Beyond these common benchmarks, vulnerability, API-misuse, crash repair, and several task-specific benchmarks remain fragmented across small or non-overlapping datasets. Table~\ref{tab:fragmented-security-protocols} lists the security/API/crash rows from the 66-system corpus, while Table~\ref{tab:benchmark-coverage-groups} records how the remaining systems fall into same-family or singleton groups such as QuixBugs, BugsInPy, RepoBugs, DS-1000, and GitBug-Java. These rows should not be read as a leaderboard; instead, they show why reported security-repair and domain-specific repair scores are harder to compare than the narrow SWE-bench Lite, HumanEval-Java, and EvalRepair-Java windows.

\begin{table}[h]
\centering
\caption{Fragmented vulnerability, API-misuse, and crash-repair protocols.}
\label{tab:fragmented-security-protocols}
\resizebox{\textwidth}{!}{
\begin{tabular}{l l l l l l}
\toprule
System & Dataset or task scope & Base model & Metric & Result & Ref. \\
\midrule
VulMaster & CVEFixes & CodeT5 & EM & 20.00\% & \cite{zhou2024out} \\
Vul-R2 & PrimeVul & Qwen-14B-Instruct-1M & exact match & 24.83\% & \cite{wen2025vul} \\
APPATCH & Zero-Day & Claude 3.5 Sonnet & F1 & 36.46\% & \cite{nong2025appatch} \\
SAN2PATCH & VulnLoc & GPT-4o & success rate & 79.49\% (31/39) & \cite{kim2025logs} \\
PredicateFix & CVE static-analysis alarms & Claude 3.5 Sonnet & correct repairs & 81/117 (69.2\%) & \cite{xiao2025predicatefixrepairingstaticanalysis} \\
LLM4CVE & CVEFixes & Llama 3 70B & CodeBLEU similarity & +20.00\% & \cite{fakih2025llm4cve} \\
VulDebugger & 50 ExtractFix/ARVO cases & GPT-4o & repair accuracy & 60.00\% (30/50) & \cite{liu2025agent} \\
PailGen & Big-Vul\_CVEfixes & DeepSeek-Coder V2 & EM & 23.23\% & \cite{shao2026pailgen} \\
ACFix & 118 access-control vulnerabilities & GPT-4 & success rate & 94.92\% & \cite{zhang2025acfix} \\
Dr.Fix & Java-Method API-misuse repair & GPT-4o & EM & 96.00\% & \cite{zhuo2026apimisuse} \\
IntDiagSolver & 30 environment-related crash bugs & DeepSeek-R1 & repair accuracy & 86.7\% & \cite{du2026environmentcrash} \\
\bottomrule
\end{tabular}}
\end{table}

\begin{table}[h]
\centering
\caption{Evaluation-reliability risk coding used for RQ3. Counts are corpus-level protocol attributes, not quality scores.}
\label{tab:evaluation-risk-coding}
\resizebox{\textwidth}{!}{
\begin{tabular}{p{4.0cm} p{6.8cm} p{1.4cm} p{5.0cm}}
\toprule
\textbf{Risk or reporting factor} & \textbf{Coding criterion} & \textbf{\# systems} & \textbf{Interpretation} \\
\midrule
Perfect fault localization assumption & Paper assumes the fault location is provided or evaluates under a comparable localized setting. & 10/66 & Scores isolate patch generation but should not be read as end-to-end repair success. \\
Train/test-split-style evaluation & Paper reports a train/test split, temporal split, or benchmark construction that separates evaluation examples from training or tuning data. & 7/66 & Split reporting improves auditability but does not by itself eliminate all pretraining leakage risks. \\
Narrowed repair scope & Evaluation is restricted to single-function, single-hunk, code-region, or function-level repair. & 6/66 & Results support capability claims under narrowed scope rather than repository-scale generality. \\
Explicit contamination or leakage discussion & Paper explicitly discusses possible data contamination, leakage, benchmark overlap, or related mitigation concerns. & 36/66 & Contamination awareness is common but unevenly operationalized across systems. \\
Active leakage mitigation or diagnostic control & Paper reports a concrete measure such as decontamination, temporal or train/test split design, benchmark construction to avoid overlap, dataset exclusion, or leakage analysis tied to evaluation. & 28/66 & Concrete controls remain less common than discussion-only acknowledgments. \\
Hidden, manual, or independent validation beyond public tests & Paper reports hidden tests, additional generated tests, manual/expert validation, sanitizer or exploit validation, LLM-gated review, or another independent acceptance check. & 39/66 & Extra validation mitigates weak public-test oracles but varies in strength and reproducibility. \\
\bottomrule
\end{tabular}}
\end{table}

The per-system labels and corpus-level counts behind Table~\ref{tab:evaluation-risk-coding} are released in \path{artifact/evaluation_reliability_by_system.csv} and \path{artifact/evaluation_reliability_risk_coding.csv}. The pair-attrition path, the sole model-controlled pair, and the associated definitions and caveats are released in \path{artifact/protocol_comparability_attrition.csv}, \path{artifact/protocol_comparability_pairwise.csv}, and \path{artifact/protocol_comparability_evidence.md}.

\begin{rqbox}{RQ3}
Across the reported results of the 66 surveyed systems, there are 2,145 unordered system pairs, of which 100 share a recorded benchmark label. Of those 100, 32 fall within three explicitly bounded comparison windows, four also share a normalized base-model label, and one is model-controlled under our stated rule: it shares the benchmark variant, metric and candidate budget, normalized base model, and fault-localization and oracle assumptions. This sole model-controlled pair, MORepair versus Luo et al. on EvalRepair-Java, lies within Fine-Tuning; no model-controlled cross-paradigm comparison exists. Benchmark-name overlap therefore does not support a paradigm leaderboard.

SWE-bench Lite, EvalRepair-Java, and HumanEval-Java support only bounded questions, while Defects4J remains split by localization, scope, and search protocol. The assurance coding shows that 36/66 systems discuss leakage or contamination, 28/66 report active controls, and 39/66 use validation beyond ordinary public tests. Future studies should report benchmark variants, denominators, model snapshots, localization assumptions, validation depth, leakage controls, and repeated-run variance; until then, narrow gains are evidence of local progress rather than general paradigm superiority.
\end{rqbox}

\section{Open Challenges and Practical Implications}
\label{sec:challenges}

The challenges below are grounded in recurring limitations exposed by the surveyed systems and evaluation protocols, rather than in generic expectations about LLMs. Table~\ref{tab:readiness-profile} first decomposes each challenge into evidence-backed sub-challenges, and the following paragraphs unpack why these sub-challenges matter for future repair systems.

\begin{table*}[t]
\scriptsize
\caption{Evidence profile behind the five open challenges.}
\label{tab:readiness-profile}
\renewcommand{\arraystretch}{0.95}
\resizebox{\textwidth}{!}{%
\begin{tabular}{p{2.6cm} p{3.4cm} p{6.1cm} p{4.1cm}}
\toprule
\textbf{Challenge area} & \textbf{Sub-challenge surfaced by the corpus} & \textbf{Evidence in the 66-system corpus} & \textbf{Practical implication} \\
\midrule
Training and deployment cost
& Model access and reproducibility
& 38/66 systems use proprietary/API models, 27/66 use open or academic models, and 1/66 does not specify the base model. The split is paradigm-skewed: 19/21 fine-tuning systems use open or academic models, whereas 37/45 prompting, procedural, and agentic systems use proprietary/API models.
& Report model source, model snapshot, and deployment constraints alongside repair accuracy. \\
& Iterative inference budget
& 28/66 systems are procedural or agentic. Their subtypes include 8 scripted tool loops, 7 tool-augmented agent systems, 4 test-feedback-loop workflows, and 3 self-controlled systems; representative examples include ChatRepair, Agentless, SWE-Search, and Learn-by-Interact~\cite{xia2024automated,xia2025demystifying,antoniades2025swe,su2025learn}.
& Report token budget, repeated-call budget, latency, and success per unit cost. \\
\midrule
Evaluation reliability
& Scope and localization assumptions
& 10/66 systems explicitly assume perfect fault localization; 7/66 use train/test-split-style evaluation; and 6/66 narrow the task to single-function, single-hunk, code-region, or function-level repair.
& Mark localization assumptions, task scope, and end-to-end validity before comparing pass@$k$ or task-specific repair metrics. \\
& Contamination reporting and control
& 36/66 systems explicitly discuss benchmark, pretraining, train/test, temporal, or corpus contamination/leakage risk; 28/66 report active mitigation or diagnostic control.
& Separate discussion-only transparency from concrete leakage mitigation or diagnostic controls. \\
& Oracle strength and extra validation
& 39/66 systems use hidden, held-out, or additional tests, manual/expert review, LLM/human review, static/runtime validation, sanitizer/exploit reproduction, or comparable independent validation beyond ordinary public tests.
& Report extra validation as a protocol attribute rather than a directly comparable score. \\
& Outcome comparability
& 21/66 systems report pass@1, 28/66 report larger pass@$k$ candidate-budget metrics, and 17/66 use single-candidate or non-pass@$k$ task-specific metrics. Only 13/66 systems fall into model-controlled, protocol-aligned, or limited within-benchmark comparison tiers; 53/66 are snapshot-only rows.
& Use protocol-aligned reporting templates, explicit denominators, stronger or hidden tests, and contamination-aware splits. \\
\midrule
Workflow robustness
& Repository-scale state space
& 13/66 systems target repository-level issue resolution, including 8 agentic systems. Representative workflows include AutoCodeRover, RepairAgent, SWE-Search, and OpenHands~\cite{zhang2024autocoderover,bouzenia2025repairagent,antoniades2025swe,wang2024openhands}.
& Evaluate multi-file, multi-hunk, and project-wide dependencies as first-class repair settings. \\
& Control reliability
& 28/66 systems are procedural or agentic and therefore organize repair as a multi-step scripted or model-controlled loop.
& Report stopping rules, tool traces, failure recovery, and run-to-run stability. \\
\midrule
Retrieval and context management
& Evidence selection
& The taxonomy table records explicit retrieval-enriched prompting/procedural rows for 7 systems, including PailGen, KGCompass, and PATCH~\cite{shao2026pailgen,yang2025enhancing,zhang2025patch}.
& Measure context recall, retrieval failure modes, and context-budget trade-offs. \\
& Structured and interactive evidence
& The taxonomy table records explicit analysis-enriched prompting/procedural rows for 10 systems, including SAN2PATCH, PredicateFix, and Dr.Fix~\cite{kim2025logs,xiao2025predicatefixrepairingstaticanalysis,zhuo2026apimisuse}; it also records 4 test-feedback loops and 3 human-feedback loops.
& Preserve structured analysis facts and combine retrieval, tests, and feedback selectively. \\
\midrule
Patch assurance and deployment trust
& Fragmented safety evidence
& Security/API/crash repair covers 11 systems, but these rows use disjoint datasets and metric families, including exact match/EM, F1, CodeBLEU similarity, correct repairs, success rate, and repair accuracy.
& Build comparable safety-oriented repair protocols. \\
& Limited explicit acceptance gates
& 10/66 systems use test feedback, human feedback, or LLM-gated review as the recorded control subtype; 56/66 do not expose such a gate as their main repair-control mechanism.
& Add acceptance gates that verify patches, rationales, and side effects before deployment. \\
\bottomrule
\end{tabular}}
\end{table*}

The corpus signals, counts, source files, and representative systems behind Table~\ref{tab:readiness-profile} are released in \path{artifact/challenge_evidence_profile.csv}.

\textbf{Training and deployment cost.} The cost block in Table~\ref{tab:readiness-profile} separates two measurable cost sources: model access and iterative inference budget. In the corpus, 19/21 fine-tuning systems use open or academic models, whereas 37/45 prompting, procedural, and agentic systems use proprietary/API models. This split makes cost and reproducibility central deployment concerns rather than incidental implementation details. For the 21 fine-tuning systems, deployment assessment depends on whether papers report training data, model adaptation, and training resources. At inference time, 28/66 procedural or agentic systems expose repair cost to loop design, tool use, test feedback, or model-controlled stopping decisions; their subtypes include 8 scripted tool loops, 7 tool-augmented agent systems, 4 test-feedback-loop workflows, and 3 self-controlled systems~\cite{xia2024automated,xia2025demystifying,antoniades2025swe,su2025learn}. For these loop-oriented systems, each additional repair round increases token, tool, or test-execution cost, so cost reporting should include loop budget and stopping criteria rather than final accuracy alone.

\textbf{Evaluation reliability.} The evaluation block separates scope assumptions, leakage reporting, oracle strength, and outcome comparability. Perfect fault localization, train/test splits, and single-function or single-hunk settings are useful for isolating specific capabilities, but they should not be taken as definitive evidence of end-to-end repair. Benchmarks may also overlap with pretraining data, which causes \emph{data leakage} and inflated pass@$k$ or task-specific repair metrics. Weak test suites let a patch pass if all tests succeed, and underspecified test suites can miss semantic faults of LLM-generated patches. This leads to \emph{patch overfitting}: the LLM produces a plausible patch that passes all available tests yet still contains hidden defects, a long-standing APR concern that becomes harder to audit when LLM training corpora and repository-level benchmarks are only partially transparent~\cite {smith2015cure,petke2024patch}. In the corpus, 36/66 systems explicitly discuss leakage or contamination risks, 28/66 report active mitigation or diagnostic control, and 39/66 use some validation beyond ordinary public tests. These checks improve interpretability, but their heterogeneous strength and reporting formats mean they should be treated as protocol attributes rather than directly comparable scores.

\textbf{Workflow robustness.} The workflow block separates repository-scale state space from control reliability. Repository-level systems are closer to real deployment than localized benchmarks, but they also expose more sources of instability. Minor changes to prompt wording, sampling temperature, or random seeds can alter the generated patch and turn a successful repair into a failure. In agentic workflows, this randomness can accumulate across steps, so two runs of the same system can follow different tool sequences and yield different results~\cite{bouzenia2025repairagent,antoniades2025swe,su2025learn}. Language models may also produce invalid commands, ignore provided rules, or loop for too long, making the total running time unpredictable. These issues make it hard to guarantee repeatable outcomes or keep running costs within fixed limits in real-world pipelines.

\textbf{Retrieval and context management.} The context block separates evidence selection from structured and interactive evidence use. Retrieval, analysis, tests, and human feedback are not isolated sub-paradigms, but recurring auxiliary evidence sources. This makes context quality a cross-cutting bottleneck. The 7 retrieval-enriched prompting/procedural systems in the corpus rely on retrieved code, patches, patterns, or graph context, but such retrieval can miss key files or functions and gives no recall guarantee~\cite{yang2025enhancing,zhang2025patch,shao2026pailgen}. Even when relevant snippets are found, the added context can push the prompt beyond the LLM's context limitation, limiting the effectiveness of retrieval augmentation. Recent adaptive memory systems, such as A-MEM, store each interaction as a structured note, maintain a link graph, and supply only a compact top-$k$ context, thereby improving recall while staying within the token budget~\cite{xu2025mem}. The 10 analysis-enriched prompting/procedural systems generate diagnostics and traces directly from program analysis; however, structured data is still commonly serialized as prompt text, so the model may not fully exploit its logical structure.

\textbf{Patch assurance and deployment trust.} The assurance block separates fragmented safety evidence from limited explicit acceptance gates. Security, API-misuse, and crash repair cover 11 systems in the corpus, but their datasets and metrics remain disjoint, ranging from exact match/EM and F1 to CodeBLEU similarity, correct repairs, success rate, and repair accuracy. This fragmentation makes it harder to accumulate assurance evidence than in common benchmark families. The 10 systems whose recorded control subtypes are test feedback, human feedback, or LLM-gated review show one route toward explicit acceptance gates, but the remaining 56 systems do not expose such a gate as their primary control mechanism. A central open problem is how to report patch-acceptance evidence before deployment, using comparable datasets, explicit oracles, and lightweight static or dynamic checks rather than broad trust claims.

\begin{rqbox}{RQ4}
Across paradigms, five evidence-backed bottlenecks stand out for semantic correctness beyond tests and for repository-level multi-hunk repair: (i) evaluation settings that still rely on perfect localization, train/test splits, narrowed scopes, or heterogeneous leakage and oracle controls, (ii) multi-step workflows whose time and token usage can grow quickly with project size, (iii) strong dependence on proprietary/API models in prompting, procedural, and agentic systems, (iv) retrieval and analysis pipelines that can supply noisy or incomplete context, and (v) limited explicit acceptance gates for patch assurance and deployment readiness.

We see three concrete opportunities. First, build benchmarks and test suites that explicitly target large-scale, multi-hunk changes, with stronger oracles, hidden tests, contamination-aware splits, and clearly marked localization assumptions, so that progress reflects real semantic correctness rather than overfitting narrow unit suites. Second, design cost-aware repair policies that treat success per unit budget as an objective, by reporting model provenance and token budgets, reusing intermediate analysis results, capping the number of LLM calls, and falling back to simpler scripted pipelines when a bug appears easy, thereby controlling latency and variability on large repositories. Third, separate generation from acceptance more sharply by combining independent validation signals, such as stronger tests, static or dynamic analysis, sanitizer or exploit reproduction, human review, and calibrated LLM-gated review checks, so that final patch decisions are supported by explicit acceptance evidence rather than by the generator model alone.
\end{rqbox}

\section{Threats to Validity}
\label{sec:threats}

\subsection{Internal validity}
Study identification, selection, and coding threaten internal validity; we mitigate these threats with a structured pipeline. We search multiple digital libraries with calibrated queries and citation chasing, apply explicit inclusion and exclusion criteria, and combine independent screening and coding checks with adjudication for borderline cases. All search scripts and results are released in our replication package, enabling others to inspect our decisions and recompute statistics with the same notebooks.

\subsection{External validity}
Narrow corpora and artificial evaluation settings threaten external validity.
This survey intentionally builds a criteria-bounded corpus rather than an exhaustive census of all LLM-related repair claims.
We focus on the software engineering, AI/NLP, and security venue families listed in Section~\ref{sec:method}, using CCF Category A as one operational search lens rather than a quality ground truth. Because another ranking perspective could change the peripheral archival set, we mitigate this threat through multi-library retrieval, citation snowballing, topical full-text screening, and separate reporting of arXiv-only systems with concrete LLM-based repair pipelines and reproducible benchmark evidence.
The resulting 66-system corpus should therefore be interpreted as a criteria-bounded evidence base, not as a field-wide census.
Records outside the venue/version criteria, tools evaluated only on proprietary data, and systems without extractable benchmark evidence fall outside our scope.

We log such boundary records and assess their analytical impact; for example, Prometheus~\cite{chen2025prometheus} is cited as a relevant boundary case rather than included in the 66-system corpus because it maps to existing repository-level agentic/retrieval-augmented dimensions and does not create a new protocol-aligned comparison window.
Accordingly, count-based observations in this paper describe the surveyed corpus and should not be read as prevalence estimates for all LLM-based repair research.

Benchmark external validity is also limited by heterogeneous protocols. We preserve the published numeric scores and protocol notes, but we do not normalize different $k$ values, model snapshots, retrieval budgets, fault-localization assumptions, or wall-clock budgets across studies. We also do not perform a joint reevaluation. This preserves faithfulness to prior work but limits cross-paper comparability. To mitigate this limitation, Section~\ref{sec:bench-trends} separates protocol-aligned comparison windows from benchmark-family snapshots and annotates benchmark scope, metric definitions, oracle strength, and fault-localization assumptions alongside each reported score.

\subsection{Construct validity}
Ambiguous definitions and measurements threaten construct validity. We mitigate these issues by operationalizing constructs precisely and enforcing a stable taxonomy codebook. We code each system along multiple dimensions: parameter adaptation, runtime control authority, generation pattern, auxiliary evidence sources, and repair scope/deployment scenario. The four primary taxonomy paradigms used in summary tables are assigned by the ordered primary-locus rule over parameter adaptation and runtime control authority, while retrieval, analysis, test feedback, human feedback, and domain knowledge remain as separate auxiliary evidence tags. We apply this scheme consistently across all studies, release the per-system coding sheets, and discuss borderline cases to make the coding choices auditable. These checks assess the reproducibility of our coding decisions rather than community consensus on the taxonomy; broader community assessment of the taxonomy remains future work.

\section{Conclusion}
\label{sec:conclusion}

LLM-based software repair has evolved from early single-prompt experiments to a set of distinct paradigms that combine different forms of control over the repair loop with different degrees of model adaptation. This survey provides a structured view of the field by proposing a taxonomy based on control authority and parameter adaptation, with auxiliary context sources such as retrieval and analysis recorded as separate evidence layers; mapping 66 representative systems into this framework; and consolidating their benchmarks, evaluation assumptions, and pass@$k$ definitions into a common set of tables.

Taken together, the taxonomy, corpus, and evaluation summary clarify the tradeoffs among fine-tuning, prompting, procedural pipelines, and agentic frameworks, and show how design choices and evaluation protocols jointly shape reported repair success. The synthesis points to the most important directions for future work: stronger, less contaminated benchmarks; more transparent, comparable reporting of pass@$k$ and resource budgets; more selective, repository-aware retrieval and analysis pipelines; and cost-aware planning for multi-step workflows. Addressing these issues is central to turning LLM-based repair from promising prototypes into dependable infrastructure in everyday software engineering practice.

\begin{acks}
This work was partly supported by the Scientific Research Start-up Fund of Yanshan University (Grant 002/8190889). This work was partly supported by the Research Council of Finland (grant No. 376966).
\end{acks}

\bibliographystyle{ACM-Reference-Format}
\bibliography{sample-base}

@article{zhang2025acfix,
  author       = {Lyuye Zhang and
                  Kaixuan Li and
                  Kairan Sun and
                  Daoyuan Wu and
                  Ye Liu and
                  Haoye Tian and
                  Yang Liu},
  title        = {ACFix: Guiding LLMs With Mined Common {RBAC} Practices for Context-Aware
                  Repair of Access Control Vulnerabilities in Smart Contracts},
  journal      = {{IEEE} Trans. Software Eng.},
  volume       = {51},
  number       = {9},
  pages        = {2512--2532},
  year         = {2025},
  url          = {https://doi.org/10.1109/TSE.2025.3590108},
  doi          = {10.1109/TSE.2025.3590108},
  bibsource    = {dblp computer science bibliography, https://dblp.org}
}

@article{goyal2026optimizing,
  title={Optimizing Software Fault Prediction Using Atomic Orbital Search for Sustainable Software Development},
  author={Goyal, Somya Rakesh and Kumar, Sunil},
  journal={Journal of Computational and Cognitive Engineering},
  volume={5},
  number={1},
  pages={91--102},
  year={2026}
}

@article{abdollahi2026benchmarking,
author = {Abdollahi, Mohammad and Zhang, Ruixin and Shiri Harzevili, Nima and Shin, Jiho and Wang, Song and Hemmati, Hadi},
title = {Surveying the Benchmarking Landscape of Large Language Models in Code Intelligence},
year = {2026},
publisher = {Association for Computing Machinery},
address = {New York, NY, USA},
issn = {1049-331X},
url = {https://doi.org/10.1145/3800957},
doi = {10.1145/3800957},
note = {Just Accepted},
journal = {ACM Trans. Softw. Eng. Methodol.},
month = mar
}

@article{shao2026pailgen,
author = {Shao, Miaomiao and Ding, Yuxin and Gao, Cuiyun and Wang, Junru and Zhu, Guoqing},
title = {Fix Pattern-Aware Vulnerability Patch Generation via In-Context Learning},
year = {2026},
publisher = {Association for Computing Machinery},
address = {New York, NY, USA},
issn = {1049-331X},
url = {https://doi.org/10.1145/3796512},
doi = {10.1145/3796512},
note = {Just Accepted},
journal = {ACM Trans. Softw. Eng. Methodol.},
month = feb
}

@article{du2026environmentcrash,
author = {Du, Xueying and Liu, Mingwei and Wang, Hanlin and Li, Juntao and Peng, Xin and Lou, Yiling},
title = {Exploring Large Language Models in Resolving Environment-Related Crash Bugs: Localizing and Repairing},
year = {2026},
publisher = {Association for Computing Machinery},
address = {New York, NY, USA},
issn = {1049-331X},
url = {https://doi.org/10.1145/3788866},
doi = {10.1145/3788866},
journal = {ACM Trans. Softw. Eng. Methodol.},
month = jan
}

@ARTICLE{zhuo2026apimisuse,
  author={Zhuo, Terry Yue and He, Junda and Sun, Jiamou and Xing, Zhenchang and Lo, David and Grundy, John and Du, Xiaoning},
  journal={IEEE Transactions on Software Engineering}, 
  title={Identifying and Mitigating API Misuse in Large Language Models}, 
  year={2026},
  volume={52},
  number={3},
  pages={855-873},
  doi={10.1109/TSE.2026.3651566}}

@inproceedings{smith2015cure,
  author       = {Edward K. Smith and
                  Earl T. Barr and
                  Claire {Le Goues} and
                  Yuriy Brun},
  editor       = {Elisabetta Di Nitto and
                  Mark Harman and
                  Patrick Heymans},
  title        = {Is the cure worse than the disease? overfitting in automated program
                  repair},
  booktitle    = {Proceedings of the 2015 10th Joint Meeting on Foundations of Software
                  Engineering, {ESEC/FSE} 2015, Bergamo, Italy, August 30 - September
                  4, 2015},
  pages        = {532--543},
  publisher    = {{ACM}},
  year         = {2015},
  url          = {https://doi.org/10.1145/2786805.2786825},
  doi          = {10.1145/2786805.2786825},
  bibsource    = {dblp computer science bibliography, https://dblp.org}
}

@inproceedings{humbatova2020taxonomy,
  author       = {Nargiz Humbatova and
                  Gunel Jahangirova and
                  Gabriele Bavota and
                  Vincenzo Riccio and
                  Andrea Stocco and
                  Paolo Tonella},
  editor       = {Gregg Rothermel and
                  Doo{-}Hwan Bae},
  title        = {Taxonomy of real faults in deep learning systems},
  booktitle    = {{ICSE} '20: 42nd International Conference on Software Engineering,
                  Seoul, South Korea, 27 June - 19 July, 2020},
  pages        = {1110--1121},
  publisher    = {{ACM}},
  year         = {2020},
  url          = {https://doi.org/10.1145/3377811.3380395},
  doi          = {10.1145/3377811.3380395},
  bibsource    = {dblp computer science bibliography, https://dblp.org}
}

@misc{hui2024qwen25codertechnicalreport,
      title={Qwen2.5-Coder Technical Report}, 
      author={Binyuan Hui and Jian Yang and Zeyu Cui and Jiaxi Yang and Dayiheng Liu and Lei Zhang and Tianyu Liu and Jiajun Zhang and Bowen Yu and Keming Lu and Kai Dang and Yang Fan and Yichang Zhang and An Yang and Rui Men and Fei Huang and Bo Zheng and Yibo Miao and Shanghaoran Quan and Yunlong Feng and Xingzhang Ren and Xuancheng Ren and Jingren Zhou and Junyang Lin},
      year={2024},
      eprint={2409.12186},
      archivePrefix={arXiv},
      primaryClass={cs.CL},
      url={https://arxiv.org/abs/2409.12186}, 
}

@inproceedings{he2023large,
  title={Large language models for code: Security hardening and adversarial testing},
  author={He, Jingxuan and Vechev, Martin},
  booktitle={Proceedings of the 2023 ACM SIGSAC Conference on Computer and Communications Security},
  pages={1865--1879},
  year={2023}
}

@inproceedings{ding2025vulnerability,
  title={Vulnerability Detection with Code Language Models: How Far are We?},
  author={Ding, Yangruibo and Fu, Yanjun and Ibrahim, Omniyyah and Sitawarin, Chawin and Chen, Xinyun and Alomair, Basel and Wagner, David and Ray, Baishakhi and Chen, Yizheng},
  booktitle={2025 IEEE/ACM 47th International Conference on Software Engineering (ICSE)},
  pages={1729--1741},
  year={2025},
  organization={IEEE}
}

@inproceedings{wen2025vul,
    author = {Wen, Xin-Cheng and Lin, Zirui and Yang, Yijun and Gao, Cuiyun and Ye, Deheng},
    title = {Vul-R2: A Reasoning LLM for Automated Vulnerability Repair},
    year = {2025},
    publisher = {IEEE Press},
    url = {https://doi.org/10.1109/ASE63991.2025.00011},
    doi = {10.1109/ASE63991.2025.00011},
    booktitle = {2025 40th IEEE/ACM International Conference on Automated Software Engineering (ASE)},
    pages = {26--38},
    numpages = {13},
    location = {Seoul, Korea, Republic of}
}

@online{TutorCode,
  author = {TutorCode},
  title = {TutorCode API Documentation},
  year = {2024},
  url = {https://tutorcode.org/docs/}
}

@article{chen2025prometheus,
  author       = {Zimin Chen and
                  Yue Pan and
                  Siyu Lu and
                  Jiayi Xu and
                  Claire {Le Goues} and
                  Martin Monperrus and
                  He Ye},
  title        = {Prometheus: Unified Knowledge Graphs for Issue Resolution in Multilingual
                  Codebases},
  journal      = {CoRR},
  volume       = {abs/2507.19942},
  year         = {2025},
  url          = {https://doi.org/10.48550/arXiv.2507.19942},
  doi          = {10.48550/ARXIV.2507.19942},
  eprinttype   = {arXiv},
  eprint       = {2507.19942},
  bibsource    = {dblp computer science bibliography, https://dblp.org}
}

@inproceedings{sobania2023analysis,
  title={An analysis of the automatic bug fixing performance of chatgpt},
  author={Sobania, Dominik and Briesch, Martin and Hanna, Carol and Petke, Justyna},
  booktitle={2023 IEEE/ACM International Workshop on Automated Program Repair (APR)},
  pages={23--30},
  year={2023},
  organization={IEEE}
}

@article{zhou2024large,
  title={Large language model for vulnerability detection and repair: Literature review and the road ahead},
  author={Zhou, Xin and Cao, Sicong and Sun, Xiaobing and Lo, David},
  journal={ACM Transactions on Software Engineering and Methodology},
  year={2025},
  publisher={ACM New York, NY}
}

@article{vaswani2017attention,
  title={Attention is all you need},
  author={Vaswani, Ashish and Shazeer, Noam and Parmar, Niki and Uszkoreit, Jakob and Jones, Llion and Gomez, Aidan N and Kaiser, {\L}ukasz and Polosukhin, Illia},
  journal={Advances in neural information processing systems},
  volume={30},
  year={2017}
}

@inproceedings{levenshtein1966binary,
  title={Binary codes capable of correcting deletions, insertions, and reversals},
  author={Levenshtein, Vladimir I and others},
  booktitle={Soviet physics doklady},
  volume={10},
  number={8},
  pages={707--710},
  year={1966},
  organization={Soviet Union}
}

@article{damerau1964technique,
  title={A technique for computer detection and correction of spelling errors},
  author={Damerau, Fred J},
  journal={Communications of the ACM},
  volume={7},
  number={3},
  pages={171--176},
  year={1964},
  publisher={ACM New York, NY, USA}
}

@inproceedings{cambronero2025abstain,
  title={Abstain and Validate: A Dual-LLM Policy for Reducing Noise in Agentic Program Repair},
  author={Cambronero, Jos{\'e} and Tufano, Michele and Shi, Sherry and Wei, Renyao and Uy, Grant and Cheng, Runxiang and Liu, Chin-Jung and Pan, Shiying and Chandra, Satish and Rondon, Patrick},
  booktitle={2026 IEEE/ACM 48th International Conference on Software Engineering: Software Engineering in Practice (ICSE-SEIP)},
  year={2026},
  publisher={IEEE/ACM}
}

@inproceedings{bui2022vul4j,
  title={Vul4j: A dataset of reproducible java vulnerabilities geared towards the study of program repair techniques},
  author={Bui, Quang-Cuong and Scandariato, Riccardo and Ferreyra, Nicol{\'a}s E D{\'\i}az},
  booktitle={Proceedings of the 19th International Conference on Mining Software Repositories},
  pages={464--468},
  year={2022}
}

@inproceedings{ruan2025specrover,
  title={SpecRover: Code Intent Extraction via LLMs},
  author={Ruan, Haifeng and Zhang, Yuntong and Roychoudhury, Abhik},
  booktitle={2025 IEEE/ACM 47th International Conference on Software Engineering (ICSE)},
  pages={963--974},
  year={2025},
  organization={IEEE}
}

@misc{hu2025tsaprtreesearchframework,
      title={TSAPR: A Tree Search Framework For Automated Program Repair}, 
      author={Haichuan Hu and Ye Shang and Weifeng Sun and Quanjun Zhang},
      year={2025},
      eprint={2507.01827},
      archivePrefix={arXiv},
      primaryClass={cs.SE},
      url={https://arxiv.org/abs/2507.01827}, 
}

@article{xu2025mem,
  title={A-mem: Agentic memory for llm agents},
  author={Xu, Wujiang and Mei, Kai and Gao, Hang and Tan, Juntao and Liang, Zujie and Zhang, Yongfeng},
  journal={arXiv preprint arXiv:2502.12110},
  year={2025}
}

@inproceedings{su2025learn,
  author       = {Hongjin Su and
                  Ruoxi Sun and
                  Jinsung Yoon and
                  Pengcheng Yin and
                  Tao Yu and
                  Sercan {\"{O}}. Arik},
  title        = {Learn-by-interact: {A} Data-Centric Framework For Self-Adaptive Agents
                  in Realistic Environments},
  booktitle    = {The Thirteenth International Conference on Learning Representations,
                  {ICLR} 2025, Singapore, April 24-28, 2025},
  publisher    = {OpenReview.net},
  year         = {2025},
  url          = {https://openreview.net/forum?id=3UKOzGWCVY},
  bibsource    = {dblp computer science bibliography, https://dblp.org}
}

@inproceedings{wang2024openhands,
  title={{OpenHands}: An Open Platform for {AI} Software Developers as Generalist Agents},
  author={Wang, Xingyao and Li, Boxuan and Song, Yufan and Xu, Frank F and Tang, Xiangru and Zhuge, Mingchen and Pan, Jiayi and Song, Yueqi and Li, Bowen and Singh, Jaskirat and others},
  booktitle={The Thirteenth International Conference on Learning Representations},
  year={2025}
}

@misc{dolcetti2025helpingllmsimprovecode,
      author       = {Greta Dolcetti and
                  Vincenzo Arceri and
                  Eleonora Iotti and
                  Sergio Maffeis and
                  Agostino Cortesi and
                  Enea Zaffanella},
  title        = {Helping LLMs improve code generation using feedback from testing and
                  static analysis},
  journal      = {Discov. Artif. Intell.},
  volume       = {6},
  number       = {1},
  pages        = {314},
  year         = {2026},
  url          = {https://doi.org/10.1007/s44163-026-01009-5},
  doi          = {10.1007/S44163-026-01009-5},
  bibsource    = {dblp computer science bibliography, https://dblp.org}
}

@inproceedings{kim2025logs,
  title={Logs In, Patches Out: Automated Vulnerability Repair via $\{$Tree-of-Thought$\}$$\{$LLM$\}$ Analysis},
  author={Kim, Youngjoon and Shin, Sunguk and Kim, Hyoungshick and Yoon, Jiwon},
  booktitle={34th USENIX Security Symposium (USENIX Security 25)},
  pages={4401--4419},
  year={2025}
}

@inproceedings{shen2021localizing,
  title={Localizing vulnerabilities statistically from one exploit},
  author={Shen, Shiqi and Kolluri, Aashish and Dong, Zhen and Saxena, Prateek and Roychoudhury, Abhik},
  booktitle={Proceedings of the 2021 ACM Asia Conference on Computer and Communications Security},
  pages={537--549},
  year={2021}
}

@misc{lewis2021retrievalaugmentedgenerationknowledgeintensivenlp,
      title={Retrieval-Augmented Generation for Knowledge-Intensive NLP Tasks}, 
      author={Patrick Lewis and Ethan Perez and Aleksandra Piktus and Fabio Petroni and Vladimir Karpukhin and Naman Goyal and Heinrich Küttler and Mike Lewis and Wen-tau Yih and Tim Rocktäschel and Sebastian Riedel and Douwe Kiela},
      year={2021},
      eprint={2005.11401},
      archivePrefix={arXiv},
      primaryClass={cs.CL},
      url={https://arxiv.org/abs/2005.11401}, 
}

@inproceedings{zhang2024autocoderover,
  title={Autocoderover: Autonomous program improvement},
  author={Zhang, Yuntong and Ruan, Haifeng and Fan, Zhiyu and Roychoudhury, Abhik},
  booktitle={Proceedings of the 33rd ACM SIGSOFT International Symposium on Software Testing and Analysis},
  pages={1592--1604},
  year={2024}
}

@article{yang2024swe,
  title={Swe-agent: Agent-computer interfaces enable automated software engineering},
  author={Yang, John and Jimenez, Carlos and Wettig, Alexander and Lieret, Kilian and Yao, Shunyu and Narasimhan, Karthik and Press, Ofir},
  journal={Advances in Neural Information Processing Systems},
  volume={37},
  pages={50528--50652},
  year={2024}
}

@inproceedings{yang2025swem,
  title={SWE-bench Multimodal: Do AI Systems Generalize to Visual Software Domains?},
  author={Yang, John and Jimenez, Carlos E and Zhang, Alex L and Lieret, Kilian and Yang, Joyce and Wu, Xindi and Press, Ori and Muennighoff, Niklas and Synnaeve, Gabriel and Narasimhan, Karthik R and others},
  booktitle={The Thirteenth International Conference on Learning Representations},
  year={2025}
}

@inproceedings{yin2024thinkrepair,
  title={Thinkrepair: Self-directed automated program repair},
  author={Yin, Xin and Ni, Chao and Wang, Shaohua and Li, Zhenhao and Zeng, Limin and Yang, Xiaohu},
  booktitle={Proceedings of the 33rd ACM SIGSOFT International Symposium on Software Testing and Analysis},
  pages={1274--1286},
  year={2024}
}

@inproceedings{xiao2025predicatefixrepairingstaticanalysis,
      title={PredicateFix: Repairing Static Analysis Alerts with Bridging Predicates}, 
      author={Yuan-An Xiao and Weixuan Wang and Dong Liu and Junwei Zhou and Shengyu Cheng and Yingfei Xiong},
      booktitle={Proceedings of the 48th IEEE/ACM International Conference on Software Engineering},
      year={2026},
      doi={10.1145/3744916.3773159},
      publisher={Association for Computing Machinery} 
}

@article{silva2025repairllama,
  title={Repairllama: Efficient representations and fine-tuned adapters for program repair},
  author={Silva, Andr{\'e} and Fang, Sen and Monperrus, Martin},
  journal={IEEE Transactions on Software Engineering},
  year={2025},
  publisher={IEEE}
}

@inproceedings{li2024exploring,
  title={Exploring parameter-efficient fine-tuning of large language model on automated program repair},
  author={Li, Guochang and Zhi, Chen and Chen, Jialiang and Han, Junxiao and Deng, Shuiguang},
  booktitle={Proceedings of the 39th IEEE/ACM International Conference on Automated Software Engineering},
  pages={719--731},
  year={2024}
}

@inproceedings{xia2022less,
  title={Less training, more repairing please: revisiting automated program repair via zero-shot learning},
  author={Xia, Chunqiu Steven and Zhang, Lingming},
  booktitle={Proceedings of the 30th ACM Joint European Software Engineering Conference and Symposium on the Foundations of Software Engineering},
  pages={959--971},
  year={2022}
}

@article{yu2025dapo,
  title={Dapo: An open-source llm reinforcement learning system at scale},
  author={Yu, Qiying and Zhang, Zheng and Zhu, Ruofei and Yuan, Yufeng and Zuo, Xiaochen and Yue, Yu and Dai, Weinan and Fan, Tiantian and Liu, Gaohong and Liu, Lingjun and others},
  journal={arXiv preprint arXiv:2503.14476},
  year={2025}
}

@article{rafailov2023direct,
  title={Direct preference optimization: Your language model is secretly a reward model},
  author={Rafailov, Rafael and Sharma, Archit and Mitchell, Eric and Manning, Christopher D and Ermon, Stefano and Finn, Chelsea},
  journal={Advances in Neural Information Processing Systems},
  volume={36},
  pages={53728--53741},
  year={2023}
}

@article{shao2024deepseekmath,
  title={Deepseekmath: Pushing the limits of mathematical reasoning in open language models},
  author={Shao, Zhihong and Wang, Peiyi and Zhu, Qihao and Xu, Runxin and Song, Junxiao and Bi, Xiao and Zhang, Haowei and Zhang, Mingchuan and Li, YK and Wu, Y and others},
  journal={arXiv preprint arXiv:2402.03300},
  year={2024}
}

@article{zhao2024novel,
  title={A Novel Approach for Automated Design Information Mining from Issue Logs},
  author={Zhao, Jiuang and Yang, Zitian and Zhang, Li and Lian, Xiaoli and Yang, Donghao},
  journal={arXiv preprint arXiv:2405.19623},
  year={2024}
}

@inproceedings{guo2022unixcoder,
  title={UniXcoder: Unified Cross-Modal Pre-training for Code Representation},
  author={Guo, Daya and Lu, Shuai and Duan, Nan and Wang, Yanlin and Zhou, Ming and Yin, Jian},
  booktitle={Proceedings of the 60th Annual Meeting of the Association for Computational Linguistics (Volume 1: Long Papers)},
  pages={7212--7225},
  year={2022}
}

@inproceedings{ahmad2021unified,
  title={Unified Pre-training for Program Understanding and Generation},
  author={Ahmad, Wasi and Chakraborty, Saikat and Ray, Baishakhi and Chang, Kai-Wei},
  booktitle={Proceedings of the 2021 Conference of the North American Chapter of the Association for Computational Linguistics: Human Language Technologies},
  pages={2655--2668},
  year={2021}
}

@inproceedings{feng2020codebert,
  title={CodeBERT: A Pre-Trained Model for Programming and Natural Languages},
  author={Feng, Zhangyin and Guo, Daya and Tang, Duyu and Duan, Nan and Feng, Xiaocheng and Gong, Ming and Shou, Linjun and Qin, Bing and Liu, Ting and Jiang, Daxin and others},
  booktitle={Findings of the Association for Computational Linguistics: EMNLP 2020},
  pages={1536--1547},
  year={2020}
}

@inproceedings{huang2023empirical,
  title={An empirical study on fine-tuning large language models of code for automated program repair},
  author={Huang, Kai and Meng, Xiangxin and Zhang, Jian and Liu, Yang and Wang, Wenjie and Li, Shuhao and Zhang, Yuqing},
  booktitle={2023 38th IEEE/ACM International Conference on Automated Software Engineering (ASE)},
  pages={1162--1174},
  year={2023},
  organization={IEEE}
}

@inproceedings{friedincoder,
  title={InCoder: A Generative Model for Code Infilling and Synthesis},
  author={Fried, Daniel and Aghajanyan, Armen and Lin, Jessy and Wang, Sida and Wallace, Eric and Shi, Freda and Zhong, Ruiqi and Yih, Scott and Zettlemoyer, Luke and Lewis, Mike},
  booktitle={The Eleventh International Conference on Learning Representations},
  year={2022}
}

@inproceedings{wei2023copiloting,
  title={Copiloting the copilots: Fusing large language models with completion engines for automated program repair},
  author={Wei, Yuxiang and Xia, Chunqiu Steven and Zhang, Lingming},
  booktitle={Proceedings of the 31st ACM Joint European Software Engineering Conference and Symposium on the Foundations of Software Engineering},
  pages={172--184},
  year={2023}
}

@article{le2015manybugs,
  title={The ManyBugs and IntroClass benchmarks for automated repair of C programs},
  author={Le Goues, Claire and Holtschulte, Neal and Smith, Edward K and Brun, Yuriy and Devanbu, Premkumar and Forrest, Stephanie and Weimer, Westley},
  journal={IEEE Transactions on Software Engineering},
  volume={41},
  number={12},
  pages={1236--1256},
  year={2015},
  publisher={IEEE}
}

@inproceedings{xia2023automated,
  title={Automated program repair in the era of large pre-trained language models},
  author={Xia, Chunqiu Steven and Wei, Yuxiang and Zhang, Lingming},
  booktitle={2023 IEEE/ACM 45th International Conference on Software Engineering (ICSE)},
  pages={1482--1494},
  year={2023},
  organization={IEEE}
}

@article{luo2025fine,
  title={When Fine-Tuning LLMs Meets Data Privacy: An Empirical Study of Federated Learning in LLM-Based Program Repair},
  author={Luo, Wenqiang and Keung, Jacky and Yang, Boyang and Ye, He and Le Goues, Claire and Bissyande, Tegawende F and Tian, Haoye and Le, Xuan Bach D},
  journal={ACM Transactions on Software Engineering and Methodology},
  year={2025},
  publisher={ACM New York, NY}
}

@inproceedings{hollowayrole,
  title={On the Role of Context Granularity in {LLM}-Driven Program Repair},
  author={Holloway, Tyler and Elenberg, Ethan R.},
  booktitle={NeurIPS 2024 Workshop on Machine Learning for Systems},
  year={2024},
  url={https://openreview.net/forum?id=mSRlcYkpOq}
}

@article{wang2025solved,
  author       = {You Wang and
                  Michael Pradel and
                  Zhongxin Liu},
  title        = {Are "Solved Issues" in SWE-bench Really Solved Correctly?
                  An Empirical Study},
  journal      = {CoRR},
  volume       = {abs/2503.15223},
  year         = {2025},
  url          = {https://doi.org/10.48550/arXiv.2503.15223},
  doi          = {10.48550/ARXIV.2503.15223},
  eprinttype   = {arXiv},
  eprint       = {2503.15223},
  bibsource    = {dblp computer science bibliography, https://dblp.org}
}

@article{parasaram2024fact,
  title={The fact selection problem in LLM-based program repair},
  author={Parasaram, Nikhil and Yan, Huijie and Yang, Boyu and Flahy, Zineb and Qudsi, Abriele and Ziaber, Damian and Barr, Earl and Mechtaev, Sergey},
  journal={arXiv preprint arXiv:2404.05520},
  year={2024}
}

@article{yangmorepair,
  title={MORepair: Teaching LLMs to Repair Code via Multi-Objective Fine-Tuning},
  author={Yang, Boyang and Tian, Haoye and Ren, Jiadong and Zhang, Hongyu and Klein, Jacques and Bissyande, Tegawende and Le Goues, Claire and Jin, Shunfu},
  journal={ACM Transactions on Software Engineering and Methodology},
  publisher={ACM New York, NY},
  year={2025}
}

@inproceedings{petke2024patch,
  title={The patch overfitting problem in automated program repair: practical magnitude and a baseline for realistic benchmarking},
  author={Petke, Justyna and Martinez, Matias and Kechagia, Maria and Aleti, Aldeida and Sarro, Federica},
  booktitle={Companion Proceedings of the 32nd ACM International Conference on the Foundations of Software Engineering},
  pages={452--456},
  year={2024}
}

@inproceedings{fan2023automated,
  title={Automated repair of programs from large language models},
  author={Fan, Zhiyu and Gao, Xiang and Mirchev, Martin and Roychoudhury, Abhik and Tan, Shin Hwei},
  booktitle={2023 IEEE/ACM 45th International Conference on Software Engineering (ICSE)},
  pages={1469--1481},
  year={2023},
  organization={IEEE}
}

@article{yang2024narrepair,
  title={Narrepair: Non-autoregressive code generation model for automatic program repair},
  author={Yang, Zhenyu and Yang, Zhen and Yu, Zhongxing},
  journal={arXiv preprint arXiv:2406.16526},
  year={2024}
}

@inproceedings{takerngsaksiri2025human,
  title={Human-in-the-loop software development agents},
  author={Takerngsaksiri, Wannita and Pasuksmit, Jirat and Thongtanunam, Patanamon and Tantithamthavorn, Chakkrit and Zhang, Ruixiong and Jiang, Fan and Li, Jing and Cook, Evan and Chen, Kun and Wu, Ming},
  booktitle={2025 IEEE/ACM 47th International Conference on Software Engineering: Software Engineering in Practice (ICSE-SEIP)},
  pages={342--352},
  year={2025},
  organization={IEEE}
}

@article{geethal2023human,
  title={Human-in-the-Loop Automatic Program Repair},
  author={Geethal, Charaka and B{\"o}hme, Marcel and Pham, Van-Thuan},
  journal={IEEE Transactions on Software Engineering},
  volume={49},
  number={10},
  pages={4526--4549},
  year={2023},
  publisher={IEEE}
}

@inproceedings{just2014defects4j,
  title={Defects4J: A database of existing faults to enable controlled testing studies for Java programs},
  author={Just, Ren{\'e} and Jalali, Darioush and Ernst, Michael D},
  booktitle={Proceedings of the 2014 international symposium on software testing and analysis},
  pages={437--440},
  year={2014}
}

@article{liu2022few,
  title={Few-shot parameter-efficient fine-tuning is better and cheaper than in-context learning},
  author={Liu, Haokun and Tam, Derek and Muqeeth, Mohammed and Mohta, Jay and Huang, Tenghao and Bansal, Mohit and Raffel, Colin A},
  journal={Advances in Neural Information Processing Systems},
  volume={35},
  pages={1950--1965},
  year={2022}
}

@article{jain2023staticfixer,
  title={Staticfixer: From static analysis to static repair},
  author={Jain, Naman and Gandhi, Shubham and Sonwane, Atharv and Kanade, Aditya and Natarajan, Nagarajan and Parthasarathy, Suresh and Rajamani, Sriram and Sharma, Rahul},
  journal={arXiv preprint arXiv:2307.12465},
  year={2023}
}

@article{haque2025systematic,
  title={A Systematic Literature Review of Parameter-Efficient Fine-Tuning for Large Code Models},
  author={Haque, Md Zahidul and Afrin, Saima and Mastropaolo, Antonio},
  journal={arXiv preprint arXiv:2504.21569},
  year={2025}
}

@inproceedings{jiang2023impact,
  title={Impact of code language models on automated program repair},
  author={Jiang, Nan and Liu, Kevin and Lutellier, Thibaud and Tan, Lin},
  booktitle={2023 IEEE/ACM 45th International Conference on Software Engineering (ICSE)},
  pages={1430--1442},
  year={2023},
  organization={IEEE}
}

@inproceedings{yang2024cref,
  title={Cref: An llm-based conversational software repair framework for programming tutors},
  author={Yang, Boyang and Tian, Haoye and Pian, Weiguo and Yu, Haoran and Wang, Haitao and Klein, Jacques and Bissyand{\'e}, Tegawend{\'e} F and Jin, Shunfu},
  booktitle={Proceedings of the 33rd ACM SIGSOFT International Symposium on Software Testing and Analysis},
  pages={882--894},
  year={2024}
}

@article{huang2023survey,
  title={A survey on automated program repair techniques},
  author={Huang, Kai and Xu, Zhengzi and Yang, Su and Sun, Hongyu and Li, Xuejun and Yan, Zheng and Zhang, Yuqing},
  journal={arXiv preprint arXiv:2303.18184},
  year={2023}
}

@article{anand2024comprehensive,
  title={A Comprehensive Survey of AI-Driven Advancements and Techniques in Automated Program Repair and Code Generation},
  author={Anand, Avinash and Gupta, Akshit and Yadav, Nishchay and Bajaj, Shaurya},
  journal={arXiv preprint arXiv:2411.07586},
  year={2024}
}

@article{dikici2025advancements,
  title={Advancements in automated program repair: a comprehensive review},
  author={Dikici, Sena and Bilgin, Turgay Tugay},
  journal={Knowledge and Information Systems},
  pages={1--47},
  year={2025},
  publisher={Springer}
}

@inproceedings{jiang2023knod,
  title={Knod: Domain knowledge distilled tree decoder for automated program repair},
  author={Jiang, Nan and Lutellier, Thibaud and Lou, Yiling and Tan, Lin and Goldwasser, Dan and Zhang, Xiangyu},
  booktitle={2023 IEEE/ACM 45th International Conference on Software Engineering (ICSE)},
  pages={1251--1263},
  year={2023},
  organization={IEEE}
}

@inproceedings{wong2024distilrr,
  title={Investigating the transferability of code repair for low-resource programming languages},
  author={Wong, Kyle and Amayuelas, Alfonso and Pan, Liangming and Wang, William Yang},
  booktitle={Findings of the Association for Computational Linguistics: NAACL 2025},
  pages={3410--3432},
  year={2025},
  address={Albuquerque, New Mexico},
  publisher={Association for Computational Linguistics},
  year={2025}
}

@article{zhang2023survey,
  title={A survey of learning-based automated program repair},
  author={Zhang, Quanjun and Fang, Chunrong and Ma, Yuxiang and Sun, Weisong and Chen, Zhenyu},
  journal={ACM Transactions on Software Engineering and Methodology},
  volume={33},
  number={2},
  pages={1--69},
  year={2023},
  publisher={ACM New York, NY}
}

@article{winter2022let,
  title={Let’s talk with developers, not about developers: A review of automatic program repair research},
  author={Winter, Emily and Nowack, Vesna and Bowes, David and Counsell, Steve and Hall, Tracy and Haraldsson, S{\ae}mundur and Woodward, John},
  journal={IEEE Transactions on Software Engineering},
  volume={49},
  number={1},
  pages={419--436},
  year={2022},
  publisher={IEEE}
}

@inproceedings{yao2023react,
  title={React: Synergizing reasoning and acting in language models},
  author={Yao, Shunyu and Zhao, Jeffrey and Yu, Dian and Du, Nan and Shafran, Izhak and Narasimhan, Karthik and Cao, Yuan},
  booktitle={International Conference on Learning Representations (ICLR)},
  year={2023}
}

@article{roziere2023code,
  title={Code llama: Open foundation models for code},
  author={Roziere, Baptiste and Gehring, Jonas and Gloeckle, Fabian and Sootla, Sten and Gat, Itai and Tan, Xiaoqing Ellen and Adi, Yossi and Liu, Jingyu and Sauvestre, Romain and Remez, Tal and others},
  journal={arXiv preprint arXiv:2308.12950},
  year={2023}
}

@article{zhu2024deepseek,
  title={Deepseek-coder-v2: Breaking the barrier of closed-source models in code intelligence},
  author={Zhu, Qihao and Guo, Daya and Shao, Zhihong and Yang, Dejian and Wang, Peiyi and Xu, Runxin and Wu, Y and Li, Yukun and Gao, Huazuo and Ma, Shirong and others},
  journal={arXiv preprint arXiv:2406.11931},
  year={2024}
}

@article{guo2024deepseek,
  title={DeepSeek-Coder: When the Large Language Model Meets Programming--The Rise of Code Intelligence},
  author={Guo, Daya and Zhu, Qihao and Yang, Dejian and Xie, Zhenda and Dong, Kai and Zhang, Wentao and Chen, Guanting and Bi, Xiao and Wu, Yu and Li, YK and others},
  journal={arXiv preprint arXiv:2401.14196},
  year={2024}
}

@article{zhang2023critical,
  title={A critical review of large language model on software engineering: An example from chatgpt and automated program repair},
  author={Zhang, Quanjun and Zhang, Tongke and Zhai, Juan and Fang, Chunrong and Yu, Bowen and Sun, Weisong and Chen, Zhenyu},
  journal={arXiv preprint arXiv:2310.08879},
  year={2023}
}

@article{huang2024evolving,
  title={Evolving paradigms in automated program repair: Taxonomy, challenges, and opportunities},
  author={Huang, Kai and Xu, Zhengzi and Yang, Su and Sun, Hongyu and Li, Xuejun and Yan, Zheng and Zhang, Yuqing},
  journal={ACM Computing Surveys},
  volume={57},
  number={2},
  pages={1--43},
  year={2024},
  publisher={ACM New York, NY}
}

@article{zhang2024systematic,
author = {Zhang, Quanjun and Fang, Chunrong and Xie, Yang and Ma, Yuxiang and Sun, Weisong and Yang, Yun and Chen, Zhenyu},
title = {A Systematic Literature Review on Large Language Models for Automated Program Repair},
year = {2026},
publisher = {Association for Computing Machinery},
address = {New York, NY, USA},
issn = {1049-331X},
url = {https://doi.org/10.1145/3799693},
doi = {10.1145/3799693},
note = {Just Accepted},
journal = {ACM Trans. Softw. Eng. Methodol.},
month = mar
}

@inproceedings{bouzenia2025repairagent,
  title={{RepairAgent}: An Autonomous, {LLM}-Based Agent for Program Repair},
  author={Bouzenia, Islem and Devanbu, Premkumar and Pradel, Michael},
  booktitle={2025 IEEE/ACM 47th International Conference on Software Engineering (ICSE)},
  pages={2188--2200},
  year={2025},
  organization={IEEE Computer Society},
  doi={10.1109/ICSE55347.2025.00157}
}

@article{prenner2023runbugrun,
  title={RunBugRun--An Executable Dataset for Automated Program Repair},
  author={Prenner, Julian Aron and Robbes, Romain},
  journal={arXiv preprint arXiv:2304.01102},
  year={2023}
}

@misc{rachum2019pysnooper,
  title={PySnooper: Never use print for debugging again},
    author={Rachum, Ram and Hall, Alex and Yanokura, Iori and others},
    year={2019},
    month={jun},
    publisher={PyCon Israel},
    doi={10.5281/zenodo.10462459},
    url={https://github.com/cool-RR/PySnooper}
}

@article{zhou2024leveraging,
  title={Leveraging large language model for automatic patch correctness assessment},
  author={Zhou, Xin and Xu, Bowen and Kim, Kisub and Han, DongGyun and Nguyen, Hung Huu and Le-Cong, Thanh and He, Junda and Le, Bach and Lo, David},
  journal={IEEE Transactions on Software Engineering},
  year={2024},
  publisher={IEEE}
}

@inproceedings{xu2025aligning,
  title={Aligning the Objective of LLM-Based Program Repair},
  author={Xu, Junjielong and Fu, Ying and Tan, Shin Hwei and He, Pinjia},
  booktitle={2025 IEEE/ACM 47th International Conference on Software Engineering (ICSE)},
  pages={2548--2560},
  year={2025},
  organization={IEEE}
}

@article{li2024hybrid,
  title={Hybrid automated program repair by combining large language models and program analysis},
  author={Li, Fengjie and Jiang, Jiajun and Sun, Jiajun and Zhang, Hongyu},
  journal={ACM Transactions on Software Engineering and Methodology},
  year={2024},
  publisher={ACM New York, NY}
}

@inproceedings{chen2024large,
  title={When large language models confront repository-level automatic program repair: How well they done?},
  author={Chen, Yuxiao and Wu, Jingzheng and Ling, Xiang and Li, Changjiang and Rui, Zhiqing and Luo, Tianyue and Wu, Yanjun},
  booktitle={Proceedings of the 2024 IEEE/ACM 46th International Conference on Software Engineering: Companion Proceedings},
  pages={459--471},
  year={2024}
}

@misc{mistralai2024codestral,
  author = {Mistral AI},
  title = {Codestral: A State-of-the-Art Code Generation Model},
  howpublished = {\url{https://mistral.ai/news/codestral/}},
  year = {2024},
  note = {Accessed: 2025-05-19}
}

@inproceedings{lai2023ds,
  title={DS-1000: A natural and reliable benchmark for data science code generation},
  author={Lai, Yuhang and Li, Chengxi and Wang, Yiming and Zhang, Tianyi and Zhong, Ruiqi and Zettlemoyer, Luke and Yih, Wen-tau and Fried, Daniel and Wang, Sida and Yu, Tao},
  booktitle={International Conference on Machine Learning},
  pages={18319--18345},
  year={2023},
  organization={PMLR}
}

@article{grattafiori2024llama,
  title={The llama 3 herd of models},
  author={Grattafiori, Aaron and Dubey, Abhimanyu and Jauhri, Abhinav and Pandey, Abhinav and Kadian, Abhishek and Al-Dahle, Ahmad and Letman, Aiesha and Mathur, Akhil and Schelten, Alan and Vaughan, Alex and others},
  journal={arXiv preprint arXiv:2407.21783},
  year={2024}
}

@inproceedings{ouyang2025knowledge,
  title={Knowledge-Enhanced Program Repair for Data Science Code},
  author={Ouyang, Shuyin and Zhang, Jie M. and Sun, Zeyu and Penuela, Albert Merono},
  booktitle={2025 IEEE/ACM 47th International Conference on Software Engineering (ICSE)},
  pages={898--910},
  year={2025},
  organization={IEEE Computer Society},
  doi={10.1109/ICSE55347.2025.00246}
}

@article{ruiz2025art,
  title={The Art of Repair: Optimizing Iterative Program Repair with Instruction-Tuned Models},
  author={Ruiz, Fernando Vallecillos and Hort, Max and Moonen, Leon},
  journal={arXiv preprint arXiv:2505.02931},
  year={2025}
}

@inproceedings{ahmed2023majority,
  author       = {Toufique Ahmed and
                  Premkumar T. Devanbu},
  title        = {Better Patching Using {LLM} Prompting, via Self-Consistency},
  booktitle    = {38th {IEEE/ACM} International Conference on Automated Software Engineering,
                  {ASE} 2023, Luxembourg, September 11-15, 2023},
  pages        = {1742--1746},
  publisher    = {{IEEE}},
  year         = {2023},
  url          = {https://doi.org/10.1109/ASE56229.2023.00065},
  doi          = {10.1109/ASE56229.2023.00065},
  bibsource    = {dblp computer science bibliography, https://dblp.org}
}

@article{tao2024magis,
  title={Magis: Llm-based multi-agent framework for github issue resolution},
  author={Tao, Wei and Zhou, Yucheng and Wang, Yanlin and Zhang, Wenqiang and Zhang, Hongyu and Cheng, Yu},
  journal={Advances in Neural Information Processing Systems},
  volume={37},
  pages={51963--51993},
  year={2024}
}

@inproceedings{zhou2024out,
  title={Out of sight, out of mind: Better automatic vulnerability repair by broadening input ranges and sources},
  author={Zhou, Xin and Kim, Kisub and Xu, Bowen and Han, DongGyun and Lo, David},
  booktitle={Proceedings of the IEEE/ACM 46th International Conference on Software Engineering},
  pages={1--13},
  year={2024}
}

@article{islam2024llm,
  title={Llm-powered code vulnerability repair with reinforcement learning and semantic reward},
  author={Islam, Nafis Tanveer and Khoury, Joseph and Seong, Andrew and Karkevandi, Mohammad Bahrami and Parra, Gonzalo De La Torre and Bou-Harb, Elias and Najafirad, Peyman},
  journal={arXiv preprint arXiv:2401.03374},
  year={2024}
}

@article{nijkamp2023codegen2,
  title={Codegen2: Lessons for training llms on programming and natural languages},
  author={Nijkamp, Erik and Hayashi, Hiroaki and Xiong, Caiming and Savarese, Silvio and Zhou, Yingbo},
  journal={arXiv preprint arXiv:2305.02309},
  year={2023}
}

@inproceedings{nong2025appatch,
  title={$\{$APPATCH$\}$: Automated Adaptive Prompting Large Language Models for $\{$Real-World$\}$ Software Vulnerability Patching},
  author={Nong, Yu and Yang, Haoran and Cheng, Long and Hu, Hongxin and Cai, Haipeng},
  booktitle={34th USENIX Security Symposium (USENIX Security 25)},
  pages={4481--4500},
  year={2025}
}

@inproceedings{pearce2023examining,
  title={Examining zero-shot vulnerability repair with large language models},
  author={Pearce, Hammond and Tan, Benjamin and Ahmad, Baleegh and Karri, Ramesh and Dolan-Gavitt, Brendan},
  booktitle={2023 IEEE Symposium on Security and Privacy (SP)},
  pages={2339--2356},
  year={2023},
  organization={IEEE}
}

@inproceedings{khan2024xcodeeval,
  title={Xcodeeval: An execution-based large scale multilingual multitask benchmark for code understanding, generation, translation and retrieval},
  author={Khan, Mohammad Abdullah Matin and Bari, M Saiful and Long, Do and Wang, Weishi and Parvez, Md Rizwan and Joty, Shafiq},
  booktitle={Proceedings of the 62nd Annual Meeting of the Association for Computational Linguistics (Volume 1: Long Papers)},
  pages={6766--6805},
  year={2024}
}

@article{liu2025agent,
  title={Agent That Debugs: Dynamic State-Guided Vulnerability Repair},
  author={Liu, Zhengyao and Ma, Yunlong and Xu, Jingxuan and Ai, Junchen and Gao, Xiang and Sun, Hailong and Roychoudhury, Abhik},
  journal={arXiv preprint arXiv:2504.07634},
  year={2025}
}

@article{yadavally2025large,
  title={Large Language Model Critics for Execution-Free Evaluation of Code Changes},
  author={Yadavally, Aashish and Nguyen, Hoan and Callot, Laurent and Guinet, Gauthier},
  journal={arXiv preprint arXiv:2501.16655},
  year={2025}
}

@inproceedings{luo2025unlocking,
  title={Unlocking LLM Repair Capabilities Through Cross-Language Translation and Multi-Agent Refinement},
  author={Luo, Wenqiang and Keung, Jacky Wai and Yang, Boyang and Klein, Jacques and Bissyande, Tegawende F and Tian, Haoye and Le, Bach},
  booktitle={Proceedings of the 48th IEEE/ACM International Conference on Software Engineering},
  year={2026},
  publisher={Association for Computing Machinery}

}

@techreport{keele2007guidelines,
  title={Guidelines for Performing Systematic Literature Reviews in Software Engineering},
  author={Kitchenham, Barbara A. and Charters, Stuart},
  institution={Keele University and Durham University},
  number={EBSE-2007-01},
  type={EBSE Technical Report},
  year={2007}
}

@inproceedings{wohlin2014guidelines,
  title={Guidelines for snowballing in systematic literature studies and a replication in software engineering},
  author={Wohlin, Claes},
  booktitle={Proceedings of the 18th international conference on evaluation and assessment in software engineering},
  pages={1--10},
  year={2014}
}

@article{fakih2025llm4cve,
  author       = {Mohamad Fakih and
                  Rahul Dharmaji and
                  Halima Bouzidi and
                  Gustavo Quiros Araya and
                  Oluwatosin Ogundare and
                  Mst{-}Ayesha Siddika and
                  Mohammad Abdullah Al Faruque},
  title        = {{LLM4CVE:} Enabling Iterative Automated Vulnerability Repair with
                  Large Language Models},
  booktitle    = {28th Euromicro Conference on Digital System Design, {DSD} 2025, Salerno,
                  Italy, September 10-12, 2025},
  pages        = {592--599},
  publisher    = {{IEEE}},
  year         = {2025},
  url          = {https://doi.org/10.1109/DSD67783.2025.00087},
  doi          = {10.1109/DSD67783.2025.00087},
  bibsource    = {dblp computer science bibliography, https://dblp.org}
}

@incollection{de2007ql,
  author       = {Oege de Moor and
                  Damien Sereni and
                  Mathieu Verbaere and
                  Elnar Hajiyev and
                  Pavel Avgustinov and
                  Torbj{\"{o}}rn Ekman and
                  Neil Ongkingco and
                  Julian Tibble},
  editor       = {Ralf L{\"{a}}mmel and
                  Joost Visser and
                  Jo{\~{a}}o Saraiva},
  title        = {.QL: Object-Oriented Queries Made Easy},
  booktitle    = {Generative and Transformational Techniques in Software Engineering
                  II, International Summer School, {GTTSE} 2007, Braga, Portugal, July
                  2-7, 2007. Revised Papers},
  series       = {Lecture Notes in Computer Science},
  volume       = {5235},
  pages        = {78--133},
  publisher    = {Springer},
  year         = {2007},
  url          = {https://doi.org/10.1007/978-3-540-88643-3\_3},
  doi          = {10.1007/978-3-540-88643-3\_3},
  bibsource    = {dblp computer science bibliography, https://dblp.org}
}

@inproceedings{nashid2023retrieval,
  title={Retrieval-based prompt selection for code-related few-shot learning},
  author={Nashid, Noor and Sintaha, Mifta and Mesbah, Ali},
  booktitle={2023 IEEE/ACM 45th International Conference on Software Engineering (ICSE)},
  pages={2450--2462},
  year={2023},
  organization={IEEE}
}

@article{kong2025contrastrepair,
  title={Contrastrepair: Enhancing conversation-based automated program repair via contrastive test case pairs},
  author={Kong, Jiaolong and Xie, Xiaofei and Cheng, Mingfei and Liu, Shangqing and Du, Xiaoning and Guo, Qi},
  journal={ACM Transactions on Software Engineering and Methodology},
  volume={34},
  number={8},
  pages={1--31},
  year={2025},
  publisher={ACM New York, NY}
}

@article{he2025code,
  title={From code to courtroom: Llms as the new software judges},
  author={He, Junda and Shi, Jieke and Zhuo, Terry Yue and Treude, Christoph and Sun, Jiamou and Xing, Zhenchang and Du, Xiaoning and Lo, David},
  journal={arXiv preprint arXiv:2503.02246},
  year={2025}
}

@inproceedings{gao2023makes,
  title={What makes good in-context demonstrations for code intelligence tasks with llms?},
  author={Gao, Shuzheng and Wen, Xin-Cheng and Gao, Cuiyun and Wang, Wenxuan and Zhang, Hongyu and Lyu, Michael R},
  booktitle={2023 38th IEEE/ACM International Conference on Automated Software Engineering (ASE)},
  pages={761--773},
  year={2023},
  organization={IEEE}
}

@article{li2024cleanvul,
  title={CleanVul: Automatic Function-Level Vulnerability Detection in Code Commits Using LLM Heuristics},
  author={Li, Yikun and Zhang, Ting and Widyasari, Ratnadira and Tun, Yan Naing and Nguyen, Huu Hung and Bui, Tan and Irsan, Ivana Clairine and Cheng, Yiran and Lan, Xiang and Ang, Han Wei and others},
  journal={arXiv preprint arXiv:2411.17274},
  year={2024}
}

@inproceedings{antoniades2025swe,
  title={SWE-Search: Enhancing Software Agents with Monte Carlo Tree Search and Iterative Refinement},
  author={Antoniades, Antonis and {\"O}rwall, Albert and Zhang, Kexun and Xie, Yuxi and Goyal, Anirudh and Wang, William Yang},
  booktitle={The Thirteenth International Conference on Learning Representations},
  year={2025}
}

@article{yang2025enhancing,
  title={Enhancing Repository-Level Software Repair via Repository-Aware Knowledge Graphs},
  author={Yang, Boyang and Tian, Haoye and Ren, Jiadong and Jin, Shunfu and Liu, Yang and Liu, Feng and Le, Bach},
  journal={arXiv preprint arXiv:2503.21710},
  year={2025}
}

@article{xia2025demystifying,
  title={Demystifying llm-based software engineering agents},
  author={Xia, Chunqiu Steven and Deng, Yinlin and Dunn, Soren and Zhang, Lingming},
  journal={Proceedings of the ACM on Software Engineering},
  volume={2},
  number={FSE},
  pages={801--824},
  year={2025},
  publisher={ACM New York, NY, USA}
}

@misc{orwall2024moatless,
  title={Moatless Tools},
  author={{\"{O}}rwall, Albert},
  year={2024},
  publisher={GitHub},
  howpublished={\url{https://github.com/aorwall/moatless-tools}}
}

@inproceedings{tang2024code,
  title={Code repair with LLMs gives an exploration-exploitation tradeoff},
  author={Tang, Hao and Hu, Keya and Zhou, Jin Peng and Zhong, Sicheng and Zheng, Wei-Long and Si, Xujie and Ellis, Kevin},
  booktitle={Proceedings of the 38th International Conference on Neural Information Processing Systems},
  pages={117954--117996},
  year={2024}
}

@inproceedings{xia2024automated,
  title={Automated program repair via conversation: Fixing 162 out of 337 bugs for \$0.42 each using ChatGPT},
  author={Xia, Chunqiu Steven and Zhang, Lingming},
  booktitle={Proceedings of the 33rd ACM SIGSOFT International Symposium on Software Testing and Analysis},
  pages={819--831},
  year={2024}
}

@misc{EclipseJDTLS2023,
  author       = {Eclipse Foundation},
  title        = {Eclipse JDT LS},
  year         = {2023},
  howpublished = {\url{https://projects.eclipse.org/projects/eclipse.jdt.ls}},
  note         = {Accessed 12 June 2025}
}

@misc{UniverseFlyJDTLS2023,
  author       = {Eclipse Foundation and Yuxiang Wei},
  title        = {UniverseFly/eclipse.jdt.ls: Modified Eclipse JDT LS 1.0.3},
  year         = {2023},
  version      = {1.0.3},
  doi          = {10.5281/zenodo.8278193},
  howpublished = {\url{https://doi.org/10.5281/zenodo.8278193}}
}

@inproceedings{zhao2024enhancing,
    author = {Zhao, Jiuang and Yang, Donghao and Zhang, Li and Lian, Xiaoli and Yang, Zitian and Liu, Fang},
    title = {Enhancing Automated Program Repair with Solution Design},
    year = {2024},
    isbn = {9798400712487},
    publisher = {Association for Computing Machinery},
    address = {New York, NY, USA},
    url = {https://doi.org/10.1145/3691620.3695537},
    doi = {10.1145/3691620.3695537},
    booktitle = {Proceedings of the 39th IEEE/ACM International Conference on Automated Software Engineering},
    pages = {1706--1718},
    numpages = {13},
    location = {Sacramento, CA, USA},
    series = {ASE '24}
}

@inproceedings{hu2019re,
  title={Re-factoring based program repair applied to programming assignments},
  author={Hu, Yang and Ahmed, Umair Z and Mechtaev, Sergey and Leong, Ben and Roychoudhury, Abhik},
  booktitle={2019 34th IEEE/ACM International Conference on Automated Software Engineering (ASE)},
  pages={388--398},
  year={2019},
  organization={IEEE}
}

@misc{chen2021evaluatinglargelanguagemodels,
      title={Evaluating Large Language Models Trained on Code}, 
      author={Mark Chen and Jerry Tworek and Heewoo Jun and Qiming Yuan and Henrique Ponde de Oliveira Pinto and Jared Kaplan and Harri Edwards and Yuri Burda and Nicholas Joseph and Greg Brockman and Alex Ray and Raul Puri and Gretchen Krueger and Michael Petrov and Heidy Khlaaf and Girish Sastry and Pamela Mishkin and Brooke Chan and Scott Gray and Nick Ryder and Mikhail Pavlov and Alethea Power and Lukasz Kaiser and Mohammad Bavarian and Clemens Winter and Philippe Tillet and Felipe Petroski Such and Dave Cummings and Matthias Plappert and Fotios Chantzis and Elizabeth Barnes and Ariel Herbert-Voss and William Hebgen Guss and Alex Nichol and Alex Paino and Nikolas Tezak and Jie Tang and Igor Babuschkin and Suchir Balaji and Shantanu Jain and William Saunders and Christopher Hesse and Andrew N. Carr and Jan Leike and Josh Achiam and Vedant Misra and Evan Morikawa and Alec Radford and Matthew Knight and Miles Brundage and Mira Murati and Katie Mayer and Peter Welinder and Bob McGrew and Dario Amodei and Sam McCandlish and Ilya Sutskever and Wojciech Zaremba},
      year={2021},
      eprint={2107.03374},
      archivePrefix={arXiv},
      primaryClass={cs.LG},
      url={https://arxiv.org/abs/2107.03374}, 
}

@inproceedings{lin2017quixbugs,
  title={QuixBugs: A multi-lingual program repair benchmark set based on the Quixey Challenge},
  author={Lin, Derrick and Koppel, James and Chen, Angela and Solar-Lezama, Armando},
  booktitle={Proceedings Companion of the 2017 ACM SIGPLAN international conference on systems, programming, languages, and applications: software for humanity},
  pages={55--56},
  year={2017}
}

@inproceedings{prenner2022can,
  title={Can OpenAI's codex fix bugs? an evaluation on QuixBugs},
  author={Prenner, Julian Aron and Babii, Hlib and Robbes, Romain},
  booktitle={Proceedings of the Third International Workshop on Automated Program Repair, hosted by International Conference on Software Engineering (ICSE)},
  pages={69--75},
  year={2022}
}

@article{tian2023chatgpt,
  title={Is ChatGPT the ultimate programming assistant--how far is it?},
  author={Tian, Haoye and Lu, Weiqi and Li, Tsz On and Tang, Xunzhu and Cheung, Shing-Chi and Klein, Jacques and Bissyand{\'e}, Tegawend{\'e} F},
  journal={arXiv preprint arXiv:2304.11938},
  year={2023}
}

@inproceedings{huang2024template,
  title={Template-Guided Program Repair in the Era of Large Language Models},
  author={Huang, Kai and Zhang, Jian and Meng, Xiangxin and Liu, Yang},
  booktitle={2025 IEEE/ACM 47th International Conference on Software Engineering (ICSE)},
  pages={1895--1907},
  year={2025},
  organization={IEEE Computer Society},
  doi={10.1109/ICSE55347.2025.00030}
}

@misc{gpt-3,
      title={Language Models are Few-Shot Learners}, 
      author={Tom B. Brown and Benjamin Mann and Nick Ryder and Melanie Subbiah and Jared Kaplan and Prafulla Dhariwal and Arvind Neelakantan and Pranav Shyam and Girish Sastry and Amanda Askell and Sandhini Agarwal and Ariel Herbert-Voss and Gretchen Krueger and Tom Henighan and Rewon Child and Aditya Ramesh and Daniel M. Ziegler and Jeffrey Wu and Clemens Winter and Christopher Hesse and Mark Chen and Eric Sigler and Mateusz Litwin and Scott Gray and Benjamin Chess and Jack Clark and Christopher Berner and Sam McCandlish and Alec Radford and Ilya Sutskever and Dario Amodei},
      year={2020},
      eprint={2005.14165},
      archivePrefix={arXiv},
      primaryClass={cs.CL},
      url={https://arxiv.org/abs/2005.14165}, 
}

@article{gpt-2,
  title={Language models are unsupervised multitask learners},
  author={Radford, Alec and Wu, Jeffrey and Child, Rewon and Luan, David and Amodei, Dario and Sutskever, Ilya and others},
  journal={OpenAI blog},
  volume={1},
  number={8},
  pages={9},
  year={2019}
}

@inproceedings{silva2024gitbug,
  title={Gitbug-java: A reproducible benchmark of recent java bugs},
  author={Silva, Andr{\'e} and Saavedra, Nuno and Monperrus, Martin},
  booktitle={Proceedings of the 21st International Conference on Mining Software Repositories},
  pages={118--122},
  year={2024}
}

@inproceedings{chow2024pyty,
  title={Pyty: Repairing static type errors in python},
  author={Chow, Yiu Wai and Di Grazia, Luca and Pradel, Michael},
  booktitle={Proceedings of the IEEE/ACM 46th International Conference on Software Engineering},
  pages={1--13},
  year={2024}
}

@incollection{caldiera1994goal,
  title={The Goal Question Metric Approach},
  author={Basili, Victor R. and Caldiera, Gianluigi and Rombach, H. Dieter},
  booktitle={Encyclopedia of Software Engineering},
  pages={528--532},
  year={1994},
  publisher={John Wiley \& Sons}
}

@inproceedings{jin2023inferfix,
  title={Inferfix: End-to-end program repair with llms},
  author={Jin, Matthew and Shahriar, Syed and Tufano, Michele and Shi, Xin and Lu, Shuai and Sundaresan, Neel and Svyatkovskiy, Alexey},
  booktitle={Proceedings of the 31st ACM Joint European Software Engineering Conference and Symposium on the Foundations of Software Engineering},
  pages={1646--1656},
  year={2023}
}

@inproceedings{jimenezswe,
  title={SWE-bench: Can Language Models Resolve Real-world Github Issues?},
  author={Jimenez, Carlos E and Yang, John and Wettig, Alexander and Yao, Shunyu and Pei, Kexin and Press, Ofir and Narasimhan, Karthik R},
  booktitle={The Twelfth International Conference on Learning Representations},
  year={2024}
}

@inproceedings{houlsby2019parameter,
  title={Parameter-efficient transfer learning for NLP},
  author={Houlsby, Neil and Giurgiu, Andrei and Jastrzebski, Stanislaw and Morrone, Bruna and De Laroussilhe, Quentin and Gesmundo, Andrea and Attariyan, Mona and Gelly, Sylvain},
  booktitle={International conference on machine learning},
  pages={2790--2799},
  year={2019},
  organization={PMLR}
}

@article{zhang2025patch,
  title={PATCH: Empowering Large Language Model with Programmer-Intent Guidance and Collaborative-Behavior Simulation for Automatic Bug Fixing},
  author={Zhang, Yuwei and Jin, Zhi and Xing, Ying and Li, Ge and Liu, Fang and Zhu, Jiaxin and Dou, Wensheng and Wei, Jun},
  journal={ACM Transactions on Software Engineering and Methodology},
  year={2025},
  publisher={ACM New York, NY}
}

@inproceedings{widyasari2020bugsinpy,
  title={Bugsinpy: a database of existing bugs in python programs to enable controlled testing and debugging studies},
  author={Widyasari, Ratnadira and Sim, Sheng Qin and Lok, Camellia and Qi, Haodi and Phan, Jack and Tay, Qijin and Tan, Constance and Wee, Fiona and Tan, Jodie Ethelda and Yieh, Yuheng and others},
  booktitle={Proceedings of the 28th ACM joint meeting on european software engineering conference and symposium on the foundations of software engineering},
  pages={1556--1560},
  year={2020}
}

@article{tufano2019empirical,
  title={An empirical study on learning bug-fixing patches in the wild via neural machine translation},
  author={Tufano, Michele and Watson, Cody and Bavota, Gabriele and Penta, Massimiliano Di and White, Martin and Poshyvanyk, Denys},
  journal={ACM Transactions on Software Engineering and Methodology (TOSEM)},
  volume={28},
  number={4},
  pages={1--29},
  year={2019},
  publisher={ACM New York, NY, USA}
}

@inproceedings{lester2021power,
  title={The Power of Scale for Parameter-Efficient Prompt Tuning},
  author={Lester, Brian and Al-Rfou, Rami and Constant, Noah},
  booktitle={Proceedings of the 2021 Conference on Empirical Methods in Natural Language Processing},
  pages={3045--3059},
  year={2021}
}

@inproceedings{haque2025traceprompt,
  title={Towards Effectively Leveraging Execution Traces for Program Repair with Code LLMs},
  author={Haque, Mirazul and Babkin, Petr and Farmahinifarahani, Farima and Veloso, Manuela},
  booktitle={Proceedings of the 4th International Workshop on Knowledge-Augmented Methods for Natural Language Processing},
  pages={160--179},
  year={2025}
}

@article{bouzenia2023tracefixer,
  title={Tracefixer: Execution trace-driven program repair},
  author={Bouzenia, Islem and Ding, Yangruibo and Pei, Kexin and Ray, Baishakhi and Pradel, Michael},
  journal={arXiv preprint arXiv:2304.12743},
  year={2023}
}

@inproceedings{ehsani2025hierarchical,
    author = {Ehsani, Ramtin and Parra, Esteban and Haiduc, Sonia and Chatterjee, Preetha},
    title = {Hierarchical Knowledge Injection for Improving LLM-based Program Repair},
    year = {2025},
    publisher = {IEEE Press},
    url = {https://doi.org/10.1109/ASE63991.2025.00122},
    doi = {10.1109/ASE63991.2025.00122},
    booktitle = {2025 40th IEEE/ACM International Conference on Automated Software Engineering (ASE)},
    pages = {1440--1452},
    numpages = {13},
    location = {Seoul, Korea, Republic of}
}

@inproceedings{wei2025swe,
  title={Swe-rl: Advancing llm reasoning via reinforcement learning on open software evolution},
  author={Wei, Yuxiang and Duchenne, Olivier and Copet, Jade and Carbonneaux, Quentin and Zhang, Lingming and Fried, Daniel and Synnaeve, Gabriel and Singh, Rishabh and Wang, Sida I},
  booktitle={The Thirty-ninth Annual Conference on Neural Information Processing Systems},
  year={2025}
}

@inproceedings{zhao2024repair,
  title={RePair: Automated Program Repair with Process-based Feedback},
  author={Zhao, Yuze and Huang, Zhenya and Ma, Yixiao and Li, Rui and Zhang, Kai and Jiang, Hao and Liu, Qi and Zhu, Linbo and Su, Yu},
  booktitle={Findings of the Association for Computational Linguistics: ACL 2024},
  pages={16415--16429},
  year={2024}
}

@inproceedings{dai2025less,
  title={Less is More: Adaptive Program Repair with Bug Localization and Preference Learning},
  author={Dai, Zhenlong and Chen, Bingrui and Zhao, Zhuoluo and Tang, Xiu and Wu, Sai and Yao, Chang and Gao, Zhipeng and Chen, Jingyuan},
  booktitle={Proceedings of the AAAI Conference on Artificial Intelligence},
  volume={39},
  number={1},
  pages={128--136},
  year={2025}
}

@article{ye2025process,
  title={Process-Supervised Reinforcement Learning for Code Generation},
  author={Ye, Yufan and Zhang, Ting and Jiang, Wenbin and Huang, Hua},
  journal={arXiv preprint arXiv:2502.01715},
  year={2025}
}

@article{liu2023rltf,
  title={Rltf: Reinforcement learning from unit test feedback},
  author={Liu, Jiate and Zhu, Yiqin and Xiao, Kaiwen and Fu, Qiang and Han, Xiao and Yang, Wei and Ye, Deheng},
  journal={arXiv preprint arXiv:2307.04349},
  year={2023}
}

@inproceedings{dou2024stepcoder,
  title={StepCoder: Improving Code Generation with Reinforcement Learning from Compiler Feedback},
  author={Dou, Shihan and Liu, Yan and Jia, Haoxiang and Zhou, Enyu and Xiong, Limao and Shan, Junjie and Huang, Caishuang and Wang, Xiao and Fan, Xiaoran and Xi, Zhiheng and others},
  booktitle={62nd Annual Meeting of the Association for Computational Linguistics, ACL 2024},
  pages={4571--4585},
  year={2024},
  organization={Association for Computational Linguistics (ACL)}
}

@article{shojaeeexecution,
  title={Execution-based Code Generation using Deep Reinforcement Learning},
  author={Shojaee, Parshin and Jain, Aneesh and Tipirneni, Sindhu and Reddy, Chandan K},
  journal={Transactions on Machine Learning Research},
  year={2023}
}

@article{schulman2017proximal,
  author       = {John Schulman and
                  Filip Wolski and
                  Prafulla Dhariwal and
                  Alec Radford and
                  Oleg Klimov},
  title        = {Proximal Policy Optimization Algorithms},
  journal      = {CoRR},
  volume       = {abs/1707.06347},
  year         = {2017},
  url          = {http://arxiv.org/abs/1707.06347},
  eprinttype   = {arXiv},
  eprint       = {1707.06347},
  bibsource    = {dblp computer science bibliography, https://dblp.org}
}

@article{dettmers2023qlora,
  title={Qlora: Efficient finetuning of quantized llms},
  author={Dettmers, Tim and Pagnoni, Artidoro and Holtzman, Ari and Zettlemoyer, Luke},
  journal={Advances in neural information processing systems},
  volume={36},
  pages={10088--10115},
  year={2023}
}

@misc{hu2021loralowrankadaptationlarge,
      title={LoRA: Low-Rank Adaptation of Large Language Models}, 
      author={Edward J. Hu and Yelong Shen and Phillip Wallis and Zeyuan Allen-Zhu and Yuanzhi Li and Shean Wang and Lu Wang and Weizhu Chen},
      year={2021},
      eprint={2106.09685},
      archivePrefix={arXiv},
      primaryClass={cs.CL},
      url={https://arxiv.org/abs/2106.09685}, 
}

@inproceedings{jiang2024repaircat,
  title={RepairCAT: Applying Large Language Model to Fix Bugs in AI-Generated Programs},
  author={Jiang, Nan and Wu, Yi},
  booktitle={Proceedings of the 5th ACM/IEEE International Workshop on Automated Program Repair, hosted by International Conference on Software Engineering (ICSE) 2024},
  pages={58--60},
  year={2024}
}

@article{li2023starcoder,
  title={Starcoder: may the source be with you!},
  author={Li, Raymond and Allal, Loubna Ben and Zi, Yangtian and Muennighoff, Niklas and Kocetkov, Denis and Mou, Chenghao and Marone, Marc and Akiki, Christopher and Li, Jia and Chim, Jenny and others},
  journal={arXiv preprint arXiv:2305.06161},
  year={2023}
}

@article{lozhkov2024starcoder,
  title={Starcoder 2 and the stack v2: The next generation},
  author={Lozhkov, Anton and Li, Raymond and Allal, Loubna Ben and Cassano, Federico and Lamy-Poirier, Joel and Tazi, Nouamane and Tang, Ao and Pykhtar, Dmytro and Liu, Jiawei and Wei, Yuxiang and others},
  journal={arXiv preprint arXiv:2402.19173},
  year={2024}
}

@article{wang2021codet5,
  title={Codet5: Identifier-aware unified pre-trained encoder-decoder models for code understanding and generation},
  author={Wang, Yue and Wang, Weishi and Joty, Shafiq and Hoi, Steven CH},
  journal={arXiv preprint arXiv:2109.00859},
  year={2021}
}

@article{wang2023codet5+,
  title={Codet5+: Open code large language models for code understanding and generation},
  author={Wang, Yue and Le, Hung and Gotmare, Akhilesh Deepak and Bui, Nghi DQ and Li, Junnan and Hoi, Steven CH},
  journal={arXiv preprint arXiv:2305.07922},
  year={2023}
}

@misc{guo_graphcodebert_2021,
    title = {{GraphCodeBERT}: {Pre}-training {Code} {Representations} with {Data} {Flow}},
    shorttitle = {{GraphCodeBERT}},
    url = {http://arxiv.org/abs/2009.08366},
    language = {en},
    urldate = {2023-02-15},
    publisher = {arXiv},
    author = {Guo, Daya and Ren, Shuo and Lu, Shuai and Feng, Zhangyin and Tang, Duyu and Liu, Shujie and Zhou, Long and Duan, Nan and Svyatkovskiy, Alexey and Fu, Shengyu and Tufano, Michele and Deng, Shao Kun and Clement, Colin and Drain, Dawn and Sundaresan, Neel and Yin, Jian and Jiang, Daxin and Zhou, Ming},
    month = sep,
    year = {2021},
    note = {arXiv:2009.08366 [cs]},
}

\end{document}